\documentclass[
 reprint, 
 amsmath,amssymb,
 aps,prb,
 floatfix]{revtex4-2}

\usepackage{subfigure}
\usepackage[utf8]{inputenc}
\usepackage{rotating}
\usepackage{amssymb}
\usepackage{mathdots} 
\usepackage{graphicx}
\usepackage{svg} 
\usepackage{dcolumn}
\usepackage{bm}
\usepackage{xcolor}
\usepackage[colorlinks=true,
            linkcolor=black,
            citecolor=black,
            urlcolor=black]{hyperref} 
\usepackage{placeins}

\begin{document}

\title{\textbf{Standard and Bipolar Rectification in Tilted Periodically-Driven Quantum Systems}}

\author{Joaqu{\'i}n Pe{\~{n}}uela-Parra$^{1,}$}
\email{Corresponding author: jop204@pitt.edu}
\author{Juan Jos{\'e} Mendoza-Arenas$^{1,2}$}
\email{jum151@pitt.edu} 
\affiliation{\vspace{0.2cm}
$^{1}$Department of Mechanical Engineering and Materials Science, \\
University of Pittsburgh, Pittsburgh, Pennsylvania 15261, USA \\
$^{2}$Department of Physics and Astronomy, University of Pittsburgh, Pittsburgh, Pennsylvania 15261, USA}

\date{October 14, 2026}

\begin{abstract}
    Developing mechanisms to control transport in correlated quantum systems is essential for advancing quantum circuit technologies. In this work, we establish a rectification mechanism for both particle and energy flow in a fermionic system coupled to external baths, in the presence of a tilted potential modulated locally by a time-periodic drive. The drive generates particle and energy currents even in the absence of chemical or thermal bias between the baths. These drive-induced currents are enhanced at resonant frequencies at which avoided crossings occur in the Floquet quasienergy spectrum, reflecting the role of the Floquet states as intermediate channels for transport. Once a bias between the baths is introduced, the drive-induced current contribution can reinforce, compete with, or even cancel the bias-induced contribution. These effects enable rectification in the non-interacting regime, which we quantify via the rectification coefficient $\mathcal{R}$. This coefficient measures the asymmetry between transport up the tilt (forward bias) and down the tilt (reverse bias). In addition to standard rectification, where transport is reduced under reverse bias, we identify a regime of bipolar rectification, in which the current flows only up the tilt regardless of the bias polarity. At the boundary between standard and bipolar regimes, the system achieves perfect rectification, corresponding to complete suppression of the reverse-bias current. In the presence of interactions between particles, the parameter region that supports bipolar rectification is enlarged and $\mathcal{R}$ can be enhanced by many orders of magnitude. Our results establish a rectification mechanism for particle and energy transport that can be experimentally realized in condensed matter and cold atom systems.
\end{abstract}
        
\maketitle

\section{Introduction}

    When a many-body quantum system is placed between two baths at different temperatures or chemical potentials, it can reach a non-equilibrium steady state (NESS) characterized by a thermal and/or electric current from one bath to the other. This state can exhibit remarkable physical phenomena, including non-equilibrium quantum phase transitions \cite{Oliveira2024Voltage-Driven, Znidaric2010-th, chen_non-equilibrium_2026}, anomalous current scaling \cite{RevModPhys.93.025003, PhysRevLett.117.040601, https://doi.org/10.1002/andp.201600298, PhysRevE.84.051115, PhysRevLett.106.220601, PhysRevB.90.115156, ML_Tensor_Networks}, and insulating domains \cite{Landi_Introduction_Review, 10.3389/fphy.2024.1474018, PhysRevB.80.035110, PhysRevB.87.235130}. Understanding the physics of these effects is essential for the development of technologies in electronics, spintronics \cite{Hirohata2020Review, Dieny2020Opportunities}, and caloritronics \cite{Uchida2021Transport, Uchida2026Spin}. Key applications include quantum counterparts of electronic circuit components such as quantum diodes, quantum transistors \cite{10.1063/1.2191730, PhysRevB.99.035129, Chen2023Quantum}, and quantum thermal machines \cite{Bhattacharjee2021Quantum}. As device dimensions shrink to the nanometer scale, the number of relevant energy levels becomes finite and discrete, rather than forming a continuous band, and semiclassical drift–diffusion descriptions for transport cease to be adequate \cite{nazarov2009quantum, datta2005quantum, Stefanucci_van_Leeuwen_2013, Lee2023Atomistic}. In this nanoscale regime, quantum effects such as coherence and interference can modify and even enhance device performance \cite{Chen2005A, Chen2023Quantum, Chu2019Predictive}. This dual role, as both a necessity for describing nanoscale physics and a potential advantage over classical operation, motivates the study of quantum versions of electronic circuit components.
    
    In recent years, considerable effort has been devoted to the study of quantum diodes \cite{Landi_Introduction_Review, Pereira2019Thermal}. In these systems, spatial asymmetries and nonlinearities are engineered to control transport under applied chemical potential or thermal biases. As a result, current flows predominantly in one direction (forward bias) and is suppressed when the baths are exchanged (reverse bias). The efficiency and performance of a diode are quantified by its rectification coefficient $\mathcal{R}$ and its contrast coefficient $\mathcal{C}$: 
    \begin{equation}
        \mathcal{R} = - \cfrac{ \mathcal{J}_{\text{F}}}{\mathcal{J}_{\text{R}}} \quad \textrm{and} \quad \mathcal{C} = \left| \frac{\mathcal{J}_{\text{F}} + \mathcal{J}_{\text{R}}}{\mathcal{J}_{\text{F}} - \mathcal{J}_{\text{R}}} \right|
    \end{equation}
    with $\mathcal{J}_{\text{F}}$ $(\mathcal{J}_{\text{R}})$ the current in the forward (reverse) bias. For a perfect diode $\mathcal{J}_{\text{R}} = 0$ with sizeable $\mathcal{J}_{\text{F}}$. Thus, $\mathcal{R} \rightarrow \infty$ and $\mathcal{C} = 1$. 

    \begin{figure*}[t]
        \centering
        \includegraphics[width=\textwidth]{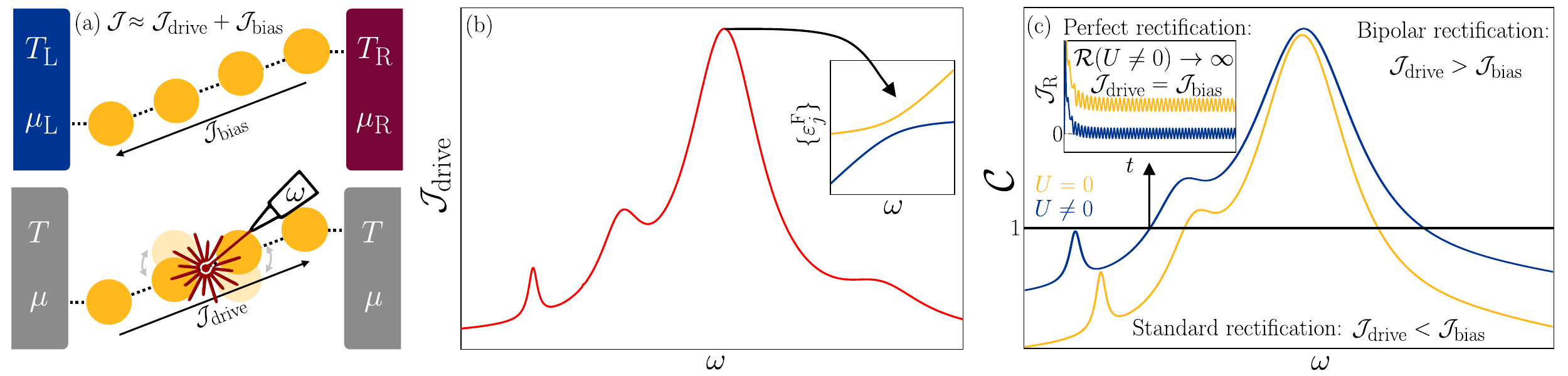}
        \caption{Qualitative illustration of the rectification mechanism and its two regimes for particle transport. (a) The two current contributions isolated separately. Top: without external driving, the chain couples to baths $L$ and $R$ at different temperatures and/or chemical potentials, so only the bias-induced current $\mathcal{J}_{\text{bias}}$ flows (from right to left in the reverse configuration). Bottom: the driven chain $S$ couples to baths at a common temperature $T$ and chemical potential $\mu$, so only the drive-induced current $\mathcal{J}_{\text{drive}}$ flows. When driving and bias act simultaneously, the total current is $\mathcal{J} \approx \mathcal{J}_{\text{drive}} + \mathcal{J}_{\text{bias}}$. (b) $\mathcal{J}_{\text{drive}}$ as a function of the frequency $\omega$ of the drive. Its peaks coincide with avoided crossings in the Floquet quasienergy spectrum $\{\varepsilon^F_j\}$ (inset). (c) Contrast coefficient $\mathcal{C}$ as a function of $\omega$, for $U=0$ (gold) and $U\neq0$ (blue). The horizontal line separates the standard regime from the bipolar regime.  The general effect of $U$ consists of shifting the resonance peaks and changing their height; this enhances $\mathcal{C}$ and widens the parameter region supporting bipolar rectification. Inset: at a fixed $\omega$ where $\mathcal{J}_{\text{drive}} \neq \mathcal{J}_{\text{bias}}$ for $U=0$, introducing interactions ($U\neq0$) can drive the reverse current $\mathcal{J}_{\text{R}}(t)$ to zero, yielding perfect rectification ($\mathcal{R}\to\infty$) at a point where it was not achieved otherwise.}
            \label{fig:summary}
    \end{figure*}
        
    Commercial semiconductor electronic diodes exhibit $\mathcal{R} \approx 10^5$ to $10^8$ \cite{RevModPhys.92.035001}. Meanwhile, other technologies remain under development, such as molecular diodes with $\mathcal{R} \approx 10^5$ \cite{chen2017molecular}, and superconducting diodes with $\mathcal{R} \approx 2 $ to $10$ \cite{Bauriedl2021Supercurrent, 10.1063/5.0283849, Golod2022Demonstration}. On the other hand, thermal diodes remain relatively underdeveloped, with most experimental realizations relying on heat transport mediated by phonons \cite{R_energy_experimental}. As a consequence, the thermal rectification achieved to date has generally remained modest. For instance, the highest experimental phonon-based thermal diode reported so far is a junction between $\textrm{Ag}_{225} \textrm{Te}_{0.9} \textrm{S}_{0.1} $ and $\textrm{Ag}_{2} \textrm{S}_{0.6} \textrm{Se}_{0.4}$ that exhibited  $\mathcal{R} = 3.3$ \cite{Hirata2025Enhanced}. An alternative approach relies on hybrid thermal diodes, in which electrons carry a large amount of heat. In such systems, much higher rectification, reaching $\mathcal{R} \approx 140$, has been obtained in structures composed of normal metals coupled to superconductors. However, their applicability is limited to sub-Kelvin temperatures \cite{martinez2015rectification}. Motivated by these experimental limitations, several theoretical works have proposed novel mechanisms capable of producing large electric and thermal rectification in quantum systems. However, these proposals often rely on complex geometries \cite{PhysRevE.99.032136, PhysRevE.103.032108, PhysRevLett.120.200603, PhysRevE.103.052143, PhysRevE.105.024120, PhysRevA.105.052605}, fine tuning of parameters \cite{tesser2022heat, PhysRevResearch.2.043343, PhysRevResearch.5.013129}, setups involving particles with different statistics \cite{PhysRevLett.123.020603, PhysRevLett.102.095503}, or idealized bias conditions such as infinite temperature and weak system–bath coupling \cite{PhysRevE.99.032116, PhysRevLett.103.140405, PhysRevE.104.054137, su2024spin, e22111311, PRX_JJ}.
    
    To achieve enhanced rectification, it is essential to develop strategies for controlling the different degrees of freedom that govern transport in quantum systems. One particularly powerful approach is the use of light-induced periodic driving, which has seen substantial development in recent years \cite{boolakee2022light, Day2024Nonperturbative, weitenberg_tailoring_2021, Ito2023Build-up, Choi2024Observation, Shan2021Giant}. By periodically modulating system parameters, one can effectively engineer new Hamiltonians and induce phenomena that have no counterpart in equilibrium, such as Floquet-engineered band structures \cite{Rudner2020-ip, Castro2022Floquet}, DC currents in the absence of static biases \cite{PhysRevA.75.063424, Koenigsmann2025Nonequilibrium}, effective interactions \cite{PhysRevLett.99.220403, PhysRevLett.115.075301, Kusari2025Periodic}, and dynamical localization \cite{Aditya2023Dynamical, Tamang2020Floquet}. These diverse non-equilibrium phenomena, together with the rapid progress in their experimental realization, motivate the exploration of periodically driven systems as a route to achieve and optimize rectification mechanisms that may be inaccessible in static configurations. 
    
    Following this approach, in this work we investigate transport through a system coupled to external baths in the presence of a tilted potential modulated locally by a time-periodic drive. Our aim is to obtain large rectification for both particle and energy flow with a setup that requires no artificial distribution of local potentials or interactions and operates even in the non-interacting regime. The simultaneous presence of both the tilted potential and its periodic driving in our model is essential. In the absence of a mismatch in particle statistics between the baths and the system \cite{Landi_Introduction_Review, PhysRevLett.102.095503, PhysRevE.80.041103}, rectification generally requires two ingredients: breaking of reflection symmetry and nonlinearities in the Hamiltonian of the system. If one of these ingredients is absent, as in the non-interacting regime with a static Hamiltonian, one generally finds $\mathcal{J}_{\text{F}} = -\mathcal{J}_{\text{R}}$, implying $\mathcal{C} = 0$ and $\mathcal{R} = 1$. We employ a tilted potential because it provides a simple and experimentally accessible way to induce spatial asymmetry. For instance, it corresponds to applying an electric field in condensed matter systems. Moreover, non-interacting dynamically-driven systems can behave outside the linear-response regime and exhibit an asymmetric response in the currents, therefore enabling finite rectification \cite{PhysRevE.99.032126}. Closest to our setup, rectification in a periodically-driven, interacting tilted lattice has been reported within a boundary-driven Lindblad description \cite{De2024Confluence}. The rectification there originates from driving the hopping at a specific interaction strength. By contrast, we obtain rectification by driving the on-site energies, both without and with interactions.
             
    Here we show that when the periodic drive is introduced, nonzero particle and energy currents emerge even in the absence of chemical or thermal bias. The resulting drive-induced current, denoted $\mathcal{J}_{\text{drive}}$ for either particle or energy transport, can be tuned via the driving frequency $\omega$, and its magnitude can be used to compensate the bias-induced current $\mathcal{J}_{\text{bias}}$. Because $\mathcal{J}_{\text{bias}}$ reverses sign with the exchange of the baths while $\mathcal{J}_{\text{drive}}$ does not, the two contributions reinforce or compete depending on the bias configuration, giving rise to rectification already in the non-interacting regime. We identify regimes of standard $(\mathcal{C} \leq 1)$ and bipolar $(\mathcal{C} > 1)$ rectification, with perfect rectification $(\mathcal{C}=1)$ occurring at their boundary. Figure~\ref{fig:summary} summarizes the competition between $\mathcal{J}_{\text{drive}}$ and $\mathcal{J}_{\text{bias}}$ that produces both standard and bipolar rectification. Panel (a) depicts the two isolated current contributions, $\mathcal{J}_{\text{drive}}$ and $\mathcal{J}_{\text{bias}}$. Panel (b) shows that $\mathcal{J}_{\text{drive}}$ is maximized at the frequencies where avoided crossings occur in the Floquet quasienergy spectrum (inset). These are the resonances at which the contrast coefficient $\mathcal{C}$ peaks. Panel (c) shows the two rectification behaviors and their robustness when interactions are included. Moreover, increasing interactions enlarge the parameter region that supports bipolar rectification. This enlargement persists at all strengths of attractive interaction and, up to a point, of repulsive interaction. For a fixed driving frequency, interactions also shift the reverse current toward zero, equivalently enhancing the rectification coefficient $\mathcal{R}$ by many orders of magnitude, as illustrated in the inset of panel (c). The same competition between particle current components illustrated in Fig.~\ref{fig:summary} applies to energy transport. 
    
    The paper is organized as follows. After introducing the model and the theoretical framework in Sec.~\ref{Model}, we present transport properties and rectification behavior in the non-interacting and interacting regimes in Secs.~\ref{Non-Interacting} and~\ref{Interacting}, respectively. Finally, conclusions and outlook are discussed in Sec.~\ref{Conclusion}. Supporting derivations, methodological details, and additional results are collected in the appendices. 

\section{Time driving Model}
\label{Model}

     We consider a system of spinless fermions subject to a tilted potential that is locally modulated by a time-periodic drive. The system Hamiltonian $\hat{H}_{\text{S}}$ is given by     
    \begin{equation}
        \hat{H}_{\text{S}}(t) = \sum_{i=1}^{D} \varepsilon_{i}(t) \hat{c}_{i}^{\dagger}\hat{c}_{i} +\sum_{i = 1}^{D-1} [t_{s}(\hat{c}_{i+1}^{\dagger}\hat{c}_{i} + \text{h.c.}) + U \hat{n}_{i}\hat{n}_{i+1}]. 
        \label{Model_EQ}
    \end{equation}  

    Here $\hat{c}_{i}^{\dagger}$ $(\hat{c}_{i})$ creates (annihilates) a fermion on site $i$, $\hat{n}_{i} = \hat{c}^{\dagger}_{i} \hat{c}_{i}$ is the particle-number operator, $\varepsilon_{i}(t)$ is the time-dependent on-site energy, $U$ is the nearest-neighbor density-density interaction, $t_{s}$ is the hopping parameter, and h.c. refers to the Hermitian conjugate. We consider a local time-dependent driving potential with frequency $\omega$ given by
    \begin{equation}
        \varepsilon_{i}(t) = \varepsilon_{i}  \quad \textrm{and} \quad \varepsilon_{j}(t) = \varepsilon_{j} \cos(\omega t).
        \label{local_driving}
    \end{equation}
    With $i \in \{1,4\}$ and $j \in\{2,3\}$ for $D =4 $, this driving corresponds to static on-site potentials at the boundaries and a time-periodic modulation of the central sites. The on-site energy of site $k$ is $\varepsilon_{k} = E \left(k - \frac{D+1}{2} \right)$ with $E$ as the tilt strength. For the driven sites, introducing a driving amplitude independent of $\varepsilon_{j}$ yields no qualitatively new phenomenology. We therefore set the amplitude equal to $\varepsilon_{j}$. This choice minimizes the number of independent parameters and ensures a smooth connection to the static tilted potential in the absence of driving $(\omega = 0)$. The shift by $\frac{D+1}{2}$ in $\varepsilon_{k}$ sets the energy reference such that the on-site potential is centered around zero, corresponding to an effective chemical potential $\mu = \frac{E(D+1)}{2}$. As a result, each driven site $j$ and the site symmetric to it about the center of the chain, at position $D+1-j$, have on-site energies of equal magnitude and opposite sign.
        
    In experimental platforms such as quantum-dot arrays or cold-atom systems, the time-dependent local potentials can be generated by externally applied electric fields, or by light locally illuminating the center of the system (see Fig.~\ref{Fig. Light Setup}). The latter is discussed in Appendix~\ref{Driving_from_light}, where we demonstrate that Eq.~\eqref{local_driving} is an effective description of a periodic train of identical Gaussian laser pulses that captures the essential physics of the light-driven setup while simplifying the Floquet analysis.
    
    \begin{figure}[t]
        \includegraphics[width=0.4\paperwidth]{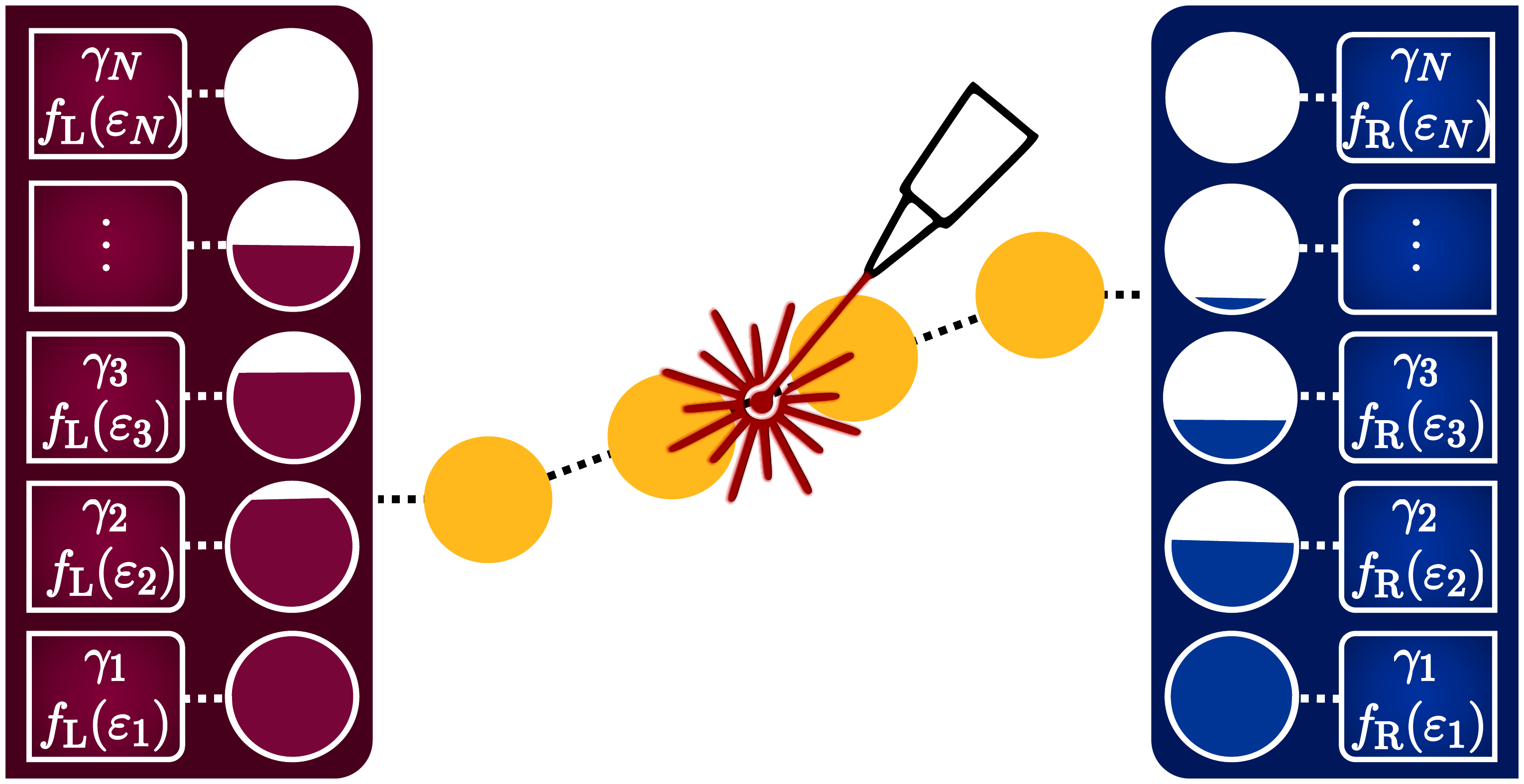}
        \caption{Schematic of the setup. The system $S$ (gold circles) is a chain coupled at its boundaries to two baths, $L$ (maroon) and $R$ (blue), each modeled using the mesoscopic-leads approach: a finite collection of modes $\varepsilon_k$, each damped towards a thermal occupation $f_{L(R)}(\varepsilon_k)$ at a rate $\gamma_{k}$, indicated by the filling of each circle. A laser (represented by the emitting tip) illuminates the central sites of $S$.}
        \label{Fig. Light Setup}
    \end{figure}      

    In the absence of time driving, a non-interacting system with a global tilt larger than the Bloch-band width
    \begin{equation}
        E(D-1) > 4|t_s|
        \label{localization_threshold}
    \end{equation}
    displays Wannier–Stark localization, and behaves as an insulator in the thermodynamic limit $(D \rightarrow \infty)$ \cite{ML_Swolak, 10.3389/fphy.2024.1474018}. This localization threshold depends jointly on the tilt strength and the system size. Namely, for a fixed chain length $D$, increasing $E$ suppresses the current even before the thermodynamic limit is reached. Once interactions are included, this localizing tendency becomes asymmetric with respect to the bias direction: current-blocking particle domains form only when the net current flows downwards with respect to $E$, while transport remains largely unaffected in the opposite direction. The interplay between $E$ and $U$ can thus lead to giant rectification of particle transport \cite{PRX_JJ}. However, this result relies on the assumption of maximal boundary bias and infinite temperature in the Lindblad equation, and its applicability to more realistic scenarios is limited. To study rectification beyond this idealized boundary-driven Lindblad bias scheme, a different framework is needed. This framework must enable the calculation of particle and energy flows through a system coupled to (possibly non-Markovian) finite-temperature baths. For this, we use the mesoscopic leads approach, also known as the extended reservoir approach \cite{ML_Swolak}. This recently proposed method is capable of tackling many-body systems at finite temperature, strong system-bath coupling, and non-quadratic interactions (in combination with tensor network techniques) \cite{ML_Tensor_Networks, ML_Swolak, Rams2020Breaking}. The application of the mesoscopic leads method replicates the well-known Landauer–Büttiker formalism in a non-interacting regime as well as the exact non-equilibrium Green's function (NEGF) technique from scattering theory \cite{ML_Subleties}. Recently, the mesoscopic leads description has been generalized to deal with time-dependent external driving without relying on perturbative techniques. This generalization offers a simple and elegant way to go beyond NEGF methods and standard weak-coupling master equations for studying non-equilibrium systems \cite{ML_Fast_Driving}.

\subsection{Mesoscopic leads approach}
\label{ML}

\subsubsection{Quantum master equation}

     We consider a system with Hamiltonian $\hat{H}_\text{S}$ coupled to one infinite bath $\alpha$ on each side, with $\alpha = \{\text{L}, \text{R} \}$ for left and right. The extended system+baths Hamiltonian is
    \begin{equation}
    \begin{aligned}
        \hat{H} & = \hat{H}_{\text{S}} + \sum_{\alpha = \text{L}, \text{R}} \hat{H}_{\alpha} + \hat{H}_{\text{S}, \alpha} \\
        & = \hat{H}_{\text{S}} +\sum_{\alpha = \text{L}, \text{R}}\sum_{m = 1}^{\infty} [\omega_{m}^{\alpha} \hat{b}_{m,\alpha}^{\dagger} \hat{b}_{m,\alpha} + ( \lambda_{m}^{\alpha} \hat{c}_{p}^{\dagger} \hat{b}_{m,\alpha} + h.c.) ].
    \end{aligned}
    \label{H_beforeML}
    \end{equation}

    Here, $\hat{b}^{\dagger}_{m,\alpha}$ ($\hat{b}_{m,\alpha}$) creates (annihilates) a fermion with energy $\omega_{m}^{\alpha}$ on bath $\alpha$, subindex $p$ denotes the system site connected to the bath (for a chain of size $D$, $p=1$ for $\alpha = \text{L}$ and $p = D$ for $\alpha = \text{R}$), and $\lambda_{m}^{\alpha}$ is the coupling between the system and mode $m$ of bath $\alpha$. 

    As each bath contains an infinite number of modes, simulating the dynamics of the system under the full Hamiltonian in Eq.~\eqref{H_beforeML} is in general unfeasible. A standard solution for this problem is to trace out the degrees of freedom of the baths. Instead of using a standard (usually heuristic) Lindblad equation where each bath is simply represented by the action of a jump operator, the mesoscopic leads formulation keeps a finite collection of damped modes in the process, as shown in Fig.~\ref{Fig. Light Setup}. At the cost of having an extended system with a larger number of sites, we gain access to the system+baths entanglement and non-Markovian physics. Such effects are usually ignored in standard boundary-driven Lindblad equations or NEGF calculations.   

    Each mesoscopic lead consists of $N$ modes, with energies $\varepsilon_{1}, \ldots , \varepsilon_{N}$. Mode $k$ of each lead is tunneling-coupled to the corresponding site $p$ of the system with an effective hopping $\kappa_{kp}$ and damped toward thermal equilibrium with a damping rate $\gamma_{k}$. These parameters can be different for each lead; however, we assume identical on-site energies, damping rates, and effective hopping parameters for both leads. Then, $\hat{H}$ is approximated as
    \begin{equation}
    \begin{aligned}
        \hat{H} & = \hat{H}_{\text{S}} + \sum_{\alpha = \text{L}, \text{R}} \hat{H}_{\alpha} + \hat{H}_{\text{S}, \alpha}  \\
        & = \hat{H}_{\text{S}} + \sum_{\alpha = \text{L}, \text{R}}\sum_{k=1}^{N} [\varepsilon_{k} \hat{a}_{k, \alpha}^{\dagger} \hat{a}_{k, \alpha} + (\kappa_{kp} \hat{a}_{k, \alpha}^{\dagger} \hat{c}_{p} + h.c.)].
    \end{aligned}
    \label{H_ML}
    \end{equation}
    Here, $\hat{a}^{\dagger}_{k,\alpha}$ ($\hat{a}_{k,\alpha}$) creates (annihilates) a fermion with energy $\varepsilon_{k}$ in mode $k$ of lead $\alpha$.

    The time evolution of the density matrix $\hat{\rho}(t)$ that describes this extended system, and which captures the damping of the lead modes, is given by a Lindblad-type quantum master equation:
    \begin{equation}
        \frac{d \hat{\rho}(t)}{dt} = - i [\hat{H}, \hat{\rho}(t)] + \sum_{\alpha = \text{L}, \text{R}}\mathcal{L}_{\alpha} \{ \hat{\rho} (t) \}.
        \label{QME}
    \end{equation}     
    Here, square brackets denote the commutator of two operators, $[\hat{A},\hat{B}] = \hat{A}\hat{B} - \hat{B}\hat{A}$. The dissipator $\mathcal{L}_{\alpha} \{ \hat{\rho} \}$ is given by
    \begin{equation}
        \begin{aligned}
        \mathcal{L}_\alpha\{\hat{\rho}\}= & \sum_{k=1}^N \gamma_k\left(1-f_{\alpha}(\varepsilon_{k})\right)\left[\hat{a}_{k , \alpha} \hat{\rho} \hat{a}_{k , \alpha}^{\dagger}-\frac{1}{2}\left\{\hat{a}_{k , \alpha}^{\dagger} \hat{a}_{k , \alpha}, \hat{\rho}\right\}\right] \\
        & +\sum_{k=1}^N \gamma_k f_{\alpha}(\varepsilon_{k})\left[\hat{a}_{k , \alpha}^{\dagger} \hat{\rho} \hat{a}_{k , \alpha}-\frac{1}{2}\left\{\hat{a}_{k , \alpha} \hat{a}_{k , \alpha}^{\dagger}, \hat{\rho}\right\}\right],
        \end{aligned}
        \label{Dissipators}
    \end{equation}
    where curly braces denote the anticommutator, $\{\hat{A},\hat{B}\} = \hat{A}\hat{B} + \hat{B}\hat{A}$. The dissipator thermalizes the lead energy modes towards the Fermi-Dirac distribution $f_{\alpha}(\omega) = (e^{\beta_{\alpha}(\omega - \mu_{\alpha})}+1)^{-1}$ with the inverse temperature $\beta_{\alpha} = T_{\alpha}^{-1}$ and chemical potential $\mu_{\alpha}$ of bath $\alpha$.

    \subsubsection{Effective spectral density}

    It is well known that the dynamics of the system is controlled by the spectral density of the bath $\mathcal{J}(\omega) = 2 \pi \sum_{m=1}^{\infty} | \lambda_{m}|^{2} \delta(\omega - \omega_{m})$ \cite{stefanini_is_2025}. The mesoscopic leads approach is based on the idea that $\mathcal{J}(\omega)$ can be approximated by a sum of Lorentzian spectral densities as
    \begin{equation}
        \mathcal{J}^{\textrm{eff}} (\omega) = \sum_{k = 1}^{N} \frac{|\kappa_{kp}|^{2} \gamma_{k}}{(\omega -\varepsilon_{k})^{2} + (\gamma_{k}/2)^{2}} .
        \label{Lorentzian_J}
    \end{equation}

    Thus, we approximate an infinite bath by a collection of modes (a mesoscopic lead) where each mode is independently connected to a residual bath described by a Lorentzian spectral density $\mathcal{J}^{\textrm{eff}}_{k} (\omega) =|\kappa_{kp}|^{2} \gamma_{k} /[(\omega -\varepsilon_{k})^{2} + (\gamma_{k}/2)^{2}]$. The parameters $\varepsilon_{k}, \gamma_{k}$ and $\kappa_{kp}$ should be chosen so $\mathcal{J}(\omega) \approx \mathcal{J}^{\textrm{eff}} (\omega)$. Typically, the spectral density vanishes for bath modes whose energy is very far from the system eigenvalues. Thus, we can assume $\mathcal{J}(\omega) = 0 $ if the mode energies lie outside an energy window that covers the eigenvalues of $\hat{H}_{\text{S}}$, namely, $\omega \notin [-W, W]$. The standard strategy for choosing the mesoscopic lead parameters is to start with an arbitrary collection of $N$ modes with energies $\varepsilon_{k} \in [-W,W]$, obtained for example from a linear discretization, and then define $\gamma_{k} = \varepsilon_{k+1} - \varepsilon_{k}$ and $\kappa_{kp} = \sqrt{\mathcal{J}(\varepsilon_{k}) \gamma_{k}/2\pi}$ \cite{ML_Tensor_Networks, PhysRevB.94.155142}. For large enough $N$, $\mathcal{J}(\omega) \approx \mathcal{J}^{\textrm{eff}} (\omega)$. In our setup (Fig.~\ref{Fig. Light Setup}), we consider a flat spectral density for both baths: $\mathcal{J}(\omega ) = \Gamma $ for $\omega \in [-W, W]$ and zero everywhere else. Unless otherwise stated, we fix  $\Gamma = |t_{s}|$ and $W =8 |t_{s}|$. From now on, we set $t_{s} = 1$ and express all parameters in units of this energy scale.    

    \subsubsection{NESS observables}
    
    To calculate the transport properties of the system, we obtain the particle $(\mathcal{J}^{P}_{\alpha \rightarrow \text{S}})$ and energy $(\mathcal{J}^{E}_{\alpha \rightarrow \text{S}})$ currents from the lead $\alpha$ into the system using the corresponding continuity equations (see Appendix~\ref{All_Currents} for details \cite{ML_Fast_Driving}). We define the currents as the rates at which particles and energy are transferred from the lead to the system, and can be written as    
    \begin{subequations}
    \label{J}
    \begin{align}
    \mathcal{J}^{P}_{\alpha \to \mathrm{S}} 
    &= i \sum_{k=1}^{N} \kappa_{kp} 
    \langle \hat{a}_{k , \alpha}^{\dagger}\hat{c}_{p} - \hat{c}_{p}^{\dagger}\hat{a}_{k , \alpha} \rangle \textrm{, and}
    \label{J:a} \\
    \mathcal{J}^{E}_{\alpha \to \mathrm{S}} 
    &=
    \begin{aligned}[t]
    &i \sum_{k=1}^{N} \varepsilon_{k} \kappa_{kp} 
    \langle \hat{a}_{k , \alpha}^{\dagger}\hat{c}_{p} - \hat{c}_{p}^{\dagger}\hat{a}_{k , \alpha} \rangle \\
    &\quad - \frac{1}{2} \sum_{k=1}^{N} \gamma_{k} 
    \langle \kappa_{kp} \hat{c}_{p}^{\dagger}\hat{a}_{k , \alpha} 
    + \kappa_{kp}^{*}\hat{a}_{k , \alpha}^{\dagger}\hat{c}_{p} \rangle .
    \end{aligned}
    \label{J:b}
    \end{align}
    \end{subequations}        

    In the NESS ($\lim_{t\rightarrow \infty} \hat{\rho}(t)$) of a time-independent Hamiltonian, the currents are spatially homogeneous, satisfying $\mathcal{J}^{P,E}_{\text{L} \rightarrow \text{S}} = -\mathcal{J}^{P,E}_{\text{R} \rightarrow \text{S}}$. However, for a time-periodic system $\hat{H}_{\text{S}}(t) = \hat{H}_{\text{S}}(T+t)$, a stationary solution satisfying $d \hat{\rho}(t)/dt = 0$ generally does not exist. Instead, the long-time physics is described by a periodic NESS, for which $\hat{\rho}(t) = \hat{\rho}(T+t) $. We denote by $\bar{\mathcal{O}}$ the cycle-averaged value of $\mathcal{O}$ after averaging over several periods of time. In this regime, the relation $\bar{\mathcal{J}}^{P}_{\text{L} \rightarrow \text{S}} = -\bar{\mathcal{J}}^{P}_{\text{R} \rightarrow \text{S}}$ holds, whereas the instantaneous currents $\mathcal{J}^{P}_{\text{L} \rightarrow \text{S}}(t)$ and $-\mathcal{J}^{P}_{\text{R} \rightarrow \text{S}}(t)$ differ. By contrast, the analogous relation does not hold for the energy currents $\bar{\mathcal{J}}^{E}_{\alpha \rightarrow \text{S}}$, because the energy flow includes a contribution from the work done by the drive \cite{ML_Fast_Driving}. Nonetheless, energy rectification can still be addressed by focusing on the flow from a single lead. 
    
    Finally, we also consider thermodynamic quantities, which characterize the energetic cost and irreversibility of the driven dynamics. In particular, we define the change in the internal energy of the system,
    \begin{equation}
        \delta U(t) = \langle \hat{H}_{\text{S}}(t) + \sum_{\alpha} \hat{H}_{S, \alpha} (t) \rangle  - \langle \hat{H}_{\text{S}}(0) + \sum_{\alpha} \hat{H}_{S, \alpha} (0) \rangle,
        \label{System_internal_energy_change}
    \end{equation}
    and the entropy production rate of the system
    \begin{equation}
        \dot{\Sigma}(t) = \dot{S}_{S}(t) - \sum_{\alpha}\beta_{\alpha} \mathcal{J}_{\alpha \to S}^{Q}.
        \label{System_entropy_production_rate}
    \end{equation}

    Here $S_{\text{S}}(t) = -\textrm{Tr}(\hat{\rho}_{\text{S}}(t) \log (\hat{\rho}_{\text{S}}(t)))$ is the von-Neumann entropy of the system, $\hat{\rho}_{\text{S}}$ is the reduced density matrix of the system, and $\mathcal{J}_{\alpha \to \text{S}}^{Q} = \mathcal{J}_{\alpha \to \text{S}}^{E} - \mu_{\alpha} \mathcal{J}_{\alpha \to \text{S}}^{P}$ is the heat current flowing from the lead $\alpha$ into the system.

    At this point, we have all the ingredients needed to study the dynamics and transport properties of our model. Since the Hamiltonian is time-periodic, it is useful to briefly introduce Floquet theory. Incorporating this framework clarifies how periodic driving restructures the spectrum, induces resonant transport channels, and thus provides the intuition needed to interpret the results discussed in the following sections.
    
    \subsection{Floquet Theory}
    \label{Floquet}

    \subsubsection{Quasienergies and the stroboscopic picture}

    The dynamics of a closed (isolated) quantum system subject to a time-dependent drive is governed by the time-dependent Schrödinger equation
    \begin{equation}
          \cfrac{d}{dt} | \psi (t) \rangle = - i \hat{H}(t) |\psi (t) \rangle.
        \label{Schrodinger}
    \end{equation}
    
    Here $| \psi (t) \rangle$ is the quantum state and $\hat{H}(t)$ is the Hamiltonian at a given time $t$. The formal solution of Eq.~\eqref{Schrodinger} is given by $|\psi(t) \rangle =  \hat{\mathcal{U}} (t, t_{0}) | \psi(t_{0}) \rangle$ where the time evolution operator from time $t_0$ to time $t$ is defined as
    \begin{equation}
     \hat{\mathcal{U}}(t, t_{0}) = \mathcal{T} \exp \left( -i \int_{t_{0}}^{t} dt' \hat{H}(t') \right),
     \label{Propagator_Ht}
    \end{equation}
    
    with $\mathcal{T}$ representing the time-ordered product. For a time-periodic system, the dynamics governed by Eq.~\eqref{Propagator_Ht} can be analyzed using Floquet theory, which is the temporal analog of Bloch theory in solid-state physics \cite{Floquet_Tsuji}. If $\hat{H}(t)$ possesses time periodicity, i.e., $\hat{H}(t + T) = \hat{H}(t)$ with period $T = 2 \pi / \omega$, the Floquet theorem states that the dynamics can be separated into a stroboscopic part that describes the long-time dynamics and a micromotion part with information about the short-time scale \cite{ikeda_nonequilibrium_2021}. The basic idea consists of splitting the time interval from $t_{0}$ to $t$ into $t - t_{0} = m T + \tau$, where $m \in \mathbb{N}$ and $\tau \in [0, T)$ is the residual time. The time evolution operator can be written as $\hat{\mathcal{U}}(t, t_{0}) =  \hat{\mathcal{U}}(t_{0} + \tau, t_{0})(\hat{\mathcal{U}}_{\text{F}})^{m}$ with $\hat{\mathcal{U}}_{\text{F}} = \hat{\mathcal{U}}(t_{0} + T, t_{0}) $ the Floquet evolution operator. At stroboscopic times $(\tau = 0, t = t_{0} + mT )$, the micromotion term $\hat{\mathcal{U}}(t_{0} + \tau, t_{0})$ reduces to the identity, so the time evolution operator simplifies to $\hat{\mathcal{U}}(t_{0}  + mT, t_{0}) = (\hat{\mathcal{U}}_{\text{F}})^m $. This motivates the definition of the stroboscopic time-independent effective Hamiltonian
    \begin{equation}
       \hat{H}_{\text{eff}} = \frac{i}{T}\ln ( \hat{\mathcal{U}}_{\text{F}} ).
       \label{H_eff}
    \end{equation}
    
    Since $\hat{\mathcal{U}}_{\text{F}}$ is unitary, its eigenvalues are complex numbers that lie on the unit circle \cite{viebahn_introduction_2020}.  These eigenvalues, known as Floquet multipliers, can be written as $\{ e^{-i \varepsilon_{j}^{\text{F}} T}\}$. The real numbers $\{ \varepsilon_{j}^{\text{F}} \}$ receive the name of Floquet quasienergies. From the definition of $\hat{H}_{\text{eff}}$ it is clear that the quasienergies are also its eigenvalues:
    \begin{equation}
        \hat{H}_{\text{eff}}| \varepsilon_{j}^{\text{F}} \rangle = \varepsilon_{j}^{\text{F}} | \varepsilon_{j}^{\text{F}}\rangle.
        \label{quasienergies}
    \end{equation}

    Expanding the initial state as $| \psi(t_{0}) \rangle = \sum_{j} c_{j} |\varepsilon_{j}^{\text{F}} \rangle $ and defining the Floquet modes as $|\phi_{j}(t) \rangle = e^{i \varepsilon_{j}^{\text{F}}t} \hat{\mathcal{U}}(t, 0) | \varepsilon_{j}^{\text{F}} \rangle$, we rewrite the general solution of Eq.~\eqref{Schrodinger} as
     \begin{equation}
        | \psi (t) \rangle = \sum_{j} c_{j }e^{- i \varepsilon_{j}^{\text{F}}t} |\phi_{j} (t) \rangle . 
        \label{Floquet state}
    \end{equation}
    
    Here the Floquet modes inherit the periodicity of the system, i.e., $| \phi_{j} (t)\rangle = | \phi_{j} (t + T)\rangle $, and satisfy the instantaneous Floquet-Schrödinger equation
    \begin{equation}
    \left(H(t) - i\frac{d}{dt} \right)   | \phi_{j} (t)\rangle = \varepsilon^{\text{F}}_{j} | \phi_{j} (t)\rangle .
    \label{eq:floquet_schrodinger}
    \end{equation}
    A subtle but essential point is that the quasienergies are defined only modulo the driving frequency $\omega$. For convenience, one typically restricts the quasienergies to the interval $- \omega/2   <  \varepsilon_{j}^{\text{F}} \leq \omega/2$, which plays the role of a Floquet first Brillouin zone \cite{Floquet_Tutorial}.

    \subsubsection{The Floquet Hamiltonian in Sambe space}
    
    This analogy with solid-state physics is not accidental: the time periodicity leads to a band structure in quasienergy. This structure emerges clearly once we move to the infinite-dimensional Fourier (Sambe) space, the temporal analogue of reciprocal space. As both $|\phi_{j} (t) \rangle$ and $H(t)$ share the same period, they admit Fourier expansions: $  | \phi_{j}(t) \rangle = \sum_{n} e^{-  i \omega n t} |\phi_{j}^{n}\rangle$ and $\hat{H}(t) = \sum_{n} e^{-  i \omega n t } \hat{H}_{n} $ with components $| \phi_{j}^{n} \rangle = T^{-1} \int_{0}^{T} dt e^{  i \omega n t} |\phi_{j} (t) \rangle$ and $\hat{H}_{n} = T^{-1} \int_{0}^{T} dt e^{  i \omega n t} \hat{H}(t)$. Substituting these expansions into Eq.~\eqref{Schrodinger}, we convert the time-dependent Schrödinger equation into a time-independent equation \cite{Floquet_Tsuji},
    \begin{equation}
        \sum_{n} (  \hat{H}_{m -n} - m \omega \delta_{nm}) |\phi_{j}^{n} \rangle = \varepsilon_{j}^{\text{F}} | \phi_{j}^{m} \rangle,
        \label{HF_before_extending}
    \end{equation}
    which can be recast as an eigenvalue problem by treating the momentum number $n \in \mathbb{Z}$ as a quantum number and extending the Fourier components of the Floquet modes as $|\phi_{j}^{n} \rangle \rightarrow |\phi_{j}, n \rangle = | \phi_{j}^{n} \rangle \otimes |n\rangle$. Here the tensor product makes explicit that the extended Floquet modes $\{ |\phi_{j}, n \rangle \}$ exist in an enlarged Hilbert space, namely the Sambe space. Let $\mathbb{T}$ denote the space generated by the Fourier modes of functions with period $T$. For example, a function $f(t) = \sum_{n} f_{n} e^{-in \omega t} $ is represented as $ (\cdots, f_{-1}, f_{0}, f_{1}, \cdots ) \in \mathbb{T}$. Thus, if $\mathbb{H}$ is the Hilbert space of the system, then $|\phi_{j}, n \rangle \in \mathbb{H} \otimes \mathbb{T}$. Within this description, the Floquet Hamiltonian is defined through the matrix elements $( \hat{\mathcal{H}}_{\text{F}} )_{mn} = \hat{H}_{m -n} - m \omega \delta_{nm}$, and the eigenvalue problem can be written as

    \setlength{\arraycolsep}{1pt}
    \begin{equation}
    \begin{pmatrix}
    \ddots & \vdots & \vdots & \vdots &  \\
    \cdots & \hat{H}_0+\omega & \hat{H}_{-1} & \hat{H}_{-2} & \cdots \\
    \cdots & \hat{H}_{1} & \hat{H}_0  & \hat{H}_{-1} & \cdots \\
    \cdots & \hat{H}_{2} & \hat{H}_{1} & \hat{H}_0 - \omega & \cdots \\
     & \vdots & \vdots & \vdots & \ddots
    \end{pmatrix}
    \begin{pmatrix}
    \vdots\\
    |\phi_j^{-1}\rangle\\
    |\phi_j^{0}\rangle\\
    |\phi_j^{1}\rangle\\
    \vdots
    \end{pmatrix}
    =
    \varepsilon_{j}^{\text{F}}
    \begin{pmatrix}
    \vdots\\
    |\phi_j^{-1}\rangle\\
    |\phi_j^{0}\rangle\\
    |\phi_j^{1}\rangle\\
    \vdots
    \end{pmatrix}.
    \end{equation}

    In this representation, the band structure of the quasienergies becomes transparent: the spectrum of $\hat{\mathcal{H}}_{\text{F}}$ consists of the ladder of bands $\{ \varepsilon_{j}^{\text{F}} - n \omega\}$, where the sector $n = 0$ corresponds to the first Brillouin zone associated to $\hat{H}_{\text{eff}}$. The advantage of this Hamiltonian is that it is fully time-independent while still reproducing the exact dynamics at all times, in contrast to $\hat{H}_{\text{eff}}$, which is exact only at stroboscopic times. Thus, the time independence is achieved at the price of enlarging the Hilbert space. By definition, $\hat{\mathcal{H}}_{\text{F}}$ is infinite-dimensional. In practice, the Sambe space is truncated by retaining a finite number of harmonics, increasing their number until finding convergence in the simulations. 
    
    \subsubsection{Transition probabilities and Floquet channels}
    
    Working in the Sambe space provides several practical advantages over the stroboscopic description based on the effective Hamiltonian of Eq.~\eqref{H_eff}. In particular, it enables the computation of transition probabilities between physical states under periodic driving, making transparent how the drive modifies transport pathways. Suppose the system is initialized in a state $|\alpha \rangle \in \mathbb{H}$ at time $t = t_0$. The probability of finding it in another state $| \beta \rangle \in \mathbb{H}$ after a measurement at time $t$, denoted by $P_{\alpha \rightarrow \beta}(t-t_0)$, can be obtained by evolving the static initial state $|\alpha, 0 \rangle$ and projecting onto the state $|\beta, n \rangle$. The projection must include an implicit sum over all Fourier (momentum) sectors, since the measurement does not resolve the harmonic index $n$ \cite{, Floquet_Tutorial}. This yields
    \begin{equation}
         P_{\alpha \rightarrow \beta} (t - t_{0}) = \sum_{n} |\langle \beta, n| e^{-i \hat{\mathcal{H}}_{\text{F}}(t-t_{0})}| \alpha, 0 \rangle|^{2} .
    \end{equation}    

    We diagonalize the Floquet Hamiltonian as $\hat{\mathcal{H}}_{\text{F}} = \sum_{{j}} \lambda^{\text{F}}_{j}|\lambda^{\text{F}}_{j} \rangle \langle \lambda^{\text{F}}_{j} |$. Since $\hat{H}(t)$ is Hermitian for all $t$, $\hat{\mathcal{H}}_{\text{F}}$ is Hermitian as well, and its eigenstates $\{ | \lambda_{j}^{\text{F}} \rangle\}$ form a complete orthonormal basis of the Sambe space. Inserting this completeness relation into $P_{\alpha \rightarrow \beta}(t-t_0)$ and defining $A_{\beta n}^{\lambda^{\text{F}}_{j}} = \langle \beta, n| \lambda^{\text{F}}_{j} \rangle \langle \lambda^{\text{F}}_{j} | \alpha, 0 \rangle$, we find that the probability takes the form
    \begin{equation}
        P_{\alpha \rightarrow \beta} (t - t_{0}) = \sum_{n, j, j'} A_{\beta n}^{\lambda^{\text{F}}_{j}} \left[ A_{\beta n}^{\lambda^{\text{F}}_{j'}} \right]^{*} e^{-i (\lambda^{\text{F}}_{j} - \lambda^{\text{F}}_{j'})(t-t_{0})} .
    \end{equation} 
    
    In the long-time limit, all oscillatory terms average to zero unless $\lambda^{\text{F}}_{j} = \lambda^{\text{F}}_{j'}$, leaving the cycle-averaged transition probability
    \begin{equation}
        \bar{P}_{\alpha \rightarrow \beta} = \sum_{n, j} | \langle\beta , n| \lambda^{\text{F}}_{j} \rangle \langle\lambda^{\text{F}}_{j}|\alpha, 0\rangle |^2  .
        \label{Transition_probability}
    \end{equation}

    This expression makes explicit that the periodic drive generates a set of Floquet states $\{ | \lambda^{\text{F}}_{j} \rangle \}$ that act as intermediate channels connecting different physical states. These channels enable transitions that may be strongly suppressed, or even forbidden, in the absence of driving. Because the quasienergy spectrum and its eigenstates depend explicitly on the driving frequency, the overlap between the Floquet states and the physical states connecting the two leads varies with $\omega$. As we show in the next section, this frequency dependence produces a regime in which transport is enhanced, reflecting an optimal overlap between the drive-induced Floquet channels and the intrinsic pathways for conduction of the system.

\section{Non-interacting systems} 
\label{Non-Interacting}

    In the non-interacting regime ($U=0$), the Hamiltonian is quadratic, and the dynamics can be solved exactly. Because the evolution preserves Gaussianity, the full many-body problem reduces to the time evolution of the single-particle density matrix $\mathbf{C}(t)$, which is governed by the Lyapunov equation described in Eq.~\eqref{Lyapunov}. This allows us to access the full-time evolution and to efficiently treat large system sizes, as is required within the mesoscopic leads approach to obtain an accurate representation of the spectral densities of the baths, $\mathcal{J}(\omega) \approx \mathcal{J}^{\mathrm{eff}}(\omega)$. As a result, the non-interacting regime provides a controlled setting to analyze transport properties and to isolate the role of the driving frequency in the emergence of rectification effects. 
    
    Throughout this section, we will use bold notation in the Hamiltonian  (e.g., $\mathbf{H}$) to denote its single-particle matrix representation defined in Eq.~\eqref{single-particle_H}. Although this may resemble a restriction to the single-particle symmetry sector, the NESS remains a many-body mixed state, and our calculations fully account for the particle-number non-conserving dissipation of Eq.~\eqref{Dissipators}.

    For $N > 40$, the effective spectral density approximates the spectral density of the bath well, and the currents are insensitive to $N$ (see Appendix~\ref{About_parameters}). We nonetheless use $N = 140$ in this section because the smaller spacing between lead modes sharpens the peaks of lead-mode-resolved quantities at a manageable computational cost.
                
\subsection{Currents controlled by frequency}
\label{Currents controlled by frequency}

    \begin{figure}[t]
        \centering
        \includegraphics[width=0.4 \paperwidth]{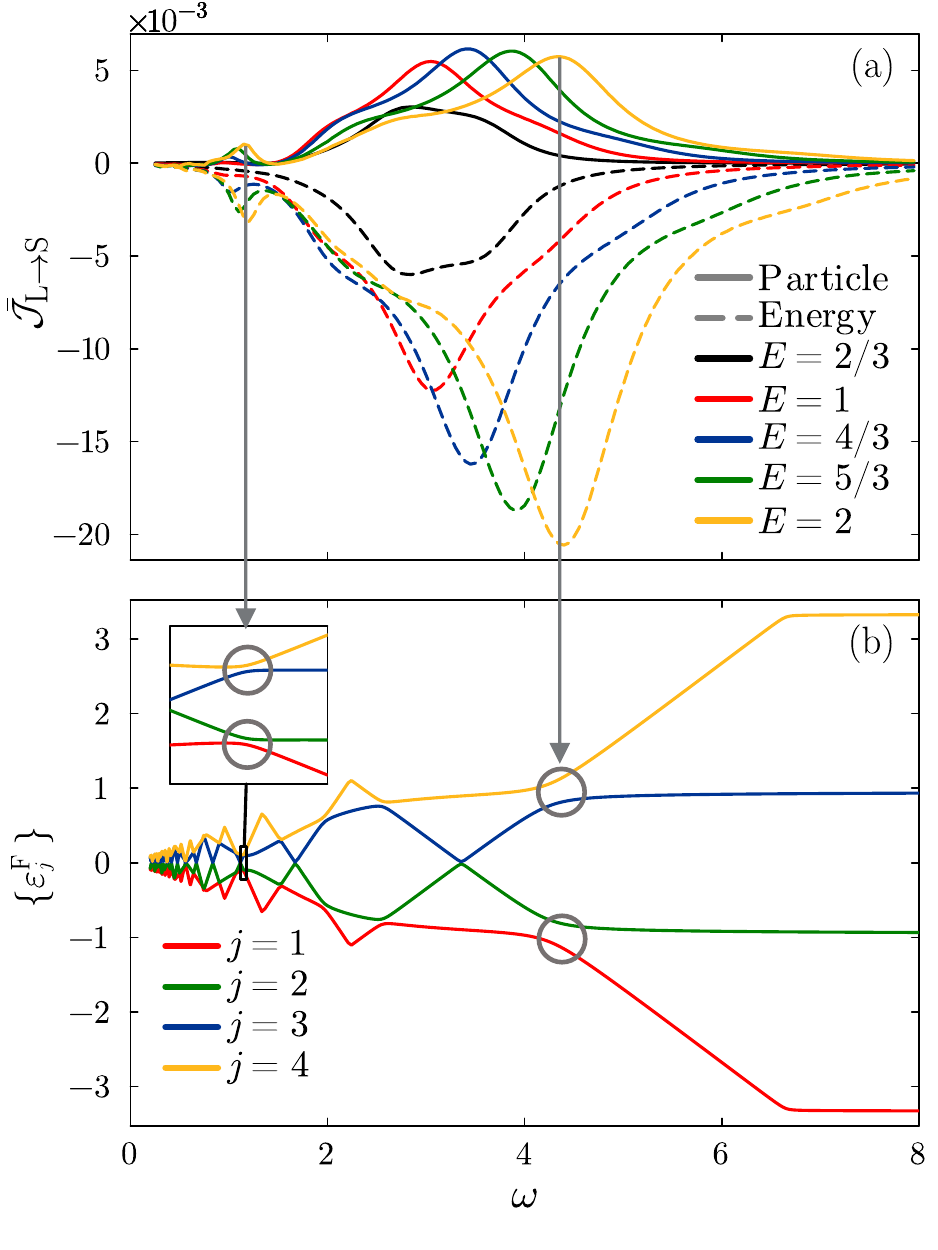}
        \caption{Frequency dependence of transport. (a) Particle and energy currents $\bar{\mathcal{J}}^{P,E}_{\text{L} \rightarrow \text{S}}$ as a function of $\omega$ without bias considering different values of $E$. Here $\mu_{\text{L}, \text{R}} = 1$, $T_{\text{L}, \text{R}} = 10$, $N=140$, and $D=4$. For $E = 2$, the peaks are located at $\omega \approx \{ 1.15, 4.35\}$. (b) Quasienergy spectrum of the stroboscopic time-independent effective Hamiltonian $\mathbf{H_{\text{eff}}} = i \ln({\mathcal{U}}_{\text{F}})/T$ for $E = 2$. The arrows signal the avoided crossings that occur at the resonance frequencies where the current is maximized.}
        \label{Fig: J vs ω without Bias}
    \end{figure}

    The oscillation of the system on-site energies induces particle and energy currents even when the two leads have identical temperatures ($\Delta T = T_{\text{R}} - T_{\text{L}} = 0$) and chemical potentials ($\Delta \mu = \mu_{\text{L}} - \mu_{\text{R}} = 0$). This is shown for $E>0$ in Fig.~\ref{Fig: J vs ω without Bias}(a), where $\mathcal{J}^{P} > 0$ corresponds to particle flow from L to S and $\mathcal{J}^{E} < 0$ to energy flow from S to L. The direction of flow is determined by the interplay between the static tilt and the periodic driving.
    
    In the absence of periodic driving, both the particle and energy currents vanish. We verified numerically that in this non-interacting regime the system relaxes to the equilibrium thermal state of the extended single-particle Hamiltonian at the common temperature and chemical potential of the leads, consistent with the equilibrium limit expected from Landauer–Büttiker/NEGF techniques. Once the time driving is switched on, the on-site energies of the central sites oscillate in time, continuously modifying their instantaneous occupations and preventing complete thermalization of the chain. By contrast, the first and last sites are not driven and remain directly coupled to the leads. Although the Lindblad dissipation acts only on the lead fermionic sites and does not directly thermalize the boundary sites of the chain, the coherent hopping between the leads and the chain continuously drives these sites to equilibrate toward the Fermi-Dirac distribution of the leads. As a consequence, the site with the more negative on-site energy is more populated, while the site with the more positive energy is less populated. This asymmetry sets a preferred direction for the non-equilibrium response induced by the driving. Reversing the tilt $E \rightarrow -E$ exchanges the roles of the two boundary sites, thereby reversing the occupation imbalance imposed by the reservoirs and flipping the direction of the cycle-averaged current. For reference, the $E=0$ case yields zero cycle-averaged particle and energy currents for all driving frequencies due to the inversion symmetry of the system, and is therefore omitted in Fig.~\ref{Fig: J vs ω without Bias}(a). Similarly, when all sites are driven, the system retains inversion symmetry even if $E>0$, so that no preferred transport direction is selected and the cycle-averaged current vanishes.

    The frequency of the time-periodic drive can be used to enhance and optimize the current, providing a useful handle to control transport properties. As shown in Fig.~\ref{Fig: J vs ω without Bias}(a), both particle and energy currents display two pronounced resonant structures (local maxima for $\mathcal{J}^{P}_{ \text{L} \rightarrow \text{S}}$ and local minima for $\mathcal{J}^{E}_{ \text{L} \rightarrow \text{S}}$). The qualitative behavior is the same for all values of $E$: Two resonances separated by an intermediate shoulder-like feature, followed by a suppression of the current in the high-frequency regime. Both resonances shift to higher frequencies as $E$ increases. Understanding the origin of these resonances is the main goal of the remainder of this section. We first show that the resonance frequencies coincide with avoided crossings in the Floquet quasienergy spectrum of the isolated driven system. We then demonstrate that these avoided crossings enhance the transition probability between neighboring sites, providing the microscopic mechanism responsible for the enhancement of transport.
    
    \begin{figure}[t]
        \includegraphics[width=0.4 \paperwidth]{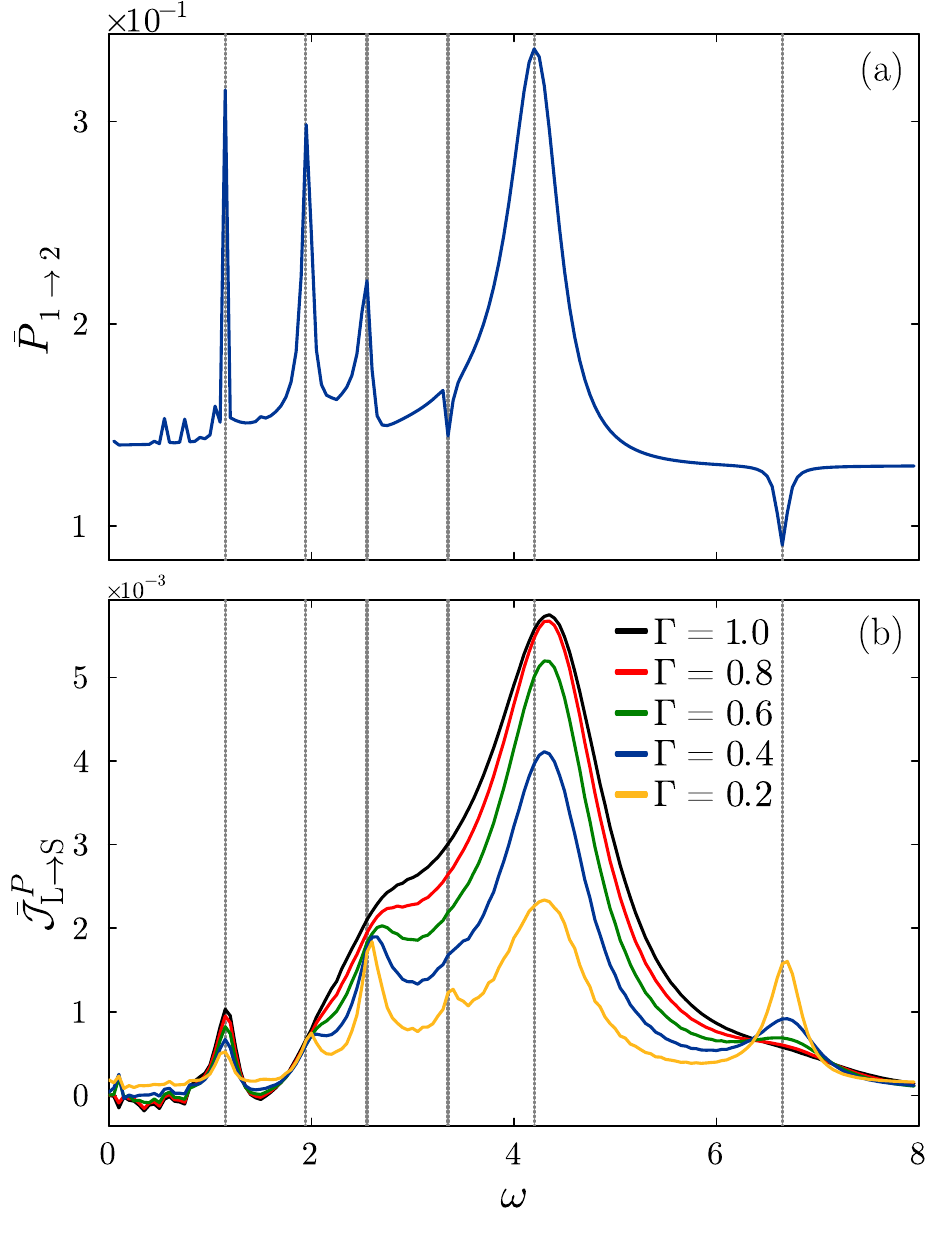}
        \caption{Microscopic origin of transport enhancement for $E=2$ and $D=4$. (a) Cycle-averaged transition probability between states localized at physical sites 1 and 2, $\bar{P}_{1\to2}$, as a function of $\omega$. (b) Particle current $\bar{\mathcal{J}}^{P}_{\text{L}\to\text{S}}$ as a function of $\omega$ for several values of the system--bath coupling $\Gamma$, with $\mu_{\text{L,R}}=1$, $T_{\text{L,R}}=10$, and $N=140$. Vertical lines mark the resonances present in $\bar{P}_{1\to2}$, at $\omega\approx\{1.15, 1.95, 2.55, 3.35, 4.2, 6.65\}$; each of these becomes a resolved peak in the current as $\Gamma$ decreases.  All are avoided crossings except $\omega\approx6.65$, which instead arises from a single-photon transition between eigenstates of $\mathbf{H}_0$.}            
        \label{Fig: Floquet}
    \end{figure}   

    The Floquet analysis of the isolated driven system is motivated by the observation that the resonance frequencies are largely determined by the properties of the system itself. Indeed, under biased conditions, the heuristic Lindblad description can already capture the resonant structure of the current, with only small quantitative shifts (see Appendix~\ref{Lindblad}). The quasienergy spectrum obtained from Eq.~\eqref{quasienergies} reveals that the resonant frequencies approximately coincide with avoided crossings in the quasienergy bands; see Fig.~\ref{Fig: J vs ω without Bias}(b). A similar association between current peaks and avoided crossings in the Floquet spectrum was reported in Ref.~\cite{PhysRevA.75.063424} for periodically driven quantum ratchets. While the Floquet quasienergy spectrum identifies the resonance frequencies, it does not explain why transport is enhanced. To establish this connection, we compute the cycle-averaged transition probability between states localized at neighboring sites. Specifically, from Eq.~\eqref{Transition_probability} we obtain the probability for a fermion to move from site 1 to site 2. As shown in Fig.~\ref{Fig: Floquet}(a), this probability shows resonances at the same frequencies that generate avoided crossings in the quasienergy spectrum. The transition probabilities between the other neighboring sites exhibit resonances at the same frequencies, so we focus on $P_{1 \rightarrow 2}$ as our diagnostic quantity. This quantity is directly connected to the correlation $\langle \hat{c}_{1}^{\dagger}\hat{c}_{2} \rangle$, which describes the coherent hopping between sites 1 and 2. Moreover, this correlation governs the particle current inside the system, whose operator is $\hat{\mathcal{J}}_{1 \rightarrow 2}^{P} = i t_{s}( \hat{c}_{1}^{\dagger}\hat{c}_{2}  - \hat{c}_{2}^{\dagger}\hat{c}_{1} )$. In Appendix~\ref{app:dissipative_spectrum}, we show that, once the boundary dissipation is included, this correlation (and hence the current) develops Lorentzian resonances centered at the avoided crossings.

    The following picture emerges: avoided crossings in the Floquet quasienergy spectrum enhance transport by promoting strong mixing between Floquet states, which in turn increases transition amplitudes between neighboring sites of the system. The closed-system transition probability $\bar{P}_{1\to2}$ neglects the dissipation induced by the leads. As a result, it does not capture why the current shows resonances at only some of the frequencies of the avoided crossings. Physically, the coupling to the leads gives each Floquet channel a finite lifetime, which broadens its contribution to the current into a resonance of finite width in $\omega$ (Appendix~\ref{app:dissipative_spectrum}). When several such resonances are spaced more closely than their combined widths, they merge into a single unresolved feature (see Appendix~\ref{app:lorentzian} for further details). This is why decreasing the system-bath coupling strength $\Gamma$, and thus narrowing each resonance, allows previously overlapping channels to separate into distinct peaks, as shown in Fig.~\ref{Fig: Floquet}(b). As expected, decreasing $\Gamma$ also reduces the overall magnitude of the current. In brief, $\Gamma$ sets the width of the resonances and thereby controls how many of them appear as distinct peaks in the current as a function of the frequency.
        
    An additional feature of the quasienergy spectrum is the presence of pairs of positive and negative eigenvalues for a given frequency $\omega$. This structure originates from a dynamical chiral symmetry of the system \cite{asboth_chiral_2014}. The symmetry is generated by a local, unitary, and Hermitian operator $\mathbf{\Pi}$ satisfying $\{ \mathbf{H}_{\text{eff}}, \mathbf{\Pi} \} = 0$ and $\mathbf{\Pi} \mathbf{\mathcal{U}}_{\text{F}} \mathbf{\Pi} = \mathbf{\mathcal{U}}_{\text{F}}^{-1}$. This can be readily verified by defining the chiral operator $\mathbf{\Pi} = \sum_{j} |\varepsilon_{j}^{\text{F}} \rangle \langle -\varepsilon_{j}^{\text{F}} |$. This construction shows that $\mathbf{\Pi}$ has doubly degenerate eigenstates with eigenvalues $\pm 1$, corresponding to the two possible chiralities of the system. However, the quasienergy eigenstates $|\varepsilon_j^{\text{F}}\rangle$ are not themselves eigenstates of $\mathbf{\Pi}$: since $\mathbf{\Pi} |\varepsilon_{j}^{\text{F}} \rangle = |-\varepsilon_{j}^{\text{F}} \rangle$, $\mathbf{\Pi}$ exchanges each pair $(|\varepsilon_j^{\text{F}}\rangle, |-\varepsilon_j^{\text{F}}\rangle)$ so neither one carries a definite chirality. As a result, the quasienergy spectrum cannot be organized into chirality sectors, and its features in Fig.~\ref{Fig: J vs ω without Bias}(b) cannot be attributed to the chirality of the states. In principle, a real crossing could be possible at $\varepsilon_j^{\text{F}} = 0$, where a state coincides with its own chiral partner and can therefore carry a definite chirality. However, we verified numerically that all approaches between quasienergies in the spectrum are avoided crossings.
        
\subsection{Non-interacting transport channels}
\label{Non-interacting transport channels}

    \begin{figure}[t]
        \centering
        \includegraphics[width=0.4 \paperwidth]{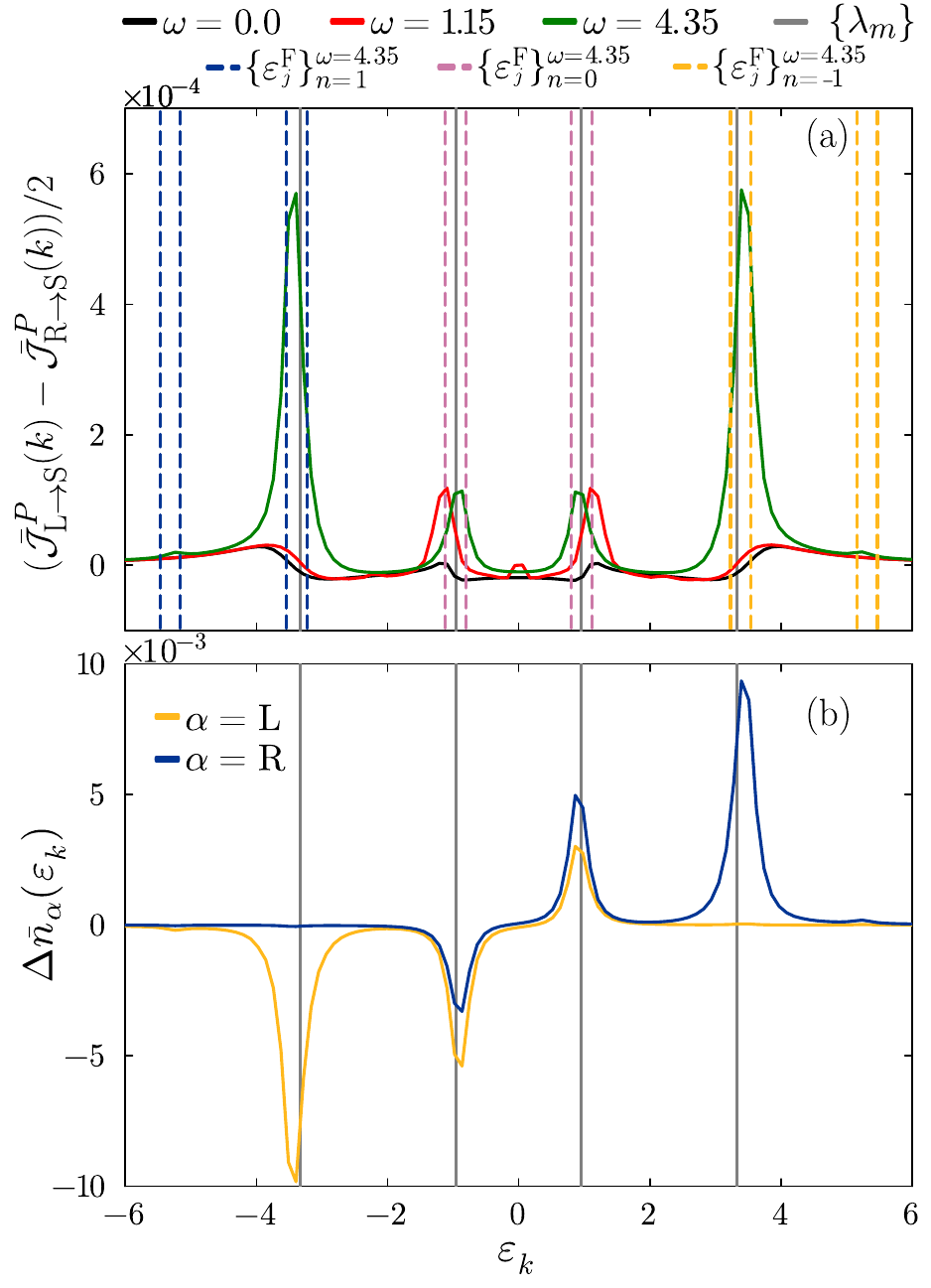}
        \caption{Energy-resolved particle current and Floquet-driven redistribution of occupations. Vertical gray solid lines correspond to the eigenvalues of $\mathbf{H}_{0}$. For both figures, $\mu_{\text{L}, \text{R}} = 1$, $T_{\text{L}, \text{R}} = 10$, $E= 2$, $D=4$, and $N =140$. (a) Particle current $\mathcal{J}^{P}$ resolved by energy modes. Dashed vertical lines $\{ \varepsilon^{\text{F}}_{j} \}_{n}^{\omega} $ indicate the four single-particle quasienergies that correspond to band $n$ of the Floquet spectrum with driving frequency $\omega$. (b) Cycle-averaged occupation deviation $\Delta n_{\alpha} (\varepsilon_{k}) $ for $\omega = 4.35$ with respect to the undriven case.}
        \label{Fig: J separated by modes}
    \end{figure}
        
    Having identified the resonant frequencies from the Floquet quasienergy spectrum, we now turn to the microscopic transport channels that these resonances activate. The avoided crossings of the Floquet quasienergy spectrum in Sec.~\ref{Currents controlled by frequency} explain the resonances at $\omega \approx 1.15$ and $4.35$ at $\Gamma = 1$. The resonance near $\omega\approx6.65$ at $\Gamma = 0.2$ in Fig.~\ref{Fig: Floquet}(b) has a different origin, since no avoided crossing accompanies it. We attribute it to single-photon transitions between eigenstates of a static Hamiltonian, a mechanism periodic driving is known to activate \cite{Floquet_Tutorial}. To demonstrate this attribution, we decompose $\mathbf{H}_{\text{S}} (t) = \mathbf{H}_{0} + \mathbf{V}(t)$, where $\mathbf{H}_{0} = T^{-1} \int_{0}^{T} dt \mathbf{H}_{\text{S}}(t)$ is the cycle-averaged single-particle Hamiltonian of the system and $\mathbf{V}(t)$ contains all non-zero Fourier harmonics of the drive. The ascending eigenvalues $\lambda_m$ and their corresponding eigenstates $|m\rangle$ of $\mathbf{H}_{0}$ satisfy $\mathbf{H}_{0} = \sum_m \lambda_m |m\rangle \langle m|$. Resonances that induce transitions between eigenstates of $\mathbf{H}_{0}$, namely $|n\rangle$ and $|m\rangle$, are expected whenever the driving frequency satisfies $k\omega \approx |\lambda_n - \lambda_m|$ for $k \in \mathbb{N}$.  At $\omega = 6.65$, the system has already entered the fast-driving regime. In a high-frequency expansion of $\mathbf{H}_{\text{eff}}$, the leading term is $\mathbf{H}_0$, with corrections appearing at order $\omega^{-p}$ ($p = 1, 2, \ldots$) that vanish as $\omega$ grows \cite{Floquet_Tsuji}. Thus,  $\mathbf{H}_{\text{eff}} \approx \mathbf{H}_0$ and the Floquet quasienergies become nearly frequency-independent, so no further avoided crossings can occur. Nevertheless, $|\lambda_1-\lambda_4| = 6.63$, i.e., both the current peak and the accompanying dip in $\bar{P}_{1\to2}$ originate from a single-photon resonance connecting the lowest- and highest-energy eigenstates of $\mathbf{H}_0$. 
    
    To better understand the peaks of $\mathcal{J}_{\text{L} \rightarrow \text{S}}^{P, E}$ and the role of drive-induced transport channels, we analyze Eqs.~\eqref{J:a} and~\eqref{J:b} by decomposing the currents into contributions from individual lead modes, i.e., $\mathcal{J}_{\text{L} \rightarrow \text{S}}^{P, E} = \sum_{k} \mathcal{J}_{\text{L} \rightarrow \text{S}}^{P, E}(k)$. This allows us to identify which modes dominate the transport, as evidenced in \ref{Fig: J separated by modes}(a). Although here we focus on particle transport, a similar analysis applies to the energy current. For interpreting the current resolved by energy modes, it is convenient to define an energy scale. The eigenvalues $\{\lambda_m\}$ of $\mathbf{H}_0$ are a natural choice, since they remain fixed for all $\omega$, unlike the quasienergies of $\mathbf{H}_{\text{eff}}$, which shift continuously with the driving frequency. From all the lead modes, the dominant contributions to the current arise from four regions centered around the eigenvalues of $\mathbf{H}_{0}$; this is indicated by vertical gray lines in Fig.~\ref{Fig: J separated by modes}. This result is consistent with the well-known behavior of static systems, where transport is expected to be dominated by lead modes resonant with the eigenstates of the system \cite{ryndyk2016theory}. In the driven case, the peaks are close to---but do not align perfectly with---the eigenvalues $\{ \lambda_m \}$, as the relevant transport channels are instead determined by the Floquet quasienergy band structure $\{ \varepsilon_{j}^{\text{F}} \}$ induced by the external drive. In particular, the frequency near $\omega \approx 4.35$ is special because the avoided crossings in multiple Floquet bands (specifically $n=\{-1,0,1\}$) align with the static eigenvalues $\{\lambda_m\}$, giving an optimal matching between undriven and driven transport channels and a maximum in the current. Regarding the width of each peak in Fig.~\ref{Fig: J separated by modes}, it is a consequence of the standard discrete–continuum hybridization mechanism familiar from impurity models such as the Anderson model \cite{Stefanucci_van_Leeuwen_2013}.  Since the lead on-site energies $\{ \varepsilon_k \}$  are linearly distributed in $[-W, W]$, the lead modes form an effective continuum that overlaps with the discrete eigenstates of the system. As a result, the latter hybridize with this continuum and acquire a finite broadening. Thus, the lead modes that contribute most strongly to the current are precisely those whose energies lie within these hybridization-broadened windows, i.e., the modes that are in resonance with the hybridized Floquet quasienergy band structure.

    This resonant activation of transport channels also modifies the occupations of the lead modes, as shown in Fig.~\ref{Fig: J separated by modes}(b). There we plot the cycle-averaged deviation $\Delta n_{\alpha}(\varepsilon_k) = \langle \hat{a}_{k , \alpha}^\dagger \hat{a}_{k , \alpha} \rangle_{\omega = 4.35} - \langle \hat{a}_{k , \alpha}^\dagger \hat{a}_{k , \alpha} \rangle_{\omega = 0}$. This quantity measures how the periodic drive changes the steady-state occupation of each lead mode relative to the undriven case, where all modes follow the Fermi–Dirac distribution and the net current is zero. The drive forces the extended system to host more (fewer) fermions in the right (left) lead. The modes with the largest deviations are precisely those that exchange the most fermions with the system during a driving cycle, and hence contribute the most to $\bar{\mathcal{J}}^{P}$. A detailed mode-resolved decomposition of the current, using an alternative definition, is presented in Appendix~\ref{All_Currents}, where the connection between the occupation deviations and the resonant enhancement of the particle current is made explicit.     
    
\subsection{Rectification mechanism} 

\begin{figure}[t]
        \centering
        \includegraphics[width=0.4 \paperwidth]{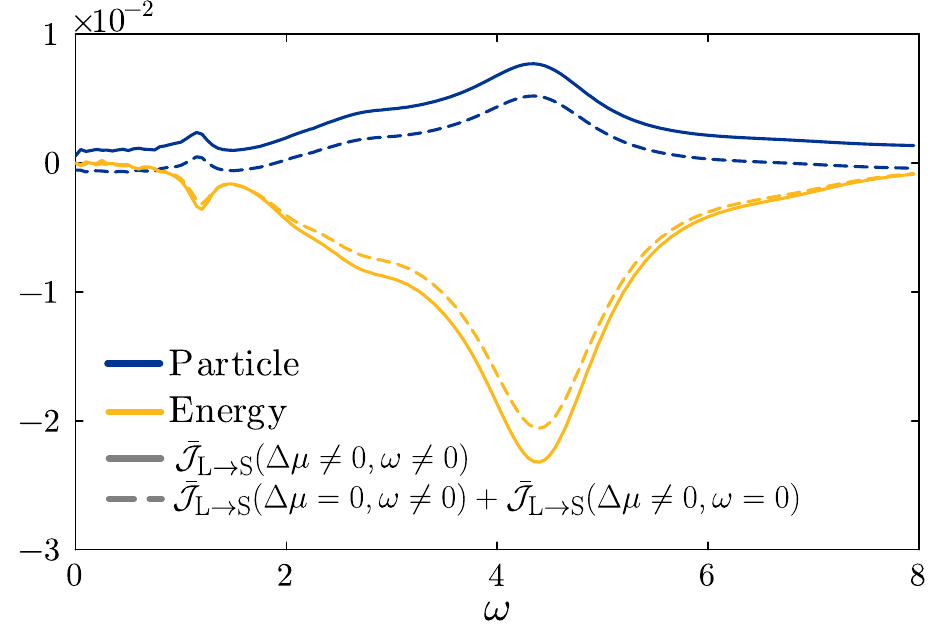}
        \caption{Interplay of bias and time driving for forward configuration. Here $\mu_{\text{L}} = 2$ and $\mu_{\text{R}} = 1$, $T_{\text{L}, \text{R}} = 10$, $E = 2$, $D=4$, and $N=140$.}
        \label{Fig: Interplay}
    \end{figure}

    When $\Delta \mu \neq 0$ or $\Delta T \neq 0$, a bias-induced contribution to the current appears. Here, rectification emerges from the interplay between the drive-induced and bias-induced contributions to the total current. If the two contributions were completely independent, the total current would be given by a simple additive decomposition. Although not exact, this decomposition provides a remarkably good description of the observed behavior, as shown in Fig.~\ref{Fig: Interplay}. For concreteness, we illustrate this for a chemical-potential bias; a thermal bias changes the calculation only through the Fermi-Dirac distribution of each lead, and hence the factors $f_{\alpha}(\varepsilon_{k})$ in Eq.~\eqref{Dissipators}, so we expect the same behavior to hold in that case. In either case, the drive-induced current is almost independent of the magnitude and sign of the bias. This already suggests that the coexistence of the two contributions to the total current will produce an asymmetric response.

    We analyze the rectification of the current from L to S, i.e., $\mathcal{J}_{ \text{F},\text{R}} = \bar{\mathcal{J}}^{P}_{\text{L} \rightarrow \text{S}}$. By definition, a positive value corresponds to net transport from L to S. In the absence of time driving, the forward configuration ($\mu_{\text{L}} > \mu_{\text{R}}$) yields a positive particle current, while the reverse configuration ($\mu_{\text{L}} < \mu_{\text{R}}$) yields a negative one. In the forward configuration, the bias-induced current flows in the same direction as the drive-induced current observed in Fig.~\ref{Fig: J vs ω without Bias}. Thus, once the external driving is turned on, the two contributions tend to reinforce each other, whereas they compete in the reverse configuration, leading to a partial cancellation and thus to rectification. For fixed chemical potential bias, the rectification is maximized at frequencies where the drive-induced current exhibits resonant peaks. For energy transport, the mechanism is analogous, with one difference. The drive-induced energy current is negative, meaning that it flows from R to L. Thus, the time driving can enhance the forward response only if the thermal bias is chosen such that the bias-induced energy current goes in the same direction. This corresponds to choosing  $T_{\text{R}} > T_{\text{L}}$ as the forward configuration.

    \begin{figure}[t]
        \centering
        \includegraphics[width=0.4 \paperwidth]{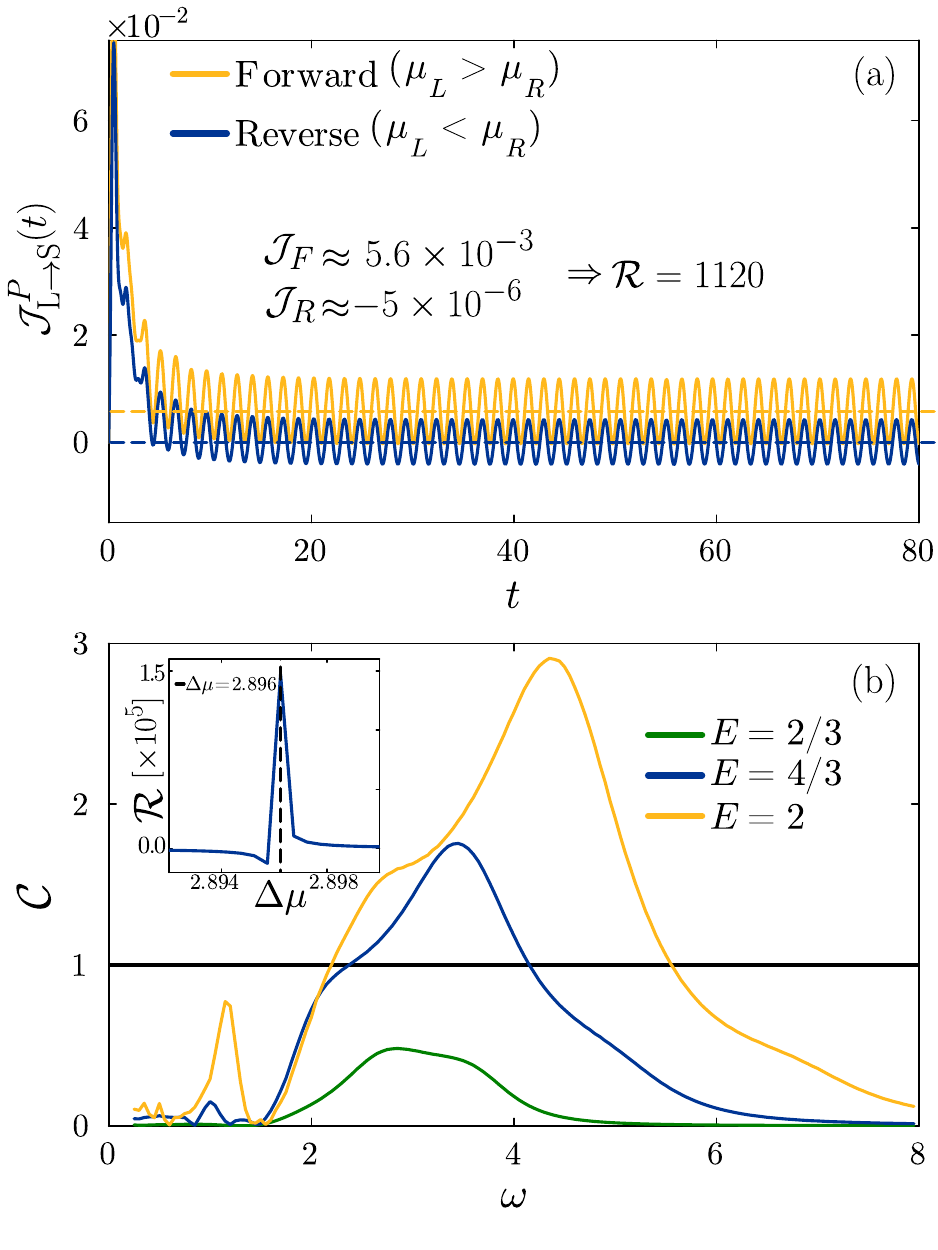}      
        \caption{Drive-induced rectification. Here $\mu_{\text{L}} = 2$ and $\mu_{\text{R}} = 1$ in forward bias, $T_{\text{L}, \text{R}} = 10$, and $D=4$, $N=140$. (a)  Particle current $\mathcal{J}_{\text{L} \rightarrow \text{S}}^{P}(t)$ for $\omega = 2.35$ and $E = 4/3$. The horizontal dashed line signals the average value in the NESS. (b) Contrast coefficient $\mathcal{C}$ for particle transport as a function of $\omega$ for various $E$. The horizontal solid line with $\mathcal{C}$ marks the boundary between standard and bipolar rectification. The inset shows the rectification coefficient $\mathcal{R}$ as a function of $\Delta \mu$ with $\omega = 4.35$, and $E = 2$.}
        \label{Fig: Rectification}
    \end{figure}
    
    The bias-induced current should increase with $\Delta \mu = \mu_{\text{L}} - \mu_{\text{R}}$, whereas the drive-induced contribution is independent of the bias. As a result, their competition is most pronounced in the low-bias regime where both contributions are of the same order of magnitude. In fact, there is a driving frequency at which the rectification is perfect, with $\mathcal{C} = 1$ and $\mathcal{R} \rightarrow \infty$. At this frequency, the drive-induced contribution cancels the bias-induced one almost exactly, so that $\mathcal{J}_{\text{R}}$ is several orders of magnitude lower than $\mathcal{J}_{\text{F}}$, as exemplified in Fig.~\ref{Fig: Rectification}(a). This frequency marks the boundary between two different rectification behaviors, since $\mathcal{C}$ changes from $\mathcal{C} < 1$ to $\mathcal{C} > 1$ across it. In particular, for low $\Delta \mu$, there is a range of frequencies $\omega$ where the drive-induced contribution is larger than the bias-induced one, and particles propagate in the same direction for both forward and reverse bias configurations. This results in $\mathcal{C} > 1$ and $\mathcal{R}< 0$; see Fig.~\ref{Fig: Rectification}(b). In this regime, the current remains sizeable for forward and reverse bias orientations, but it flows only from L to R, with no effective backward transport. This phenomenon is known as bipolar rectification and can be used to design bipolar diodes \cite{PhysRevLett.134.186301}. Outside of this region, standard rectification occurs, characterized by $\mathcal{C} \leq 1$ where the drive-induced current enhances $\mathcal{J}_{\text{F}}$ and suppresses $\mathcal{J}_{\text{R}}$. For larger $\Delta \mu$, the only possible regime is standard rectification, as the drive-induced contribution is no longer strong enough to induce transport against the bias. The diverging behavior of $\mathcal{R}$ at the boundary between the two types of rectification regimes is also evident by fixing $\omega$ and varying $\Delta \mu$, as shown in the inset of Fig.~\ref{Fig: Rectification}(b). 
    
    Having established this frequency-dependent picture, we now examine how the tilt $E$ controls the emergence of bipolar rectification. For a fixed bias, bipolar rectification emerges only once the tilt $E$ exceeds a certain threshold; the precise threshold depends on the chosen bias. At a small tilt, the response of the system remains mostly symmetric under exchange of the bias, so rectification is strongly suppressed. Increasing $E$ reduces the bias-induced current, reflecting the behavior of the corresponding static tilted lattice with Wannier–Stark localization \cite{ML_Swolak, 10.3389/fphy.2024.1474018}. Here, the time-periodic driving prevents the formation of fully current-blocking domains, and transport remains sizeable even when the static localization threshold of Eq.~\eqref{localization_threshold} is fulfilled. At the same time, increasing $E$ effectively enhances the amplitude of the periodic driving, thereby strengthening the drive-induced contribution. Consequently, as the static tilt increasingly suppresses the bias-induced current, the drive-induced contribution comes to dominate, enabling the emergence of bipolar rectification.    
    
   \begin{figure}[t]
        \centering
        \vspace{0.25 cm}
        \includegraphics[width=0.4 \paperwidth]{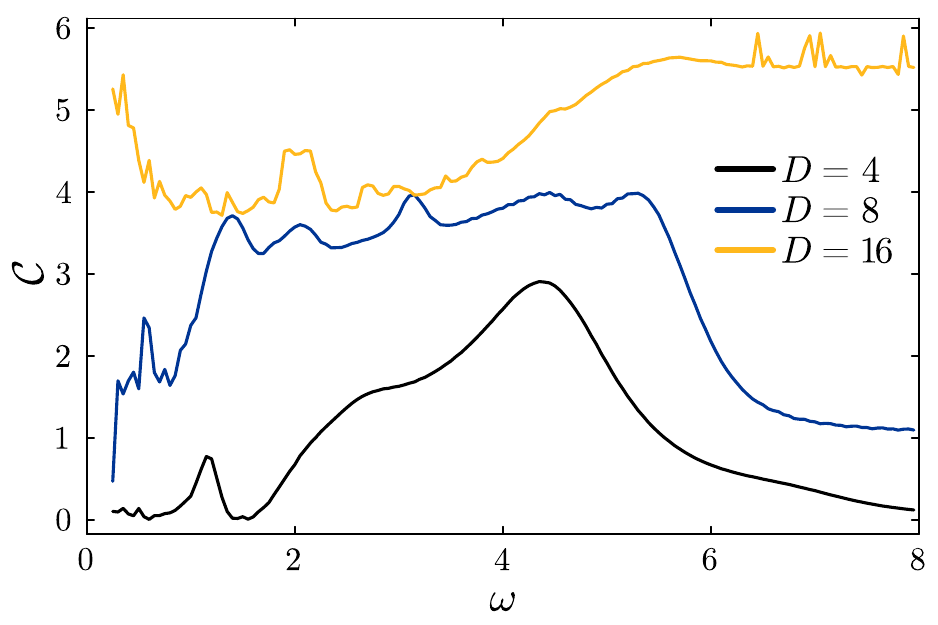}
        \caption{Contrast coefficient $\mathcal{C}$ for particle transport considering different system sizes $D$. Here $E=6/(D-1)$, $\mu_{\text{L}} = 2$ and $\mu_{\text{R}} = 1$ in forward bias, $T_{\text{L}, \text{R}} = 10$, and $N=140$.}
        \label{Fig: Larger systems}
    \end{figure}

    The emergence of bipolar and standard rectification persists for larger systems with one subtlety. As shown in Fig.~\ref{Fig: Larger systems}, we fix $\Delta \mu = 1$ and increase both the system size $D$ and the size of the driven region while keeping a fixed $50 \% $ ratio between both. For example, for $D=8$, the driven sites are $j=\{3,4,5,6\}$ in Eq.~\eqref{local_driving}. Increasing the number of driven sites enhances the drive-induced current, while the bias-induced current is suppressed as the system is pushed closer to the Wannier–Stark localized regime described in Eq.~\eqref{localization_threshold}. Consequently, the region where $\mathcal{C} > 1$ grows with system size, reflecting the increasing dominance of the drive-induced contribution in the net current. For $D=16$, the system exhibits only bipolar rectification for low $\Delta \mu$, with no standard-rectification window remaining. This occurs because, as discussed above, the bias-induced contribution decreases with system size, and we increased only the number of driven sites while keeping the bias fixed. For sufficiently large $\Delta \mu$, however, the drive-induced and bias-induced contributions will become of the same order of magnitude. Therefore, the bipolar and standard rectification behavior observed in Fig.~\ref{Fig: Rectification}(b) can be recovered. Consequently, we expect this rectification mechanism to persist in much larger systems, provided that a sufficiently large central region is driven. 

\subsection{Thermodynamic signatures of rectification}
    
\begin{figure}[t]
            \centering
            \includegraphics[width=0.4 \paperwidth]{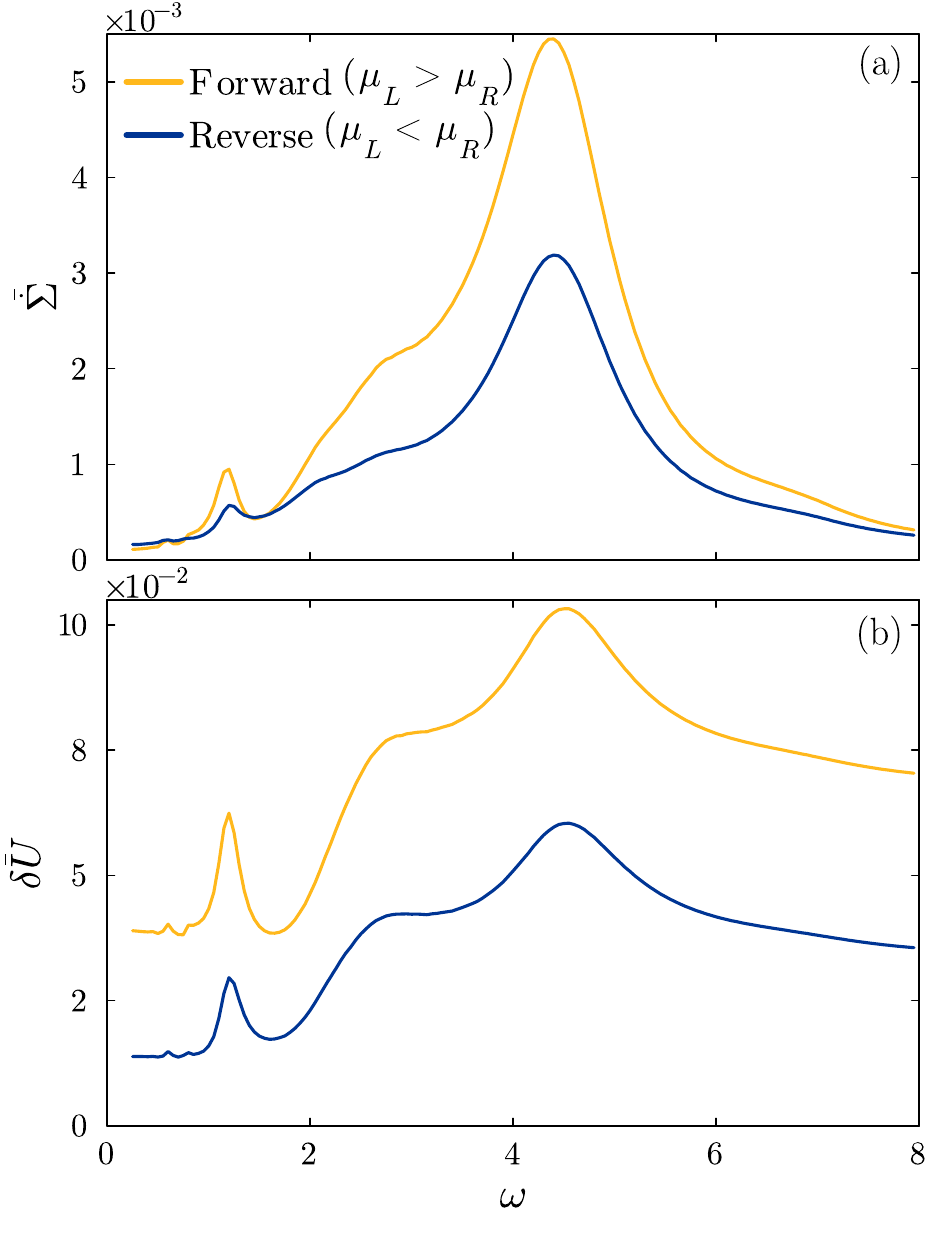}
            \caption{Cycle-averaged thermodynamic observables. For both figures, the initial state is the NESS without time driving $(\omega = 0)$, $\mu_{\text{L}} = 2$ and $\mu_{\text{R}} = 1$ in forward bias, $T_{L, R} = 10$, $E = 2$, $D=4$, and $N=140$. (a) Entropy production rate of the system $\dot{\Sigma}$. (b) Change in the internal energy of the system $\delta U$.}
            \label{Fig: Thermodynamics}
        \end{figure}

    Examining the rectification mechanism from a thermodynamic perspective is valuable. It confirms that the mechanism is fully consistent with the laws of thermodynamics, and provides additional physical insight that is not accessible from transport observables alone. Namely, analyzing the change in the internal energy $\delta U$ and the entropy production rate of the system $\dot{\Sigma}$ reveals how much energy the drive injects into the system to displace it from its static NESS and how much irreversibility is generated in the process. To isolate the contribution of the drive alone, we choose as the initial state the NESS in the absence of time driving ($\omega = 0$). The initial energy then already accounts for everything the leads have transferred to the system, so Eq.~\eqref{System_internal_energy_change} directly quantifies the change in the internal energy induced by the time-periodic drive during the time evolution. As shown in Fig.~\ref{Fig: Thermodynamics}, the cycle-average values of $\dot{\Sigma}$ and $\delta U$ are maximized simultaneously with the contrast coefficient $\mathcal{C}$. Thus, rectification is maximized when the time-periodic drive most strongly perturbs the system away from its static steady state, increasing both its energy content and its irreversibility.     
    
    Moreover, the cycle-averaged values of $\dot{\Sigma}$ and $\delta U$ are larger under forward bias, indicating that our setup acts as a rectifier with an asymmetric response not only in fermionic transport but also in energy absorption and entropy production. The peak in $\dot{\Sigma}$, previously reported for a periodically driven two-site system \cite{ML_Fast_Driving}, is consistent with our results. Here, we extend this observation by connecting it to a directional asymmetry between forward and reverse bias, and to the associated behavior of $\delta U$. By doing so, we show that the preferred transport direction (forward bias) is not only more conductive, but also more irreversible and dissipative (absorbing energy from the residual baths more efficiently).

\section{Interacting systems} 
\label{Interacting}


     In the interacting regime ($U \neq 0$), the dynamics can no longer be solved exactly via Eq.~\eqref{Lyapunov}. This follows from the Hamiltonian no longer being quadratic. The many-body density matrix $\hat{\rho}(t)$ cannot be obtained directly from the single-particle density matrix, as was possible in Gaussian systems. Instead, the dynamics must be solved for $\hat{\rho}(t)$ using Eq.~\eqref{QME}. Since modeling the mesoscopic leads requires using a large number of modes to accurately approximate the spectral densities of the baths, the full many-body evolution generated by Eq.~\eqref{QME} becomes exponentially demanding. This computational cost arises from the exponential growth of the Hilbert-space dimension with the size of the extended system. To overcome this limitation, we follow Ref.~\cite{ML_Tensor_Networks} to implement the time-evolving block decimation (TEBD) algorithm \cite{PhysRevLett.93.040502, PhysRevLett.93.207205, PhysRevLett.98.070201, RevModPhys.93.015008, PAECKEL2019167998}. This method, based on a matrix product state (MPS) representation of quantum states, provides an efficient and controlled approximation to the real-time dynamics of one-dimensional interacting systems coupled to non-interacting lead modes. Our codes are implemented using the Julia version of the ITensor library \cite{SciPostPhysCodeb.4, SciPostPhysCodeb.4-r0.3}.
    
    We consider lead sizes $N = \{40, 80\}$ and bond dimensions $\chi = \{150, 300\}$, depending on the specific scenario. We verify the convergence of the relevant observables with respect to both $N$ and $\chi$ (see Appendix~\ref{About_parameters}). The choice of both quantities, which depends on $U$, reflects a balance between numerical accuracy and computational cost. As discussed in Sec.~\ref{ML}, one of the requirements to accurately describe the effect of the macroscopic baths within the mesoscopic lead approximation is to consider an energy window for the lead modes, $\varepsilon_k \in [-W, W]$, that is sufficiently large to comprise all eigenvalues of the system Hamiltonian that may couple to the baths during the dynamics. For strong interactions, $U = \{ -5,5\}$, we extend this window to $[-2W, 2W]$. This choice comes from the fact that attractive interactions $(U<0)$ shift the eigenvalues of $\hat{H}_{\text{S}}$ toward negative energies, whereas repulsive interactions $(U>0)$ shift them toward positive values. To maintain sufficient resolution of lead-mode-resolved quantities within this enlarged window, we also increase the number of modes to $N = 80$ in these strongly interacting cases.

\subsection{General effect of interactions on transport observables}
    \label{EffectofU}

   We begin by analyzing the effect of $U$ on $\mathcal{J}^{P, E}_{\text{L} \rightarrow \text{S}} (t)$. In the absence of time-dependent driving, the coexistence of interactions and tilt can promote the formation of current-blocking domains only in one bias configuration, effectively allowing fermionic transport only if it flows ``up" the tilt \cite{PRX_JJ}. In our case, restricted to weak bias, the asymmetric transport under bias exchange relative to the tilt motivates our exploration of the interacting regime. Our drive-induced current flows up the tilt, where transport is allowed; we therefore expect interactions not to hinder, and possibly even favor transport in this configuration, while potentially suppressing transport in the reverse bias configuration.
   
    \begin{figure}[t]
        \centering
            \includegraphics[width=0.4 \paperwidth]{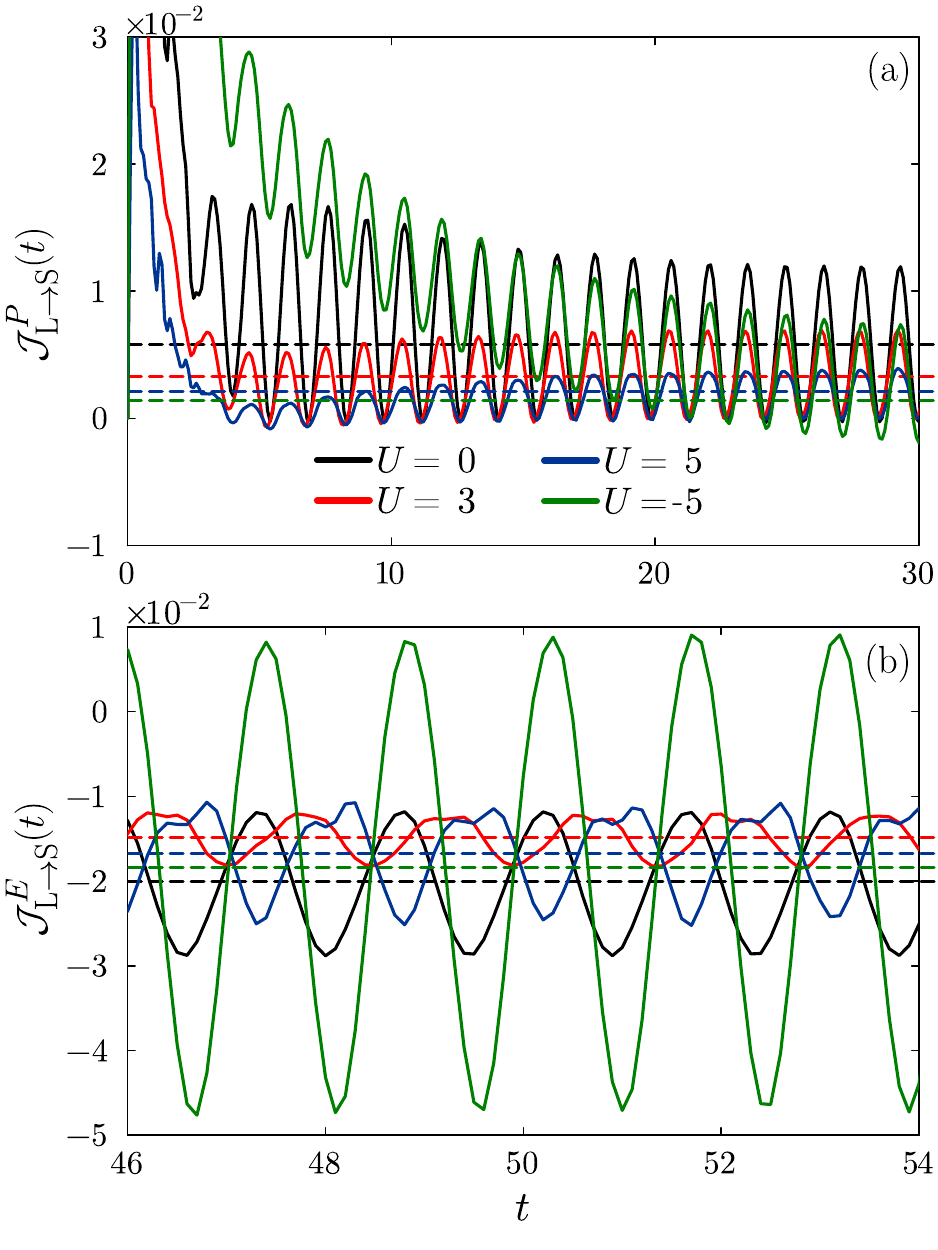}      
            \caption{Time evolution of the currents for different values of the interaction strength $U$, considering both repulsive $(U>0)$ and attractive $(U<0)$ scenarios. Here $E = 2$, $D=4$, $\omega = 4.35$, $\mu_{\text{L}, \text{R}} = 1$, $T_{\text{L}, \text{R}} = 10$, and $N=80$. The horizontal lines correspond to the cycle-averaged value of the current. (a) Particle current $\mathcal{J}_{L\rightarrow S}^{P}(t)$. (b) Energy current $\mathcal{J}_{L\rightarrow S}^{E}(t)$.}
        \label{Fig: Effect of U on JP and JE}
    \end{figure}

    As a reference point, we consider the resonance peak of the non-interacting case, $\omega = 4.35$. As shown in Fig.~\ref{Fig: Effect of U on JP and JE}, interactions shift the cycle-averaged value of the current, change the amplitude of its oscillations, and speed up (slow down) the relaxation to the periodic NESS for repulsive (attractive) interactions. While this qualitative influence is shared by particle and energy currents, the quantitative details differ.
    
    As shown in Fig.~\ref{Fig: Effect of U on JP and JE}(a), for the particle current, increasing $|U|$ decreases the oscillation amplitude and shifts the cycle-averaged value toward lower values. This holds for both repulsive and attractive interactions, although the sign of $U$ affects the magnitude of these changes. The qualitative shape of the oscillation remains unchanged. In contrast, Fig.~\ref{Fig: Effect of U on JP and JE}(b) shows that the energy current exhibits the same two effects---a shift in mean value and a change in amplitude---without the clear trend seen in $J^P$. This is, repulsive and attractive interactions do not push the mean value in a consistent trend set by $|U|$. In addition, the oscillation does not merely change in amplitude but also in form for positive interactions, indicating the emergence of higher-frequency components in the energy-current dynamics that are absent in the attractive case. This change in form is consistent with the energy current being an intrinsically more complex observable than the particle current, by involving terms that depend explicitly on $U$ and the tilted potential $E$ (see Appendix~\ref{All_Currents}).        

    The difference in relaxation speed between attractive and repulsive interactions is analyzed in Appendix~\ref{relaxation_rate}. There we show that this difference is captured by the Liouvillian gap, which determines the slowest relaxation in the long-time limit and is systematically smaller for $U<0$ than for $U>0$ across almost all driving frequencies. The appendix also proposes a microscopic mechanism for this asymmetry, based on the analysis of the energy-resolved current in Sec.~\ref{Interacting transport channels}.
           
\subsection{Interacting transport channels}
\label{Interacting transport channels}

\begin{figure}[t]
        \centering
        \includegraphics[width=0.4 \paperwidth]{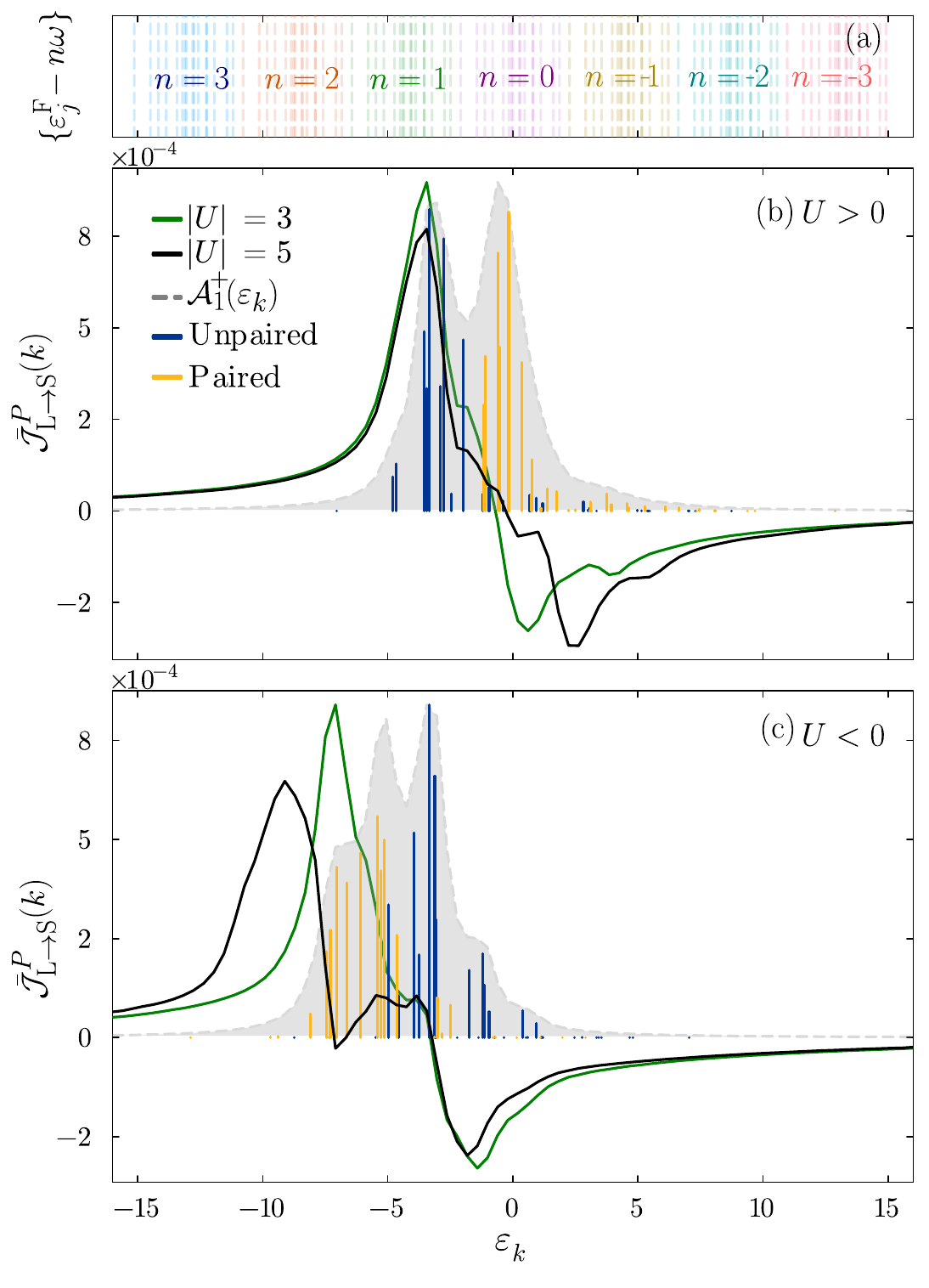}
        \caption{Comparison between the spectral function and the energy-resolved current. (a) Many-body Floquet quasienergies $\varepsilon^{\text{F}}_{j} - n\omega$ for $U=3$. (b), (c) Cycle-averaged particle current $\bar{\mathcal{J}}^{P}_{\text{L}\rightarrow\text{S}}(k)$ resolved by lead modes for (b) $U>0$ and (c) $U<0$. All panels use $\omega = 4.35$, $\mu_{\text{L}, \text{R}} = 1$, $T_{\text{L}, \text{R}} = 10$, $E = 2$, $D = 4$, and $N = 80$. Vertical lines mark the addition energies of site 1 for $|U|=3$ only, with heights proportional to $|P^{(\nu,n)}_{m}(1)|^{2}$. Blue (gold) lines mark transitions that add a fermion next to an empty (occupied) site 2, which create unpaired (paired) configurations. The dashed gray curve is $\mathcal{A}^{+}_{1}$ of Eq.~\eqref{eq:addition_energies}. The vertical lines and the gray curve are each rescaled to the maximum of the $|U|=3$ current.}    
        \label{Fig: J separated by modes (with U)}
        \end{figure}

    The analysis of Sec.~\ref{Non-interacting transport channels} links the peaks of the energy-resolved current to the single-particle eigenvalues $\{\lambda_m\}$ of the cycle-averaged Hamiltonian. In the interacting chain, the many-body Floquet quasienergies offer no direct substitute. As Fig.~\ref{Fig: J separated by modes (with U)}(a) shows, their spectrum is dense, since it superposes the spectra of all particle-number sectors, folded into a single Floquet zone of width $\omega$. A peak of the current therefore cannot be assigned to a single quasienergy.  The relevant energies follow instead from the way the leads exchange particles with the chain. Through the hopping term in Eq.~\eqref{H_ML}, each lead mode adds or removes one fermion at a time, so these energies are the differences between eigenenergies of the cycle-averaged Hamiltonian whose fermion numbers differ by one, known as addition energies. At $U=0$, the many-body eigenstates are Slater determinants of single-particle orbitals, so adding a fermion to an empty orbital $m$ costs $\lambda_m$ regardless of how many fermions the chain already holds. The addition energies hence reduce to $\{\lambda_m\}$, which makes the present analysis the direct generalization of Sec.~\ref{Non-interacting transport channels}. We therefore compare the energy-resolved current with the spectral function of the contact sites, which encodes the addition energies together with the amplitude of each transition. We again fix $\omega = 4.35$, the frequency at which the rectification is largest in the noninteracting chain.
    
    We evaluate the spectral function for the many-body cycle-averaged Hamiltonian $\hat{H}_{0} = T^{-1}\int_{0}^{T} dt\, \hat{H}_{\text{S}}(t)$, whose eigenstates have a definite number $\nu$ of fermions. Let $|\Psi_{\nu,n}\rangle$ be an eigenstate with $\nu$ fermions and energy $E_{\nu,n}$, where the index $n$ labels the eigenstates of this sector. We define the quasiparticle and quasihole amplitudes of the contact site $p$ for the reference state $|\Psi_{\nu,n}\rangle$ as $P^{(\nu,n)}_{m}(p) = \langle \Psi_{\nu,n}|\hat{c}_{p}|\Psi_{\nu+1,m}\rangle$ and $Q^{(\nu,n)}_{m}(p) = \langle \Psi_{\nu-1,m}|\hat{c}_{p}|\Psi_{\nu,n}\rangle$~\cite{Stefanucci_van_Leeuwen_2013}. The spectral function of site $p$ in the state $|\Psi_{\nu,n}\rangle$ reads
    \begin{align}
        \mathcal{A}^{(\nu,n)}_{p}(\omega) &= 2\pi \Big[ \sum_{m} |P^{(\nu,n)}_{m}(p)|^{2}\, \delta\big(\omega - [E_{\nu+1,m} - E_{\nu,n}]\big) \nonumber \\
        &\quad + \sum_{m} |Q^{(\nu,n)}_{m}(p)|^{2}\, \delta\big(\omega - [E_{\nu,n} - E_{\nu-1,m}]\big) \Big],
        \label{eq:spectral_function}
    \end{align}
    where $m$ runs over the eigenstates with $\nu+1$ fermions in the first sum and with $\nu-1$ fermions in the second.

    The first sum describes the addition of a fermion to site $p$, and the second sum describes its removal. The squared amplitude $|P^{(\nu,n)}_{m}(p)|^{2}$ is the probability that adding a fermion to site $p$ takes the chain from $|\Psi_{\nu,n}\rangle$ to $|\Psi_{\nu+1,m}\rangle$. In this transition, the lead mode that injects the fermion supplies the energy $E_{\nu+1,m} - E_{\nu,n}$. Likewise, $|Q^{(\nu,n)}_{m}(p)|^{2}$ is the probability that removing a fermion from site $p$ takes the chain to $|\Psi_{\nu-1,m}\rangle$, and the lead mode that receives the fermion absorbs the energy $E_{\nu,n} - E_{\nu-1,m}$. Together, these probabilities satisfy the sum rule $\int d\omega\, \mathcal{A}^{(\nu,n)}_{p}(\omega)/(2\pi) = 1$. Hence $\mathcal{A}^{(\nu,n)}_{p}(\omega)/(2\pi)$ is the probability density that a fermion exchanged between site $p$ and a lead mode has energy $\omega$~\cite{Stefanucci_van_Leeuwen_2013}. This interpretation connects the spectral function with the energy-resolved current. Each lead mode $k$ exchanges fermions of energy $\varepsilon_k$ with its contact site $p$. The current contribution from mode $k$ is therefore controlled by $\mathcal{A}^{(\nu,n)}_{p}(\varepsilon_k)$ and should be appreciable only if $\varepsilon_k$ lies close to a peak. The coupling to the leads sets how close, since it broadens each Dirac delta into a Lorentzian through the same discrete--continuum hybridization that broadens the peaks of Fig.~\ref{Fig: J separated by modes}~\cite{Stefanucci_van_Leeuwen_2013}.
    
    Although the peaks of the spectral function set where the current can flow, the occupations of the chain relative to those of the lead set its sign and size. We neglect coherences between eigenstates of $\hat{H}_{0}$ and write the density matrix of the chain as
        \begin{align}
            \hat{\rho}_{\text{S}} = \sum_{\nu,n} w_{\nu,n}\, |\Psi_{\nu,n}\rangle\langle\Psi_{\nu,n}|,
            \label{eq:density_matrix}
        \end{align}
    where $w_{\nu,n}$ are the weights of the mixture. The spectral function of the chain is then $\mathcal{A}_{p}(\omega) = \sum_{\nu,n} w_{\nu,n}\, \mathcal{A}^{(\nu,n)}_{p}(\omega)$. The tensor-network calculation does not give access to these weights, so we omit them and use the unweighted sum $\sum_{\nu,n} \mathcal{A}^{(\nu,n)}_{p}(\omega)$. This omission preserves the positions of the peaks, which identify the transitions with which the dominant lead modes can resonate, but not their heights. Consequently, the largest peaks of the current need not coincide with the tallest peaks of the unweighted spectral function.

    Each transition between $|\Psi_{\nu,n}\rangle$ and $|\Psi_{\nu+1,m}\rangle$ appears twice in Eq.~\eqref{eq:spectral_function}. It enters as an addition for the reference state $|\Psi_{\nu,n}\rangle$, with amplitude $P^{(\nu,n)}_{m}(p)$, and as a removal for the reference state $|\Psi_{\nu+1,m}\rangle$, with amplitude $Q^{(\nu+1,m)}_{n}(p)$. These two amplitudes are equal. To count each transition once, we keep only the additions. We model the broadening by the leads with a Lorentzian of full width $\Gamma$, which gives
        \begin{align}
            \mathcal{A}^{+}_{p}(\omega) = \sum_{\nu,n,m} |P^{(\nu,n)}_{m}(p)|^{2}\, \frac{\Gamma}{(\omega - \Delta_{mn})^{2} + \Gamma^{2}/4},
            \label{eq:addition_energies}
        \end{align}
    where $\Delta_{mn} = E_{\nu+1,m} - E_{\nu,n}$ are the addition energies. Since $\hat{H}_{0}$ conserves the number of fermions, each addition connects a sector with $\nu$ fermions to the sector with $\nu+1$ fermions. These sectors contain $\binom{D}{\nu}$ and $\binom{D}{\nu+1}$ eigenstates, respectively. Summing the products over $\nu$ gives at most $\sum_{\nu=0}^{D-1}\binom{D}{\nu}\binom{D}{\nu+1} = \binom{2D}{D-1}$ addition energies, that is, 56 for $D=4$. At $U=0$, the transitions with nonzero $P^{(\nu,n)}_{m}(p)$ collapse onto the four eigenvalues $\{\lambda_m\}$. At $U\neq0$, the interaction generically separates them. 

    We focus on the current from L to S and hence on the spectral function of site 1. The current from R to S admits an analogous analysis for site $D$. At $U\neq0$, the term $U\hat{n}_{1}\hat{n}_{2}$ makes the energy needed to add a fermion to site 1 depend on the occupation of site 2. We refer to the configuration that the added fermion creates as \emph{paired} if site 2 is occupied and as \emph{unpaired} if site 2 is empty. In the occupation basis $|n_{1} n_{2} \cdots \rangle$, an unpaired configuration $|10 \cdots \rangle$ costs approximately $\varepsilon_{1}$. A paired configuration $|11 \cdots \rangle$ costs approximately $\varepsilon_{1}+U$. The spectral function thus has two groups of peaks, shown in blue (unpaired) and gold (paired) in Fig.~\ref{Fig: J separated by modes (with U)}(b) and (c). To classify each transition, we split $\hat{c}^{\dagger}_{1} = \hat{c}^{\dagger}_{1}(1-\hat{n}_{2}) + \hat{c}^{\dagger}_{1}\hat{n}_{2}$, where the first term adds a fermion next to an empty site 2 and the second next to an occupied one. This split decomposes the amplitude $P^{(\nu,n)}_{m}(1)$ into an unpaired and a paired part. We classify the transition according to the part with the larger modulus.

    The comparison between the energy-resolved current from L to S and $\mathcal{A}^{+}_{1}$ in Fig.~\ref{Fig: J separated by modes (with U)}(b) and (c) shows that the sign of $U$ selects which group of transitions dominates the current. At $U=3$, the current peaks at $\varepsilon_k \approx -3.5$, where the transitions create mainly unpaired configurations. At $U=-3$, it peaks at $\varepsilon_k \approx -7$, where the transitions create mainly paired configurations. For both signs of $U$, the dominant group is the one with the lower addition energies. The same selection holds at $|U|=5$, for which we omit the addition energies to keep the panels readable. The coupling to the leads merges the many addition energies of each group into the wide features of panels (b) and (c). The type of configuration that the injected fermion creates also affects how the fermion propagates. This effect motivates the analysis in Appendix~\ref{relaxation_rate}, which addresses the sign asymmetry of the relaxation rate found in Sec.~\ref{EffectofU}.
    
    The addition energies also explain why the energy window of the lead modes must grow with $|U|$. For $U<0$, the addition energies of paired transitions at site 1 shift to lower values, which moves the peak of the left-lead current from $\varepsilon_k \approx -7$ at $|U|=3$ to $\varepsilon_k \approx -9$ at $|U|=5$, beyond $-W$. For $U>0$, the peak stays near $-3.5$, since the addition energies of unpaired transitions are approximately independent of $U$. Nevertheless, for $U>0$ we expect the dominant contributions to the current from R to S to shift toward positive $\varepsilon_k$, beyond $W$. We therefore use the enlarged window $[-2W, 2W]$ for $|U|=5$, which we keep symmetric for both leads.
        
\subsection{Efficiency of an interacting diode}

    \begin{figure}[t]
    
        \includegraphics[width=0.4 \paperwidth]{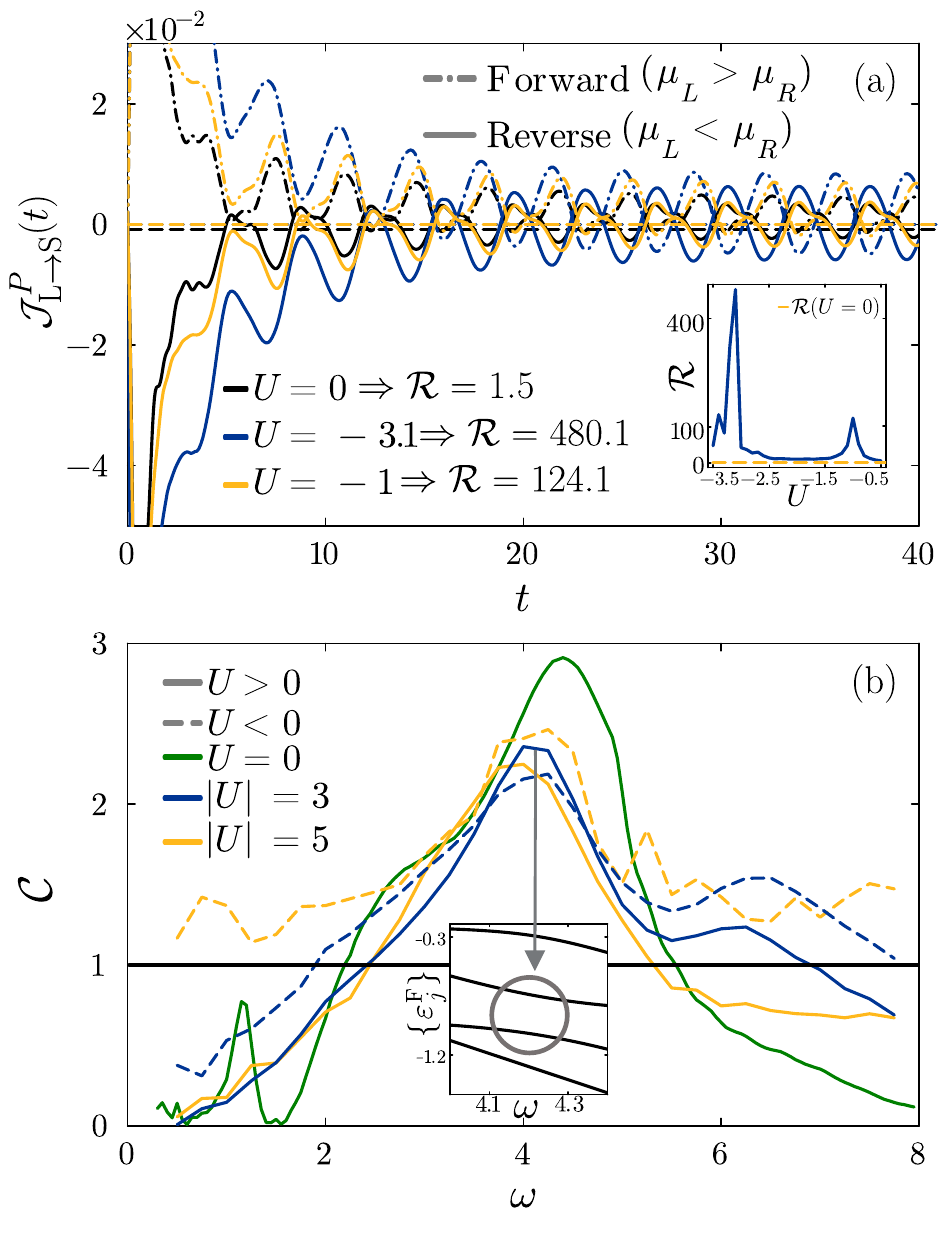}   
        \caption{Efficiency of an interacting diode. Here $E = 2$, $D=4$, $\mu_{\text{L}} = 2$ and $\mu_{\text{R}} = 1$ in forward bias, $T_{\text{L}, \text{R}} = 10$, $N = 40$ for $|U|\leq3$ and $N = 80$ otherwise. (a) Particle current $\mathcal{J}_{\text{L}\rightarrow \text{S}}^{P}(t)$ for forward (dash-dot lines) and reverse (solid lines) using $\omega = 1.75$. The horizontal dashed line signals the average value in the NESS. The rectification coefficient $\mathcal{R}$ is shown for different values of $U$. The inset shows $\mathcal{R}$ over the full range of $U$ considered. (b) Contrast coefficient $\mathcal{C}$ for particle transport and various interactions. The inset shows the quasienergy spectrum of the stroboscopic time-independent effective Hamiltonian $\hat{H}_{\text{eff}} = i \ln(\hat{\mathcal{U}}_{\text{F}})/T$ for $U = 3$. The arrow signals the avoided crossing that occurs at the resonance frequency where the current is maximized.}            
        \label{Fig: Efficiency with U}
    \end{figure}          

    Finally, we study the efficiency of the rectification mechanism by varying $\omega$ and $U$. Since $U$ modifies the spectrum of $\hat{H}_{\text{S}}(t)$, the optimal frequencies matching the avoided crossings will change, and hence the resonant peaks will not coincide with those of the non-interacting case. As exemplified in Fig.~\ref{Fig: Efficiency with U}(a) for a generic frequency $\omega = 1.75$, both the shift of the time-average values of the particle current and the change in the amplitude result in an enhancement of $\mathcal{R}$. In general, $\mathcal{R}$ will be enhanced whenever the reverse current is shifted towards zero while the forward current is displaced away from it. The inset of Fig.~\ref{Fig: Efficiency with U}(a) shows that the enhancement is robust and $\mathcal{R}$ exceeds the non-interacting reference $\mathcal{R}(U=0)$ for every $U \in [ -3.5, -0.5]$. However, the enhancement is strongly non-monotonic in $U$: it is far more pronounced near $U=-3.1$ and $U=-1$ than for other values of $U$. In addition to the enhancement of $\mathcal{R}$, the rectification mechanism itself remains robust, as shown in Fig.~\ref{Fig: Efficiency with U}(b). Apart from the expected shift in frequency, the frequency at which $\mathcal{C}$ is maximal still corresponds to avoided crossings in the Floquet quasienergy spectrum, as depicted in the inset of Fig.~\ref{Fig: Efficiency with U}(b). This result indicates that the avoided crossings still play the main role in the dynamics of the tilted driven system. 

    Our setup continues to exhibit the two distinct regimes of standard and bipolar rectification with perfect rectification at their boundary. The region displaying bipolar rectification extends over a broader range of driving frequencies. For $U = -5$, the bipolar region is extended to the full range of driving frequencies under consideration. This extension suggests that the drive-induced contribution to the current dominates over any bias-induced current already present in the undriven interacting chain. This robustness of the phenomenology is particularly noteworthy because the rectification mechanism relies on the interplay between the bias-induced and drive-induced contributions to the current. In the non-interacting regime, these contributions were nearly independent. Once $U$ is introduced, one might reasonably expect these contributions to interfere with each other, potentially leading to qualitatively different physics in which rectification is disrupted or even absent. However, our results show that the mechanism not only survives but is strengthened with $U \neq 0$, demonstrating that the observed rectification is not an artifact of neglecting interactions, but a stable feature of the driven system.   
    
\section{Conclusions and outlook}
\label{Conclusion}

    In this work, we show that a time-periodic drive locally modulating a tilted potential generates particle and energy currents through a system coupled to external baths, even when both baths have identical temperatures and chemical potentials. In the presence of a bias, the drive-induced contribution can compete with or reinforce the bias-induced contribution to the net current. This competition gives rise to two distinct rectification regimes as a function of the driving frequency: standard $(\mathcal{C} \leq 1)$ and bipolar $(\mathcal{C} > 1)$ rectification, with perfect rectification $(\mathcal{C} = 1)$ occurring at their boundary. The rectification is maximized at the resonant peaks of the drive-induced current. These peaks occur at frequencies that generate avoided crossings in the Floquet band structure and optimally align the drive-induced Floquet channels with the intrinsic pathways for conduction of the system. This behavior shows that the Floquet states act as intermediate channels that control and activate the transport. From a thermodynamic point of view, we also show that the diode not only rectifies fermionic transport but also entropy production and energy absorption. We first establish the rectification mechanism in the non-interacting case ($U=0$), where the two regimes and their connection to Floquet avoided crossings are most transparent. Introducing nearest-neighbor interactions ($U \neq 0$) shifts the mean value of the current and changes the amplitude of its oscillations. The resonant frequencies are shifted as well, since interactions modify the underlying energy spectrum of the system. Despite this, the rectification mechanism remains robust: both regimes persist, and their signatures---including their correspondence to avoided crossings in the Floquet spectrum---survive the introduction of interactions. Crucially, the mechanism is strengthened, with $U$ enhancing $\mathcal{R}$ and $\mathcal{C}$ at fixed frequency and widening the parameter region exhibiting bipolar rectification.
    
    Although we have focused on particle transport to illustrate and understand the rectification behavior, the same mechanism is expected to apply to energy transport. This is because rectification arises solely from the interplay between the drive-induced and bias-induced contributions to the current, and this competition operates in the same way for particle and energy transport, as shown in Fig.~\ref{Fig: Interplay}. We expect this behavior to hold even in the interacting regime, since Fig.~\ref{Fig: Effect of U on JP and JE}(b) shows that a finite $U$ also shifts the energy current, an effect that can bring the reverse current to zero and thereby enhance rectification.  
    
    Overall, our results show that the two regimes of bipolar and standard rectification that emerge in our setup are a robust phenomenon that persists in larger systems and can be enhanced in the presence of interactions. These features make the mechanism promising for experimental realization. In particular, tilted lattice potentials and interacting fermionic chains similar to those considered here have already been realized in trapped-ion \cite{Morong2021-yg}, cold-atom \cite{Guardado-Sanchez2020-zp, PRXQuantum.2.040325}, and superconducting-qubit platforms \cite{PhysRevLett.127.240502}. Moreover, periodic driving has been experimentally demonstrated in optical lattices with ultracold fermionic potassium atoms \cite{Bordia2017-km} and in gold-graphene-gold heterostructures illuminated with a Gaussian laser \cite{boolakee2022light}, providing a natural setting for the driven protocols studied here. Quantum computing devices \cite{doi:10.1126/science.adh9932} and atomtronic architectures \cite{RevModPhys.94.041001} may also offer suitable platforms for quantum simulation of the present scheme. In this direction, our proposal based on a periodic train of Gaussian laser pulses can be regarded as a concrete realization of bipolar rectification, a regime that has not yet been observed experimentally, to our knowledge.

    Our results open several directions for future research. An important next step is to relate the parameters of our model to those of real materials in order to assess the feasibility of observing the rectification mechanism experimentally and to identify the most promising physical realizations. More broadly, the same frequency-controlled competition that enables diode-like behavior in one dimension could be exploited to engineer other transport devices, such as transistor-like elements for particle and energy flow \cite{10.1063/1.2191730, PhysRevB.99.035129} or thermopower generators \cite{ML_Tensor_Networks, Guzman2024Key}. Finally, it would be interesting to explore the spinful case and investigate whether rectification can be decomposed into spin-resolved contributions, potentially enabling selective rectification of only one spin species.   
    
\section*{Acknowledgements} 
    We thank J. C. Egues, P. H. Penteado, and I. Franco for insightful discussions. We also thank G. T. Landi for suggesting the analysis based on energy-resolved currents. This research was supported in part by the University of Pittsburgh Center for Research Computing and Data, RRID:SCR$\_$022735, through the resources provided. Specifically, this work used the H2P cluster, which is supported by NSF award number OAC-2117681. This work used Bridges-2 GPU and Ocean at the Pittsburgh Supercomputing Center through allocation PHY260126 from the Advanced Cyberinfrastructure Coordination Ecosystem: Services \& Support (ACCESS) program, which is supported by U.S. National Science Foundation grants \#2138259, \#2138286, \#2138307, \#2137603, and \#2138296~\cite{Boerner2023ACCESS}.

\section*{Author contributions}
    J.P.P. and J.J.M.A. conceived the project and designed the research. J.P.P. implemented all numerical codes, performed the simulations, carried out the analytical derivations, prepared the figures, and wrote the manuscript. J.J.M.A. supervised the work and provided continuous feedback on the calculations, simulations, and interpretation of the results. Both authors contributed to discussions and to the final editing of the manuscript.    


\appendix{}

\section{Effective driving from a train of light pulses}
\label{Driving_from_light}

    Here we show that the driving considered in Eq.~\eqref{local_driving} is an effective description of the driving induced by a train of light pulses incident on the center of a material. Within the dipole approximation, the incident electromagnetic field couples directly to the local charge density of the material and therefore only modifies the on-site energies \cite{boolakee2022light, 10.1021/jp505771f}. The hopping and nearest-neighbor interaction terms are left unchanged. For the former, the dipole matrix elements between distinct Wannier orbitals, $\langle w_{k} | \hat{x} | w_{l} \rangle$ with $k \neq l$, decay exponentially with the ratio of the lattice constant to the orbital width. On the other hand, the nearest-neighbor interaction involves no dipole matrix element at all. We therefore omit both terms in the following discussion, where they would appear unmodified. The Hamiltonian describing the system in the presence of the light-matter coupling can thus be written as    
    \begin{equation}
                \hat{H}_{\text{S}}(t) = \sum_{i} \varepsilon_{i} \hat{c}_{i}^{\dagger}\hat{c}_{i} + |e|\sum_j [\vec{E}_{j}(t) \cdot \vec{r}_{j} ]\hat{c}_{j}^{\dagger}\hat{c}_{j}.
    \end{equation}

    Following the convention of Eq.~\eqref{local_driving}, the index $i$ $(j)$ refers to the undriven (driven) sites. Here $\vec{E}_{j}$ denotes the electric field at site $j$, $\vec{r}_{j}$ the real-space position of that site, and $e$ is the electron charge, for which take $|e| = 1$. Three length scales control the coupling. The Wannier orbitals are localized on a scale much smaller than the lattice constant, so the orbital at site $j$ samples the field at the single position $\vec{r}_{j}$. The confinement length of the field greatly exceeds the orbital width, so the field is uniform across each orbital and the dipole approximation applies~\cite{Rivera2016Shrinking}. This length need not exceed the lattice constant~\cite{Benz2016Picocavities}. The same length is much smaller than the chain, so the field is appreciable only over the central sites. We further take the field amplitude to be uniform across the illuminated region, and the potential it induces to take a common value at the two contact sites, namely sites 1 and $D$, which couple to the left and right leads. The illumination therefore leaves the contact sites unbiased. The chain lies along $\hat{x}$, so $\vec{r}_{j} = r_{j}\hat{x}$ and only the longitudinal component $\vec{E}_{j}(t) \cdot \hat{x}$ contributes. For a Gaussian laser pulse with amplitude $E_j$, frequency $\omega$, phase $\phi$, width $\sigma$, and pulse center $t_{0}$, this component is given by $E_j(t) = E_{j} e^{-(t-t_{0})^{2}/2\sigma^2}\cos(\omega t + \phi)$. Extending this expression to a train of light pulses centered at times $t_{0} = 4\pi n /\omega$, with $n \in \mathbb{Z}$, the total electric field on site $j$ becomes $\sum_{n} E_{j}(t, t_{0} = 4\pi n/\omega)$, where the second argument specifies the center time of each pulse in the train. The Hamiltonian can then be rewritten as
    \begin{equation}
        \begin{aligned}
            \hat{H}_{\text{S}}(t) & = \sum_{i} \varepsilon_{i} \hat{c}_{i}^{\dagger}\hat{c}_{i} + \sum_j [E_{j}(t) r_{j}]\hat{c}_{j}^{\dagger}\hat{c}_{j} \\
            & = \sum_{i} \varepsilon_{i} \hat{c}_{i}^{\dagger}\hat{c}_{i} + \sum_j \sum_{n}( E_{j}(t, t_{0} = 4\pi n/\omega) r_{j})\hat{c}_{j}^{\dagger}\hat{c}_{j}.
        \end{aligned}
    \end{equation}
    
    The correspondence between the two driving protocols requires three choices: $E_j r_j = \varepsilon_j$, $\phi = 0$, and $\sigma = 5/\omega$. Setting the lattice constant to unity and placing the origin at the center of the chain puts site $j$ at $r_{j} = j - (D+1)/2$. The amplitudes of Eq.~\eqref{local_driving} then read $\varepsilon_{j} = E r_{j}$, so the condition $E_j r_j = \varepsilon_j$ reduces to $E_j = E$. The field thus takes the same amplitude at every illuminated site, and the site dependence of the driving is carried entirely by the positions $r_{j}$. The choice $\phi = 0$ aligns the field of the train with $\cos(\omega t)$, matching the phase convention of Eq.~\eqref{local_driving}. The frequency dependence of $\sigma$ is essential for this correspondence: scaling $\sigma \propto 1/\omega$ keeps the overlap between neighboring pulses fixed across driving frequencies, which lets the pulse train reconstruct the harmonic profile $\cos(\omega t)$ regardless of $\omega$. We verify this by analyzing the envelope of the pulse train, $C(t) = \sum_{n} e^{-(t - 4\pi n/\omega)^{2}/2\sigma^{2}}$. The factor $E_{j} \cos(\omega t)$ is common to every pulse of the train, so it factors out of the sum:
    \begin{equation}
    \sum_{n} E_{j}(t, 4\pi n/\omega) = E_{j} \cos(\omega t) \, C(t) .
    \end{equation}
    
    Physically, $C(t)$ stays close to a constant when neighboring pulses overlap substantially. The product $E_j\cos(\omega t)\,C(t)$ approximates the continuous drive $E_j\cos(\omega t)$ rather than a train of isolated spikes. The overlap between neighboring pulses is set by the ratio $\sigma/T_{\text{p}}$, where $T_{\text{p}} = 4\pi/\omega$ is the pulse spacing. This ratio equals $\sigma \omega /4\pi$, so scaling $\sigma$ with $1/\omega$ makes it independent of the driving frequency. Since $C(t)$ has period $T_{\text{p}}$, it admits a Fourier series in the harmonics $m\Omega$, with $\Omega = 2\pi/T_{\text{p}} = \omega/2$. Its coefficients follow from the periodization identity for a Gaussian train: for $h(x) = \sum_{n} e^{-(x-n)^{2}/2\tau}$, the Fourier coefficients are $\hat{h}(k) = \sqrt{2\pi\tau}\, e^{-2\pi^{2}k^{2}\tau}$ \cite{osgood2019fourier}. Rescaling $x = t/T_{\text{p}}$ maps $h(x)$ onto $C(t)$, with $\tau = \sigma^{2}/T_{\text{p}}^{2}$. The coefficients of $C(t)$ are $c_{m} = \frac{\sqrt{2\pi}\,\sigma}{T}\, e^{-2\pi^{2}m^{2}\sigma^{2}/T_{\text{p}}^{2}} = \frac{\sqrt{2\pi}\,\sigma\omega}{4\pi}\, e^{-m^{2}\sigma^{2}\omega^{2}/8}$, where the last equality follows from substituting $T_{\text{p}} = 4\pi/\omega$. The coefficients decay as a Gaussian in $m$, with width set by $\sigma\omega$. Our choice $\sigma = 5/\omega$ approximates $4\pi/(\sqrt{2\pi}\,\omega)$ to within $0.3\%$. This value sets $c_{0} = 1$ and reduces the exponent to $\pi m^{2}$, giving
    \begin{equation}
    C(t) = 1 + 2e^{-\pi}\cos(\omega t/2) + \mathcal{O}(10^{-5}) .
    \end{equation}
    The train thus reproduces the harmonic driving up to a residual oscillation of amplitude $2e^{-\pi} \simeq 8.6\%$ at frequency $\omega/2$.
    
    \begin{figure}[t]
        \centering
        \includegraphics[width=0.4 \paperwidth]{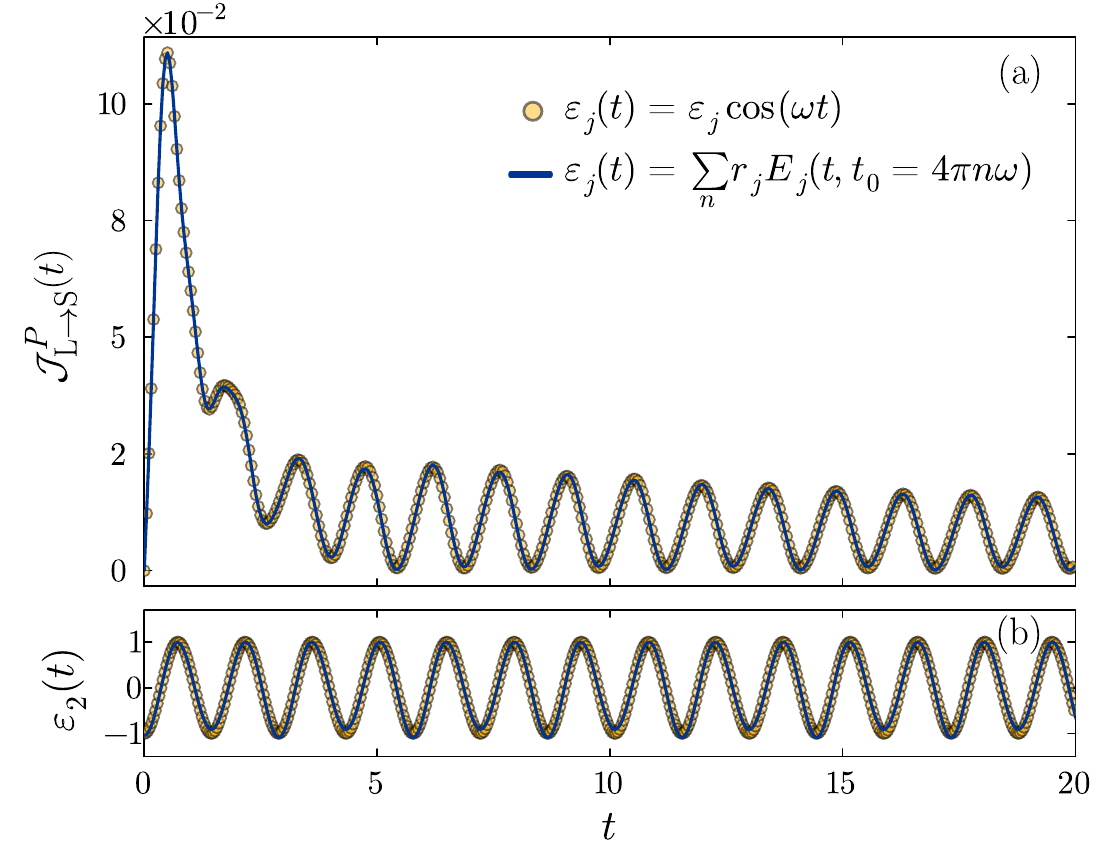}
        \caption{Harmonic driving as a sum of pulses. Here $E = 2$, $D=4$, $\omega = 4.35$, $\mu_{\text{L}} = 2$, $\mu_{\text{R}} = 1$, $T_{\text{L}, \text{R}} = 10$, and $N = 140$. (a) Particle current $\mathcal{J}_{\text{L} \rightarrow \text{S}}(t)$ for different types of periodic drivings. (b) On-site potential for site $2$ modulated by each driving.}
        \label{fig: Comparison_Light_Pulses}
    \end{figure}   

    Finally, we verify this correspondence numerically by focusing on the resonant peak of the non-interacting case, $\omega = 4.35$. As shown in Fig.~\ref{fig: Comparison_Light_Pulses}(a), the transport observables such as the particle current are nearly identical whether computed using $\varepsilon_j (t) = \sum_n E_{j}(t, 4\pi n/\omega) r_{j}$ or using the harmonic drive $\varepsilon_j (t) = \varepsilon_j \cos(\omega t)$. For completeness, Fig.~\ref{fig: Comparison_Light_Pulses}(b) shows the time profile of the on-site potential on site 2, revealing the close agreement between the two descriptions.

\section{Exact solution for non-interacting case via Lyapunov Equation}
\label{Lyapunov_Appendix}

    In the absence of interactions, any quadratic $\hat{H}(t)$ characterizing a system of $M$ sites can be described as a matrix $\mathbf{H}(t)  \in \mathbb{C}^{M \times M}$. The latter is known as the single-particle Hamiltonian, where
    \begin{equation}
        \hat{H}(t) = \sum_{i,j = 1}^{M} [\mathbf{H}(t)]_{ij} \hat{d}^{\dagger}_{i}\hat{d}_{j} .
        \label{single-particle_H}
    \end{equation}  

    We can apply this formulation to the extended system+leads Hamiltonian given by Eq.~\eqref{H_ML}. Here, $M = 2N  + D$ and $\hat{d}_i$ $(\hat{d}_i^{\dagger})$ is a fermionic annihilation (creation) operator that corresponds to left lead sites if $i \leq N$, to system sites if $N < i \leq D+N$, and to right lead sites if $i > N + D$. Thus, the system+leads single-particle Hamiltonian can be written as
    \begin{equation}
    \mathbf{H}(t) =
                    \left[
                    \begin{array}{ccc}
                    \mathbf{H_{\text{L}} } & \mathbf{H_{\text{LS}}} & \mathbf{0} \\
                    \mathbf{H^{\dagger}_{\text{LS}}} & \mathbf{H_{\text{S}}}(t) & \mathbf{H_{\text{SR}}} \\
                    \mathbf{0} & \mathbf{H^{\dagger}_{\text{RS}}} & \mathbf{H_{\text{R}}} 
                    \end{array}
                    \right].
    \end{equation}     
    where each block corresponds to the single-particle Hamiltonian of the leads $(\mathbf{H_{\alpha}})$, the system $(\mathbf{H_{S}})$, and their coupling $(\mathbf{H_{\alpha S}})$.

    If our initial state is Gaussian, the dynamics of the whole setup will remain Gaussian. Since the NESS is independent of the initial state, we are free to choose a Gaussian initial state without loss of generality. This is convenient because, for Gaussian states, the many-body density matrix $\hat{\rho}(t) \in \mathbb{C}^{2^M \times 2^M}$ can be obtained directly from the correlation matrix 
    \begin{equation}
        \mathbf{C_{ij}}(t) = \textrm{Tr}(\hat{d}^{\dagger}_{j} \hat{d}_{i} \hat{\rho}(t))  \in \mathbb{C}^{M \times M},
        \label{C_Matrix}
    \end{equation}
    which is also known as the single-particle density matrix. In fact, $\hat{\rho}(t)$ can be reconstructed from $\mathbf{C}(t)$ as $\hat{\rho}(t)~=~\exp(-\sum_{i,j = 1}^{M} [\mathbf{M}(t)]_{ij} \hat{d}^{\dagger}_{i}\hat{d}_{j})/Z$ with $\mathbf{M(t)} = \ln [\mathbf{1 - C(t)}] -\ln[\mathbf{C(t)}]$ and $Z^{-1} = \det[ \mathbf{1 - C(t)}]$ \cite{ML_Entropy_Production}. 
    
    The Lyapunov equation \cite{ML_Fast_Driving} arises as the equation of motion of $\mathbf{C}(t)$ and is given by
    \begin{equation}
        \frac{d \mathbf{C}(t)}{dt} = -[\mathbf{W}(t) \mathbf{C}(t) + \mathbf{C}(t)\mathbf{W^{\dagger}}(t)] + \mathbf{F} .
        \label{Lyapunov}
    \end{equation}

    Here $\mathbf{W} (t) = i \mathbf{H}(t) + \boldsymbol{\Upsilon}/2$,
    \begin{equation}
        \boldsymbol{\Upsilon}=\left[\begin{array}{ccc}
        \boldsymbol{\Upsilon_{\text{L}} }& \mathbf{0} & \mathbf{0} \\
        \mathbf{0} & \mathbf{0} & \mathbf{0} \\
        \mathbf{0} & \mathbf{0} & \boldsymbol{\Upsilon_{\text{R}} }
        \end{array}\right], \quad \mathbf{F}=\left[\begin{array}{ccc}
        \mathbf{F_{\text{L}} } & \mathbf{0} & \mathbf{0} \\
        \mathbf{0} & \mathbf{0} & \mathbf{0} \\
        \mathbf{0} & \mathbf{0} & \mathbf{F_{\text{R}} }
        \end{array}\right],
    \end{equation}
    $\boldsymbol{\Upsilon_{\alpha}} = \textrm{diag} \{ \gamma_{k \alpha}\}$, and $\mathbf{F_{\alpha} }= \textrm{diag} \{ \gamma_{k \alpha} f (\varepsilon_{k}) \}$. Unlike the many-body density matrix, whose dimension grows exponentially with system size, $\mathbf{C}(t)$ scales only linearly with the number of sites. Hence, Eq.~\eqref{Lyapunov} can be solved efficiently using standard Runge-Kutta methods.

\section{Alternative definitions of currents}
\label{All_Currents}
    
    The currents defined in Eqs.~\eqref{J:a} and~\eqref{J:b} are the result of 
    \begin{equation}
        \mathcal{J}_{\alpha}^{P} (t) = - \frac{d \langle \hat{N}_{\alpha} \rangle}{dt},    \quad \textrm{and } \quad  \mathcal{J}_{\alpha}^{E} (t) = -\frac{d \langle \hat{H}_{\alpha} \rangle}{dt}.
        \label{Continuity_EQ_def_currents}
    \end{equation} 
    
    Here $\hat{N}_{\alpha} = \sum_{k=1}^{\infty} \hat{b}^{\dagger}_{k}\hat{b}_{k}$ and $\hat{H}_{\alpha} = \sum_{k=1}^{\infty} \omega_k \hat{b}_{k}^{\dagger}\hat{b}_{k}$, where $b_{k}$ are the bath operators from Eq.~\eqref{H_beforeML}. Although written in terms of lead operators in Eqs.~\eqref{J:a} and~\eqref{J:b}, these currents correspond to the flow of particles and energy from the macroscopic bath into the system. We call them ``internal'' currents because they flow within the extended system formed by the system and the discrete lead modes, as opposed to the currents flowing between the leads and the residual baths, which we introduce below. This internal character means that their expectation value do not depend on the microscopic modeling of the baths. The same physics at a given time should therefore be expected if, instead of the mesoscopic-leads approach, a chain-mapping description of the baths is used \cite{ML_Fast_Driving}. 

    As an alternative to this formulation, we can define ``external'' currents corresponding to fermions flowing from the residual baths into the leads from the dissipation in Eq.~\eqref{Dissipators}. These definitions are given by $\mathcal{I}_{\alpha}^{P} = \textrm{Tr}[\hat{N} \mathcal{L}_{\alpha}(\hat{\rho})]$ for the particle current, and $\mathcal{I}_{\alpha}^{E} = \textrm{Tr}[\hat{H} \mathcal{L}_{\alpha}(\hat{\rho})]$ for the energy current. These expressions are dependent on the microscopic modeling of the bath and reduce to
    \begin{subequations}
        \label{I}
        \begin{align}
        \mathcal{I}^{P}_{\alpha}
        &= \sum_{k=1}^{N} \gamma_{k} 
        \langle f_{\alpha}(\varepsilon_{k}) - \hat{a}_{k , \alpha}^{\dagger}\hat{a}_{k , \alpha} \rangle \textrm{, and}
        \label{I:a} \\
        \mathcal{I}^{E}_{\alpha}
        &=
        \begin{aligned}[t]
        &\sum_{k=1}^{N} \gamma_{k} \varepsilon_{k} 
        \langle f_{\alpha}(\varepsilon_{k}) - \hat{a}_{k , \alpha}^{\dagger}\hat{a}_{k , \alpha} \rangle \\
        &\quad - \frac{1}{2} \sum_{k=1}^{N} \gamma_{k} 
        \langle \kappa_{kp} \hat{c}_{p}^{\dagger}\hat{a}_{k , \alpha} 
        + \kappa_{kp}^{*}\hat{a}_{k , \alpha}^{\dagger}\hat{c}_{p} \rangle .
        \end{aligned}
        \label{I:b}
        \end{align}
    \end{subequations}
    Here $\hat{a}_{k,\alpha}$ annihilates a fermion in mode $k$ of lead $\alpha$, and $\hat{c}_p$ annihilates a fermion at site $p$ of the system.

    \begin{figure}[b]
        \centering
        \includegraphics[width=0.4 \paperwidth]{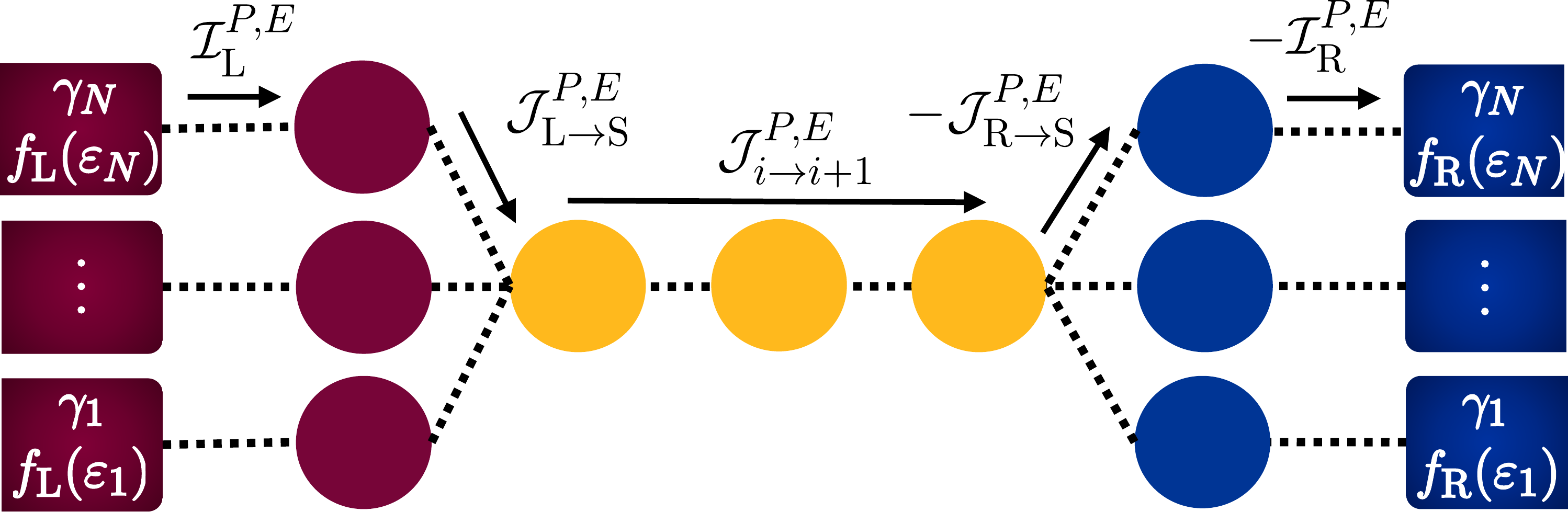}
        \caption{Schematic representation of all definitions of the currents. The arrows represent the positive direction of the currents.}
        \label{Fig: Scheme_currents}
    \end{figure}  

    On the other hand, once $\hat{H}_{\text{S}} = \sum_{j=1}^{D-1} \hat{h}_{j,j+1}$ is chosen, the particle and energy currents through the site $j$ of the system follow from the continuity equation and are given by the expectation values of $\hat{\mathcal{J}}_{j}^{P} = i [\hat{h}_{j,j+1}, \hat{n}_{j} ]$, and $\hat{\mathcal{J}}_{j}^{E} = i [\hat{h}_{j-1,j}, \hat{h}_{j, j+1} ]$ respectively. For our specific model given by Eq.~\eqref{Model_EQ}, these operators take the form
    \begin{subequations}
    \label{J_inside_sytem}
    \begin{align}
    \hat{\mathcal{J}}^{P}_{j \rightarrow j+1}
    &= i t_{s}( \hat{c}_{j}^{\dagger}\hat{c}_{j+1}  - \hat{c}_{j+1}^{\dagger}\hat{c}_{j} ) \text{, and}
    \label{J_inside_sytem:a} \\
    \hat{\mathcal{J}}^{E}_{j-1 \rightarrow j+1}
    &=
    \begin{aligned}[t]
    & i t_{s}^{2}( \hat{c}_{j-1}\hat{c}^{\dagger}_{j+1}  +  \hat{c}_{j-1}^{\dagger}\hat{c}_{j+1} ) \\
    &\quad + U (\hat{\mathcal{J}}_{j-1 \rightarrow j}^{P} \hat{n}_{j+1} + \hat{n}_{j-1} \hat{\mathcal{J}}_{j \rightarrow j+1}^{P}) \\
    &\quad + \frac{\varepsilon_{j}}{2}(\hat{\mathcal{J}}_{j-1 \rightarrow j}^{P} + \hat{\mathcal{J}}_{j \rightarrow j+1}^{P}) .
    \end{aligned}
    \label{J_inside_sytem:b}
    \end{align}
    \end{subequations} 

   \begin{figure}[t]
    \centering
        \includegraphics[width=0.4 \paperwidth]{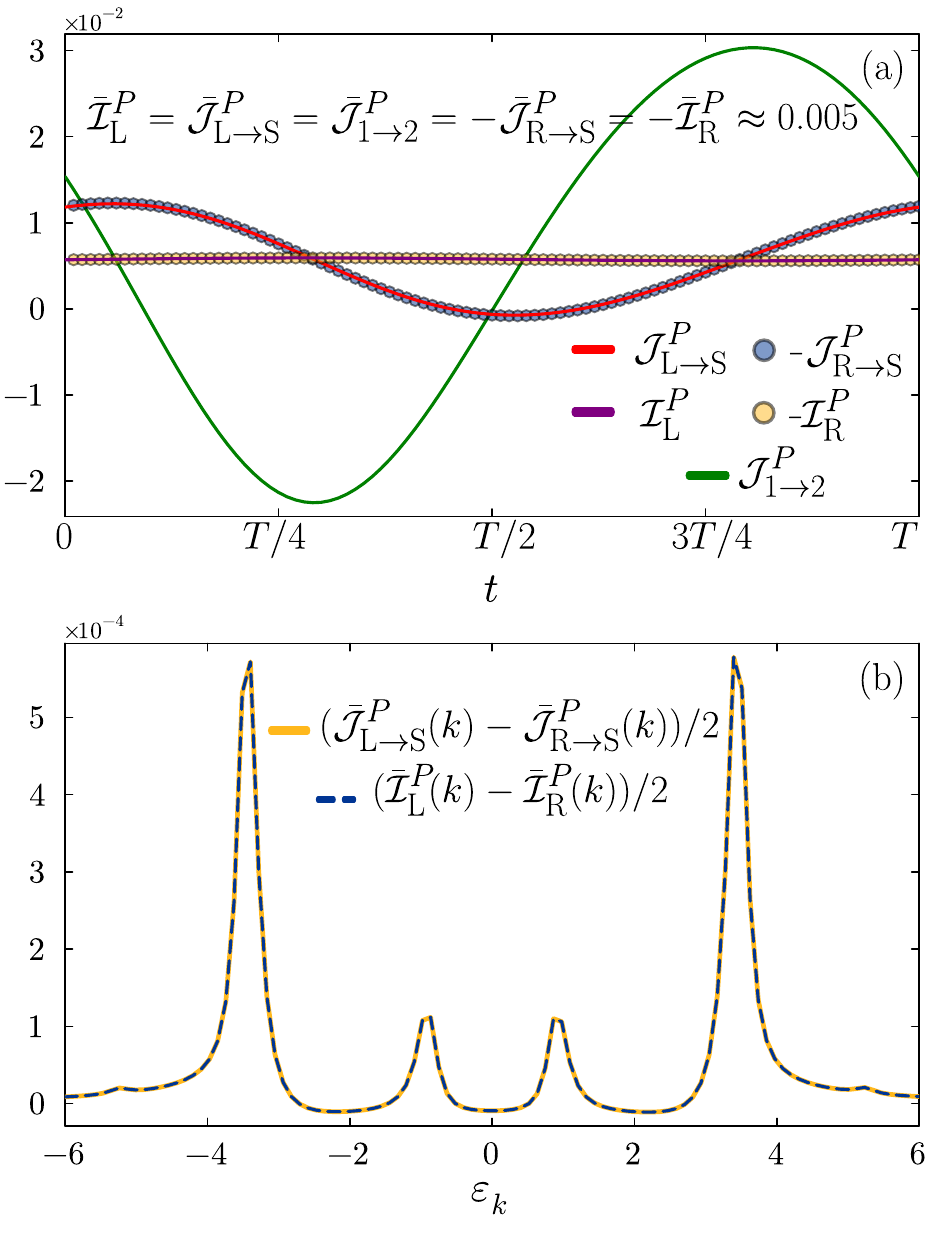}
        \caption{Comparison between transport properties calculated from different definitions of the particle current. For both figures, $\mu_{\text{L}, \text{R}} = 1$, $T_{\text{L}, \text{R}} = 10$, $E= 2$, $D=4$, and $N =140$. (a) Dynamics of all particle currents during 1 cycle after reaching the periodic NESS. (b) Cycle-averaged internal $\mathcal{J}^{P}$ and external $\mathcal{I}^{P}_{\alpha \rightarrow \text{S}}$ particle currents resolved by energy modes.}
        \label{Fig_JvsI}
    \end{figure}    

    All currents are summarized in the scheme of Fig.~\ref{Fig: Scheme_currents}. In the NESS of a time-independent Hamiltonian, we have spatial homogeneity for the current and therefore equality among all definitions:
    \begin{equation}
        \mathcal{I}_{\text{L}}^{P,E} = \mathcal{J}^{P,E}_{ \text{L} \to \text{S}} = \mathcal{J}_{i\rightarrow i+1}^{P,E}  =-\mathcal{J}^{P,E}_{ \text{R} \to \text{S}} = -\mathcal{I}_{\text{R}}^{P,E} .
    \end{equation}

    However, as discussed in Sec.~\ref{ML}, under external time-dependent driving this homogeneity property holds only for the cycle-averaged particle current. This is illustrated in Fig.~\ref{Fig_JvsI}(a), which shows that once the system reaches the periodic NESS, the cycle-averaged value of all current definitions coincides. In addition, decomposing the currents into mode-resolved contributions, as done in Fig.~\ref{Fig: J separated by modes}, can also be performed for the external currents as $\mathcal{I}_{\text{L} \rightarrow \text{S}}^{P, E} = \sum_{k} \mathcal{I}_{\text{L} \rightarrow \text{S}}^{P,E}(k)$. This decomposition yields the same peak structure that can be extracted from resolving the internal currents, as shown in Fig.~\ref{Fig_JvsI}(b). Thus, while this alternative viewpoint does not reveal new physics, it clarifies the origin of the current from the perspective of the external current. The observed deviation of the lead-mode occupations from their undriven values (that is, the corresponding Fermi-Dirac distributions of the leads) in Fig.~\ref{Fig: J separated by modes}(b) is precisely what drives the current.

\section{Numerical assessment of lead and tensor-network parameters}
\label{About_parameters}

\begin{figure}[t!]
        \includegraphics[width=0.4 \paperwidth]{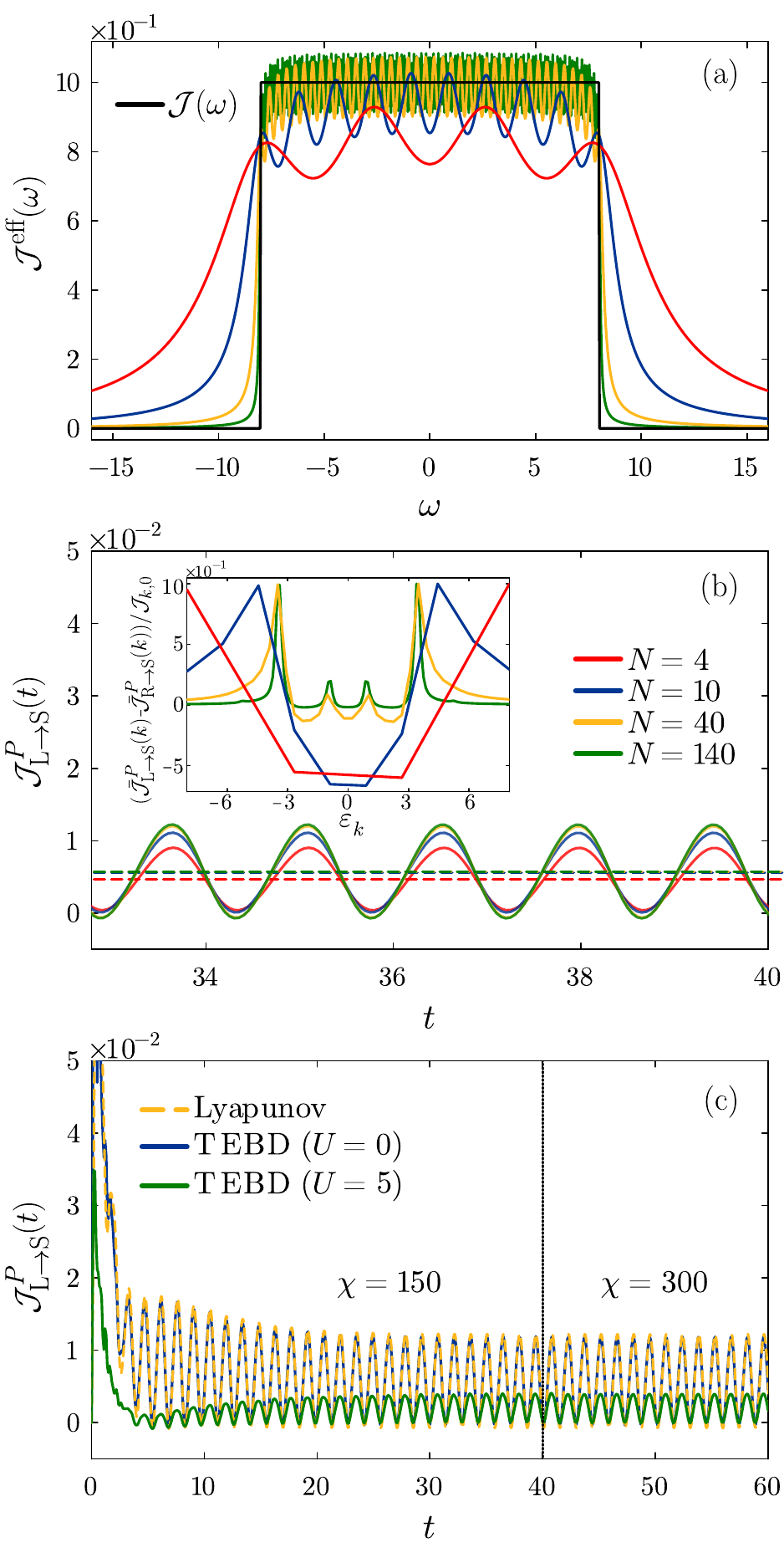}    
        \caption{Assessment of lead and tensor-network parameters.
        (a) Effective spectral density $\mathcal{J}^{\text{eff}}(\omega)$ of the mesoscopic leads for various $N$, compared with the exact spectral density $\mathcal{J}(\omega)$ of the bath. In (b) and (c), $\mu_{\text{L},\text{R}} = 1$, $T_{\text{L},\text{R}} = 10$, $E = 2$, and $D = 4$. (b) Particle current $\mathcal{J}^{P}_{\text{L} \rightarrow \text{S}}(t)$ for different $N$. The inset shows its cycle-averaged energy-mode resolution, normalized by $\mathcal{J}_{0,k} = \max_{k} (\bar{\mathcal{J}}_{\text{L}\rightarrow \text{S}}(k) - \bar{\mathcal{J}}_{\text{R}\rightarrow \text{S}}(k))$. (c) Particle current $\mathcal{J}^{P}(t)$ for different $U$, with $N = 80$ for $U = 5$ and $N = 40$ otherwise. The vertical line marks the increase of the bond dimension from $\chi = 150$ to $300$.}    
        \label{Fig: Parameters}   
    \end{figure}

    As introduced in Sec.~\ref{ML}, the mesoscopic lead approach relies on approximating the spectral density of the baths as a sum of Lorentzian spectral densities, as defined in Eq. \eqref{Lorentzian_J}. This representation depends on the lead parameters $\{\varepsilon_k,\kappa_{kp},\gamma_k\}$ with $k=1,\ldots, N$. Throughout this work we choose $\gamma_k=\varepsilon_{k+1}-\varepsilon_k$ and $\kappa_{kp}=\sqrt{\mathcal{J}(\varepsilon_k)\gamma_k/(2\pi)}$, leaving the number of lead modes $N$ as the only free parameter controlling the accuracy of the effective spectral density $\mathcal{J}^{\mathrm{eff}}(\omega)$. As shown in Fig.~\ref{Fig: Parameters}(a), the flat spectral density considered in this work is overall well reproduced for $N>40$, with the effective spectral density retaining a similar structure as $N$ is increased further.

    While the spectral density itself is well approximated, it is also important to verify that the relevant observables remain unchanged as $N$ is increased. To this end, Fig.~\ref{Fig: Parameters}(b) shows the two main observables considered in this work: the particle current $\mathcal{J}^{P}_{\text{L} \rightarrow \text{S}}$ and the cycle-averaged value of its resolution by energy modes $\mathcal{J}^{P}_{\text{L} \rightarrow \text{S}}(\varepsilon_k)$, whose sum over all modes yields the total current. No relevant changes are observed in the dynamics for $N>40$. For the energy-resolved current, increasing $N$ produces two effects. The first one is a decrease in peak magnitude, since the same cycle-averaged total current is now distributed among a larger number of modes. For this reason, the inset of Fig.~\ref{Fig: Parameters}(b) normalizes the energy-resolved current by $\mathcal{J}_0 = \max_{k} [\bar{\mathcal{J}}_{\text{L}\rightarrow \text{S}}(k) - \bar{\mathcal{J}}_{\text{R}\rightarrow \text{S}}(k)]$. Without this normalization, the magnitudes for different $N$ differ too much to compare the peak shapes. The second is an improvement in energy-mode resolution, manifested as narrower peaks. This behavior is expected since increasing $N$ reduces the spacing between neighboring lead modes, thereby allowing resonances with the system eigenstates to be resolved more accurately. 
    
    The results above demonstrate that the bath-induced dynamics is already well captured by the parameters employed throughout this work. In the interacting case, two additional parameters require analysis: the bond dimension $\chi$ of the MPS representation and the time step $dt$. To study their importance, we take as reference the resonance peak of the non-interacting case at $\omega = 4.35$ and study the dynamics of $J^{P}_{L \rightarrow S}$ for different values of $U$ using the TEBD algorithm. The results are shown in Fig.~\ref{Fig: Parameters}(c). The agreement with the analytical solution obtained from the Lyapunov equation when $U=0$ confirms that the time step $dt = 0.05$ is small enough to justify the Trotter decomposition of the time-evolution operator, and also captures the time-dependence of the Hamiltonian. To assess the effect of the bond dimension, we increase $\chi$ from $150$ to $300$ at the point marked by the vertical line in Fig.~\ref{Fig: Parameters}(c). The current remains unchanged once the system reaches its periodic NESS, confirming that $\chi=150$ already captures the dynamics. Together, these results indicate that our simulations are quantitatively reliable at the parameter values used. 

\section{Comparison with standard boundary-driven Lindblad description}
\label{Lindblad}

    In a standard boundary-driven Lindblad description, the baths are fully traced out and represented solely through dissipative jump operators acting on the system boundaries. The time evolution is still described by Eq.~\eqref{QME}, but the dissipator now takes the form
    \begin{equation}
    \begin{aligned}
            \mathcal{L}_\alpha\{\hat{\rho}\}= & \sum_{k= \{ 1, D\}} \gamma_{k}\left(1-f_{k}\right)\left[\hat{c}_k \hat{\rho} \hat{c}_k^{\dagger}-\frac{1}{2}\left\{\hat{c}_k^{\dagger} \hat{c}_k, \hat{\rho}\right\}\right] \\
            & +\sum_{k= \{ 1, D\}} \gamma_{k} f_{k}\left[\hat{c}_k^{\dagger} \hat{\rho} \hat{c}_k-\frac{1}{2}\left\{\hat{c}_k \hat{c}_k^{\dagger}, \hat{\rho}\right\}\right] .
            \end{aligned}
            \label{Lindblad_Case}
    \end{equation}

    Here $f_{k} \in [0,1]$ is the bias strength parameter, and $\gamma_{k}$ is the coupling rate between the system and the bath connected at boundary site $k$. Unless otherwise stated, we fix $\gamma_{1, D} = 1$ and vary only $f_{k}$. Unlike in Eq.~\eqref{Dissipators}, $f_{k}$ is regarded as a phenomenological parameter rather than as a Fermi--Dirac distribution~\cite{PhysRevLett.106.220601, PhysRevE.84.051115, PhysRevB.87.235130, PRX_JJ, PhysRevLett.120.200603, PhysRevE.99.032136}. This description therefore does not capture finite-temperature effects. Such effects have been approximated through two-site driving schemes~\cite{Prosen_Znidaric_2009, znidaric_thermalization_2010, mendoza-arenas_coexistence_2015, Mendoza-Arenas_Znidaric_2019}, whose accuracy decreases at low temperature. The mesoscopic-leads approach of the main text instead incorporates the temperature through the Fermi--Dirac occupations of the lead modes and remains valid at arbitrary temperature~\cite{ML_Tensor_Networks, ML_Fast_Driving}.

\begin{figure}[t]
        \centering
        \includegraphics[width=0.4 \paperwidth]{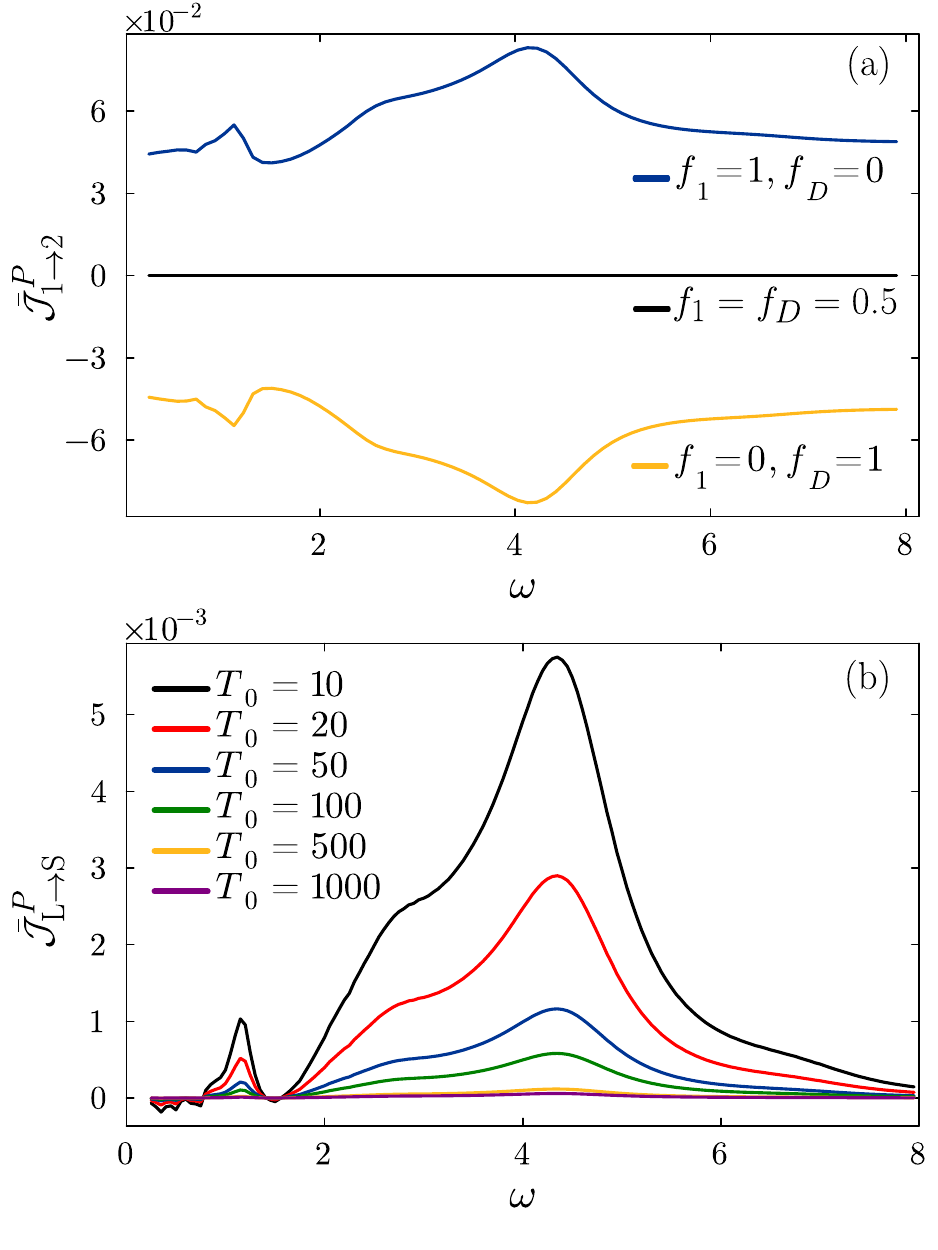}
        \caption{Infinite-temperature transport. We consider $E=2$, $U = 0$, and $D=4$. (a) Particle current from the boundary-driven Lindblad description $\bar{\mathcal{J}}^{P}_{1 \rightarrow 2}$ as a function of $\omega$, without bias $(f_1 = f_{D} = 0.5)$, under maximal bias $(f_{1} = 1, f_{D} = 0)$, and under reversed maximal bias $(f_{1} = 0, f_{D} = 1)$. For maximal bias, the extrema are located at $\omega \approx \{1.15, 4.2 \}$. (b) Particle current from the mesoscopic leads description $\bar{\mathcal{J}}^{P}_{\text{L} \rightarrow \text{S}}$ as a function of $\omega$ zero bias but varying the temperature $T_{\text{L}, \text{R}} = T_{0}$ and using $\mu_{\text{L}, \text{R}} = 1$, and $N=140$.}
        \label{Fig: ML vs Lindblad}
    \end{figure}

    In the scenario described by Eq.~\eqref{Lindblad_Case}, zero bias corresponds to injecting and removing particles at equal rates on both boundaries $(f_1 = f_{D})$. We focus on the symmetric choice $f_{1, D} = 0.5$; other values satisfying $f_1 = f_{D}$ yield the same vanishing current, although quantities such as site populations will differ. In this regime, the cycle-average current vanishes for all frequencies; see Fig.~\ref{Fig: ML vs Lindblad}(a). Physically, this reflects the implicit infinite-temperature character of this boundary-driven description, where the baths fully thermalize the system and drive it toward a NESS proportional to the identity operator. We verify this by checking that $\hat{\rho} \propto \hat{I}$ satisfies the steady-state condition of Eq.~\eqref{Lindblad_Case} only when $f_k = 0.5$. Since $[\hat{O}, \hat{I}] = 0$ for any operator $\hat{O}$, the Hamiltonian term vanishes, and we only need to verify that the dissipation part also vanishes:
    \begin{equation} 
    \begin{aligned}
            \frac{d \hat{I}}{dt} & = \sum_{k} \gamma_{k}\left(1-f_{k}\right)\left[\hat{c}_k \hat{I} \hat{c}_k^{\dagger}-\frac{1}{2}\left\{\hat{c}_k^{\dagger} \hat{c}_k, \hat{I}\right\}\right] \\
            & +\sum_{k} \gamma_{k} f_{k}\left[\hat{c}_k^{\dagger} \hat{I} \hat{c}_k-\frac{1}{2}\left\{\hat{c}_k \hat{c}_k^{\dagger}, \hat{I}\right\}\right] \\
            & = \sum_{k} \gamma_{k}\left(1-f_{k}\right) \left[\hat{c}_k, \hat{c}_k^{\dagger} \right] + \gamma_{k} f_{k } \left[\hat{c}_k^{\dagger}, \hat{c}_k \right]\\
            & = \sum_{k} \gamma_{k}\left(1-2f_{k}\right) (1- 2 \hat{n}_{k}) = 0 \Leftrightarrow f_k = 0.5.\\
            \end{aligned}
            \label{Current_zero_lindblad}
    \end{equation} 

    The last line follows from $[\hat{c}_k, \hat{c}_k^\dagger] = 1 - 2\hat{n}_k$. This confirms that $f_k = 0.5$ makes the identity a steady state of the dissipator, and consequently the current vanishes, $\langle \hat{\mathcal{J}}_{1 \rightarrow 2} \rangle \propto \textrm{Tr}(\hat{\mathcal{J}}_{1 \rightarrow 2} \hat{I}) = 0$. Similarly, as shown in Fig.~\ref{Fig: ML vs Lindblad}(b), the current also vanishes as $T \to \infty$ in the mesoscopic-lead approach. This occurs because the Fermi-Dirac distribution becomes a flat function and hence $f_{\alpha} (\varepsilon_{k}) \rightarrow 0.5$ for all lead modes. Consequently, the proof in Eq. \eqref{Current_zero_lindblad} also applies, and the NESS is proportional to the identity. This result shows that any finite current without bias and the rectification mechanism discussed in this work are both inherently finite-temperature effects. The derivation of Eq.~\eqref{Current_zero_lindblad} assumed the steady state is exactly the identity (up to a normalization). It does not rule out other, non-identity steady states with vanishing current under periodic driving. A related example is the case $f_{1,D} = \alpha$ with $\alpha \neq 0.5$ for $U=0$, which corresponds to an unbiased scenario where the NESS is not proportional to the identity. Instead, it is described by the correlation matrix $\mathbf{C}_{ij}(t) = \alpha\,\delta_{ij}$, which also results in zero current. This can be seen by substituting this form into the Lyapunov equation~\eqref{Lyapunov} as follows:
        \begin{equation}
            \begin{aligned}
                \frac{d \mathbf{C}(t)}{dt} & = -\alpha[\mathbf{W}(t)  +\mathbf{W^{\dagger}}(t)] + \mathbf{F} \\
                & =-\alpha \boldsymbol{\Upsilon} + \mathbf{F} \\
                & =\left[\begin{array}{cccc}
            f_{1} -\alpha & 0 & 0 & 0\\
            0 & 0 & 0 & 0 \\
        0 & 0 & 0 & 0\\
            0 & 0 & 0 & f_{D}-\alpha
            \end{array}\right]  = \mathbf{0} \Leftrightarrow f_{1,D} = \alpha.
                \end{aligned}
        \end{equation}
    
    Here we assumed $D=4$ and replaced the definitions
       \begin{equation}
        \boldsymbol{\Upsilon}=\left[\begin{array}{cccc}
        1 & 0 & 0 & 0 \\
        0 & 0 & 0 & 0\\
        0 & 0 & 0 & 0\\
        0 & 0 & 0 & 1
        \end{array}\right], \quad \text{and} \quad \mathbf{F}=\left[\begin{array}{cccc}
        f_{1} & 0 & 0 & 0 \\
        0 & 0 & 0 & 0 \\
        0 & 0 & 0 & 0\\
        0 & 0 & 0 & f_{D}
        \end{array}\right]. 
        \end{equation}
        
    This calculation shows that, at least for $U=0$, a vanishing current does not require the NESS to be proportional to the identity. Nevertheless, in all our simulations (even in the interacting case), the NESS at $f_k = 0.5$ converges numerically to the identity. 
            
    The absence of current at zero bias implies that the boundary-driven Lindblad approach cannot be used to study the competition between drive-induced and bias-induced current contributions or the rectification that emerges from it. In fact, Fig.~\ref{Fig: ML vs Lindblad}(a) shows that at $U=0$ the current only changes sign upon bias reversal, without any rectification. Nevertheless, the resulting current under maximal bias $(f_{1} = 1, f_{D} = 0)$ is qualitatively very similar to the one shown in Fig.~\ref{Fig: J vs ω without Bias}. Naturally, the peak positions do not coincide exactly with the mesoscopic leads result. The Lindblad description does not treat the leads, and the Hamiltonian is simply given by Eq.~\eqref{Model_EQ}, whereas the rest of the article considers the extended system+lead Hamiltonian given by Eq.~\eqref{H_ML}. Even so, the simplified Lindblad description is sufficient to approximately capture the peak locations, indicating they are primarily determined by intrinsic system properties. This explains the effectiveness of the Floquet-based analysis performed in Sec.~\ref{Non-Interacting}.

\section{Dissipative Floquet spectrum}
\label{app:dissipative_spectrum}

    This appendix explains why the current in Fig.~\ref{Fig: Floquet} peaks at the avoided crossings of the Floquet quasienergy spectrum. Using the boundary-driven Lindblad description of Appendix~\ref{Lindblad} at $U=0$, we show that dissipation turns each avoided crossing into a Lorentzian resonance of the two-point correlations. Since the current and every other quadratic observable follow from these correlations, all of them inherit this resonant structure. The Lorentzian profile also provides the starting point for Appendix~\ref{app:lorentzian}, which analyzes how neighboring resonances merge. We work within this reduced description, rather than the full mesoscopic-lead formulation, because the relevant generator is then a $D\times D$ matrix small enough to diagonalize exactly and interpret its eigenvalues directly. From the Lyapunov Eq.~\eqref{Lyapunov}, the dynamics is generated by $\mathbf{W}(t)=i\mathbf{H}(t)+\boldsymbol{\Upsilon}/2$. We define the propagator $\mathbf{G}(t,t_{0})$ of the single-particle density matrix such that $\mathbf{C}(t) = \mathbf{G}(t, t_{0}) \mathbf{C}(0) \mathbf{G}^{\dagger}(t, t_{0})$. By substituting it into the Lyapunov equation and comparing terms, we show that the propagator must satisfy $\dot{\mathbf{G}}(t) = -\mathbf{W}(t)\mathbf{G}(t)$, so that
    \begin{equation}
    \mathbf{G}(t,t_{0}) = \mathcal{T} \exp \left( -\int_{t_0}^{t} \mathbf{W}(t')\,dt' \right).
    \end{equation}

    If we identify $i \mathbf{H}^{\star}(t) = \mathbf{W}(t)$, $\mathbf{G}(t, t_0)$ takes the same form as the propagator in Eq.~\eqref{Propagator_Ht}, and since $\mathbf{H}^{\star}(t)$ inherits the periodicity of $\mathbf{W}(t)$, all of our Floquet machinery can be applied. Specifically, diagonalizing the stroboscopic generator $i\ln\big(\mathbf{G}(T,0)\big)/T$ yields a Floquet quasienergy spectrum. Since $\mathbf{H}^{\star}(t)$ is non-Hermitian, this spectrum is now complex,
    \begin{equation}
    \varepsilon_a^F = e_a(\omega) - i\,\frac{w_a(\omega)}{2},
    \label{eq:complex_quasienergy}
    \end{equation}
    where $e_a(\omega)$ plays the role of the usual Floquet quasienergy and $w_a(\omega)  \geq 0$ is an effective decay rate. The emergence of a finite imaginary part is equivalent to assigning a lifetime to the corresponding Floquet state.
    
\begin{figure}[t]
        \centering
        \includegraphics[width=0.4\paperwidth]{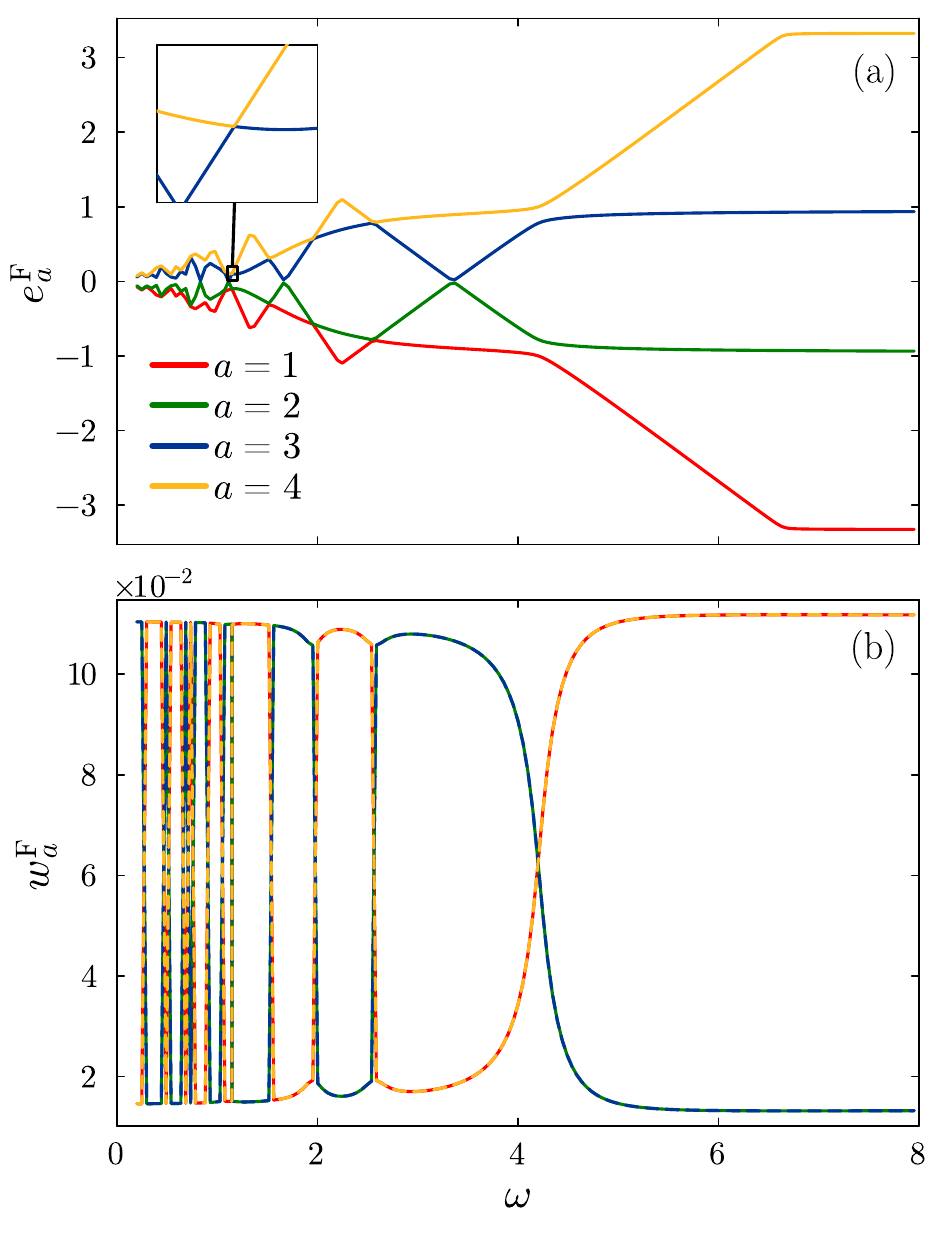}
        \caption{Dissipative spectrum. (a) Real and (b) imaginary parts of the complex Floquet quasienergies $\{\varepsilon_a^F\}$ for $\gamma_{1,D}=0.5$, $f_{1}=1$, $f_{D}=0$, $E=2$, and $D = 4$. Inset: zoom near $\omega\approx1.15$, where the gap between the real parts closes within the resolution of the plot.}
        \label{Fig: Lindblad spectrum}
    \end{figure}

    The real and imaginary parts of $\{\varepsilon_a^F\}$ for $\gamma_{1,D}=0.5$, $f_{1} = 1$, and $f_{D}=0$ are shown in Fig.~\ref{Fig: Lindblad spectrum}. The spectrum closely resembles the closed-system spectrum of Fig.~\ref{Fig: J vs ω without Bias}(b), but the avoided crossings are narrower. This resemblance confirms that the $\{\varepsilon_a^F\}$ of Eq.~\eqref{eq:complex_quasienergy} are well-defined, controlled deformations of the ordinary Floquet quasienergies, rather than a qualitatively different object introduced by the dissipation. At the isolated crossing near $\omega\approx4.2$, the two quasienergy curves cross in the imaginary part while their real parts stay separated, so the gap between them never closes. In contrast, at $\omega\approx1.15$ the gap closes within the resolution of the inset of Fig.~\ref{Fig: Lindblad spectrum}(a). 
    
    Since the dissipative spectrum nearly coincides with the closed-system spectrum, apart from narrower avoided crossings, we write $\mathbf{C}$ in the basis of these dissipative Floquet states. This choice reveals a pole structure in the correlation matrix elements, motivating the Lorentzian decomposition of the current performed in Appendix~\ref{app:lorentzian} to analyze the effect of the system-reservoir coupling $\Gamma$ on the current. Let $|\phi_a(t)\rangle$ denote the right Floquet eigenmodes of $\mathbf{H}^{\star}(t)$, satisfying Eq.~\eqref{eq:floquet_schrodinger} with $\varepsilon_a^F$ given by Eq.~\eqref{eq:complex_quasienergy}. Taking the adjoint of this relation gives $\langle\phi_a(t)|\big(\mathbf{H}^{\star\dagger}(t) + i\,d/dt\big)= \varepsilon_a^{F*}\langle \phi_a(t)|$: the adjoint of a Floquet mode $\langle\phi_a(t)|$ is automatically a left Floquet eigenmode of $\mathbf{H}^{\star\dagger}(t)$ with the conjugate eigenvalue $\varepsilon_a^{F*}$. Expanding the single-particle density matrix in this Floquet basis takes the form
    \begin{equation}
        \mathbf{C}(t) = \sum_{a,b} C_{ab}(t)\,|\phi_a(t)\rangle\langle\phi_b(t)|.
        \label{eq:C_expansion}
    \end{equation}

    Differentiating each basis element with respect to time via the product rule, and using Eq.~\eqref{eq:floquet_schrodinger} for $|\phi_a(t)\rangle$ together with its adjoint for $\langle\phi_b(t)|$, gives $\frac{d}{dt} \big(|\phi_a(t)\rangle\langle\phi_b(t)|\big) = -i\big[\mathbf{H}^{\star}(t)|\phi_a(t)\rangle\langle\phi_b(t)| - |\phi_a(t)\rangle\langle\phi_b(t)|\mathbf{H}^{\star\dagger}(t)\big] + i(\varepsilon_a^F-\varepsilon_b^{F*})|\phi_a(t)\rangle\langle\phi_b(t)|$. Multiplying by $C_{ab}(t)$ and summing over $a,b$, the bracketed term reproduces $-i\big[\mathbf{H}^{\star}(t)\mathbf{C}(t) - \mathbf{C}(t)\mathbf{H}^{\star\dagger}(t)\big]$ exactly, which cancels against the same term on the right-hand side of the Lyapunov equation $\dot{\mathbf{C}} = -i\big[\mathbf{H}^{\star}\mathbf{C} - \mathbf{C}\,\mathbf{H}^{\star\dagger}\big] + \mathbf{F}$ (which follows from $\mathbf{W}(t)=i\mathbf{H}^{\star}(t)$). Thus,
    \begin{equation}
        \sum_{a,b}\Big[\dot{C}_{ab}(t) + i(\varepsilon_a^F-\varepsilon_b^{F*})C_{ab}(t)\Big]|\phi_a(t)\rangle\langle\phi_b(t)| = \mathbf{F}(t).
        \label{eq:Cab_projected}
    \end{equation}

    Since $\mathbf{H}^{\star}(t)$ is non-Hermitian, its right Floquet eigenmodes are not orthonormal. We therefore introduce the left Floquet eigenmodes $\langle\tilde{\phi}_a(t)|$ of $\mathbf{H}^{\star}(t)$, normalized such that $\langle\tilde{\phi}_a(t)|\phi_c(t)\rangle=\delta_{ac}$, and denote their adjoints by $|\tilde{\phi}_b(t)\rangle$, so that $\langle\phi_c(t)|\tilde{\phi}_b(t)\rangle=\delta_{cb}$. Projecting Eq.~\eqref{eq:Cab_projected} onto $\langle\tilde{\phi}_a(t)|\cdots|\tilde{\phi}_b(t)\rangle$ and defining $F_{ab}(t) = \langle\tilde{\phi}_a(t)|\mathbf{F}|\tilde{\phi}_b(t)\rangle$ as the matrix elements of $\mathbf{F}$ in the Floquet basis gives
    \begin{equation}
    \dot{C}_{ab}(t) \;=\; -\Big[i(e_a-e_b)+\tfrac{1}{2}(w_a+w_b)\Big]C_{ab}(t) + F_{ab}(t),
    \label{eq:Cab_ODE}
    \end{equation}
    with $e_a$ and $w_a$ as defined in Eq.~\eqref{eq:complex_quasienergy}. This is simply the statement that a Lyapunov equation is diagonal in the eigenbasis of its own generator. 
    
    Although $\mathbf{F}$ is time-independent, its matrix elements $F_{ab}(t)$ are periodic, since the Floquet modes are. We expand them as $F_{ab}(t) = \sum_{m} F^{(m)}_{ab}\, e^{-im\omega t}$. For $w_a + w_b > 0$, the initial transient of Eq.~\eqref{eq:Cab_ODE} decays, and the solution approaches $C_{ab}(t) = \int_{-\infty}^{t} e^{-z_{ab}(t-s)} F_{ab}(s)\, ds$ with $z_{ab} = i(e_a-e_b)+\tfrac{1}{2}(w_a+w_b)$. Since $F_{ab}(s)$ oscillates, $C_{ab}(t)$ oscillates at the same harmonics. Evaluating the integral for each harmonic yields the periodic steady state
    \begin{equation}
    C_{ab}^{\mathrm{ss}}(t) \;=\; \sum_{m} \frac{F^{(m)}_{ab}\, e^{-im\omega t}}{\,i(e_a-e_b-m\omega)+\dfrac{1}{2}(w_a+w_b)\,}.
    \label{eq:Cab_lorentzian}
    \end{equation}
    
    With $\delta_m = e_a-e_b-m\omega$ and $\Lambda = (w_a+w_b)/2$, the resonant factor of each term has real part $\Lambda/(\Lambda^2+\delta_m^2)$. This expression is a Lorentzian which peaks where $|\delta_m|$ is minimal. For $m=0$, this minimum occurs at avoided crossings within a single Floquet zone. For $m\neq0$, $\delta_m$ vanishes where quasienergies of neighboring zones meet, corresponding to an $|m|$-photon transition. The resonance at $\omega\approx6.65$ is of this second type: there $e_4 - e_1 = \omega$, so the $m=1$ term of the pair $(4,1)$ resonates.
                       
\section{Impact of \texorpdfstring{$\Gamma$}{Gamma} into transport resonances}
\label{app:lorentzian} 

   \begin{figure}[t]
        \centering
        \includegraphics[width=0.38 \paperwidth]{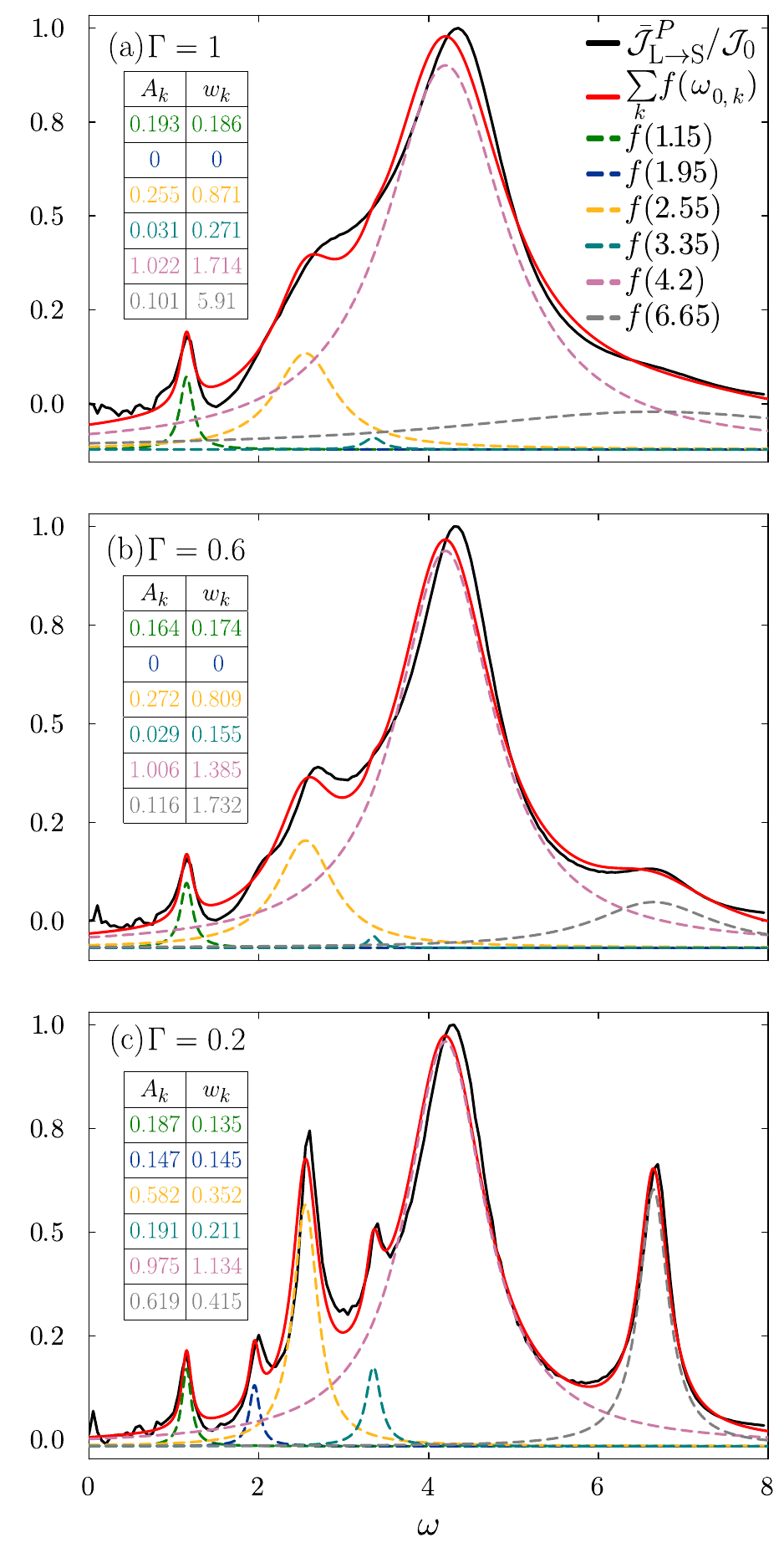}     
        \caption{Lorentzian fit of the current. Normalized particle current $\bar{\mathcal{J}}^{P}_{\text{L}\to\text{S}}(\omega)/\mathcal{J}_0$ (black) for (a) $\Gamma=1$, (b) $\Gamma=0.6$, and (c) $\Gamma=0.2$, where $\mathcal{J}_0 = \max_\omega \bar{\mathcal{J}}^{P}_{\text{L}\to\text{S}}(\omega)$ at the corresponding value of $\Gamma$. The normalization enables a direct comparison of the currents for different $\Gamma$ across the panels. Also shown are the fitted Lorentzian sum of Eq.~\eqref{eq:lorentzian_sum} (red) and its individual components (dashed) at the resonance frequencies $\omega_0\approx\{1.15, 1.95, 2.55, 3.35, 4.2, 6.65\}$. In each figure, we indicate the fitted amplitudes $A_k$ and widths $w_k$.}
        \label{fig:lorentzian_fits}
    \end{figure}    
    
    In Fig.~\ref{Fig: Floquet} it was shown that lowering $\Gamma$ reveals current resonances that are not singled out at larger $\Gamma$. To understand this behavior, we start by representing the current $\bar{\mathcal{J}}^{P}_{\text{L}\to\text{S}}(\omega)$ as a sum of Lorentzian contributions. Each Lorentzian is associated with a resonance of the current, located at an avoided crossing in the Floquet quasienergy spectrum or, for $\omega\approx6.65$, at a single-photon transition,
    
    \begin{equation}
    \bar{\mathcal{J}}^{P}_{\text{L}\to\text{S}}(\omega) \;\approx\; \sum_{k} \frac{A_k\,(w_k/2)^2}{(\omega-\omega_{0,k})^2 + (w_k/2)^2}
    \label{eq:lorentzian_sum}
    \end{equation}
    with $\omega_{0,k}$ the location of the $k$-th resonance, $A_k$ its amplitude, and $w_k/2$ its width. This decomposition is a property inherited from Eq.~\eqref{eq:Cab_lorentzian}.

    In Appendix~\ref{app:dissipative_spectrum}, we show that the widths of the Lorentzians in the correlation functions equal the imaginary parts of the quasienergies defined in Eq.~\eqref{eq:complex_quasienergy}. These imaginary parts emerge from the dissipation acting on the boundary sites. In the mesoscopic-lead formulation, the dissipation does not act directly on the system. Nevertheless, each Floquet branch can acquire a finite lifetime through hybridization with the lead modes suffering dissipation; see Fig.~\ref{Fig. Light Setup}. This hybridization also broadens the energy-mode-resolved current shown in Fig.~\ref{Fig: J separated by modes}. We therefore expect the width $w_k$ to increase with $\Gamma$.
     
    This is systematically analyzed in Fig.~\ref{fig:lorentzian_fits}, where we fit Eq.~\eqref{eq:lorentzian_sum} to $\bar{\mathcal{J}}^{P}_{\text{L}\to\text{S}}(\omega)$ for three values of $\Gamma$. The fitted widths of the resonances at $\omega_0\approx1.15$, $2.55$, $4.2$, and $6.65$ grow monotonically with $\Gamma$, consistent with the hybridization picture above. The resonances near $\omega_0\approx1.95, $ and $3.35$ behave differently: as $\Gamma$ increases, they broaden until they are fully absorbed by their neighbors and can no longer be resolved as independent features. At this point, the fit assigns zero amplitude and zero width to the resonance, since the decomposition no longer contains a separate Lorentzian term for it. For low $\Gamma$, this obstruction is removed: the resonance separates from its neighbors and re-emerges as a distinct, well-fitted peak.
         
\section{Microscopic origin of the interaction-dependent relaxation rate}
\label{relaxation_rate}

    In this appendix, we discuss the origin of the asymmetry in the relaxation rate that the system exhibits to reach the periodic NESS for repulsive and attractive interactions; see Fig.~\ref{Fig: Effect of U on JP and JE}. We use the boundary-driven Lindblad description instead of the mesoscopic leads, since it can be solved via exact diagonalization for small systems. We exploit this property to compute the spectrum of the Liouvillian superoperator $\mathcal{L}$, defined via $d\hat{\rho}(t)/dt = \mathcal{L}(\hat{\rho}(t))$ (compare with Eq.~\eqref{QME}), diagonalized in its vectorized form. This spectrum determines the time required to converge to the periodic steady state, and thereby helps explain the behavior observed in Fig.~\ref{Fig: Effect of U on JP and JE}. In general, the density matrix can be written as $\hat{\rho}(t) = \sum_{i} c_{i}(0) e^{\lambda_{i} t} \hat{\rho}_{i}$, where $\mathcal{L}\hat{\rho}_{i} = \lambda_{i}\hat{\rho}_{i}$ defines the eigenbasis and $c_{i}(0)$ is the set of coefficients obtained by projecting the initial state $\hat{\rho}(0)$ onto it. Since $\mathcal{L}$ generates a physical (completely positive, trace-preserving) evolution, its spectrum is constrained: $\mathrm{Re}[\lambda_{i}] \leq 0$ for all $i$ \cite{minganti_spectral_2018, chen_periodically_2024}, and for a periodic system there exists at least one physical mode with $\lambda_{i} = 0$, corresponding to the steady state. Sorting the eigenvalues as $0 = |\mathrm{Re}[\lambda_{0}]| \leq |\mathrm{Re}[\lambda_{1}]| \leq \dots$, the Liouvillian gap $\Delta \equiv |\mathrm{Re}[\lambda_{1}]|$ determines the slowest relaxation dynamics in the long-time limit. 

   \begin{figure}[b]
        \centering
        \includegraphics[width=0.4 \paperwidth]{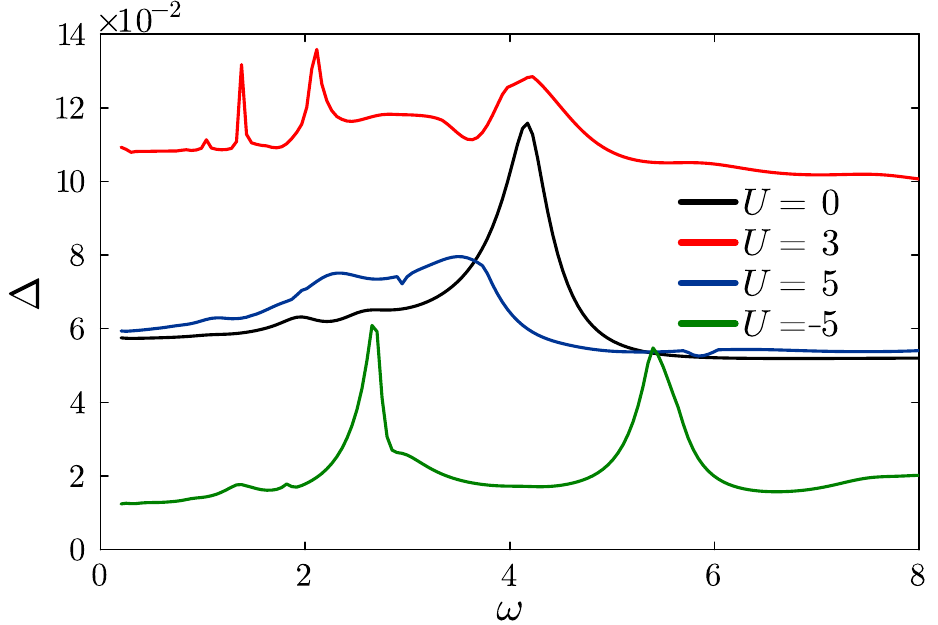}
        \caption{Liouvillian gap $\Delta = |\mathrm{Re}(\lambda_1)|$ as a function of the driving frequency $\omega$, computed from the stroboscopic effective Liouvillian $\mathcal{L}$ for different values of the interaction strength $U$. Here $D=4$, $E=2$ under maximal bias $(f_{1} = 1, f_{D}=0)$.}
        \label{Fig: Liouvillian gap}
    \end{figure}  
    \FloatBarrier
        
    By applying the Floquet formalism to construct a stroboscopic effective version of $\mathcal{L}$ in the interacting case, we find that repulsive interactions $(U>0)$ yield a larger $\Delta$ than those of attractive interactions $(U<0)$ of the same magnitude across almost the full range of frequencies; see Fig.~\ref{Fig: Liouvillian gap}. The sign asymmetry in relaxation speed observed in Fig.~\ref{Fig: Effect of U on JP and JE} therefore traces back to this asymmetry in $\Delta$, already present in the boundary-driven Lindblad description. Although $\Delta$ captures the sign asymmetry, it does not reproduce the trend of relaxation times with $|U|$ (e.g., at $\omega=4.35$, $\Delta$ is larger for $U=0$ than for $U=5$). This is expected since $\Delta$ is computed for the boundary-driven Lindblad scenario, so the comparison is qualitative. Nevertheless, the highest peaks in $\Delta$ and in the entropy production rate of Fig.~\ref{Fig: Thermodynamics}(a) occur approximately at the same frequency. A larger $\Delta$ has been associated with higher entropy production in several classes of Markovian systems \cite{Bao2025Universal, Tejero2025Asymmetric, Guo2026Information}. Our results extend this connection to periodically driven quantum systems.

    \begin{figure}[b]
        \centering
        \includegraphics[width=0.4 \paperwidth]{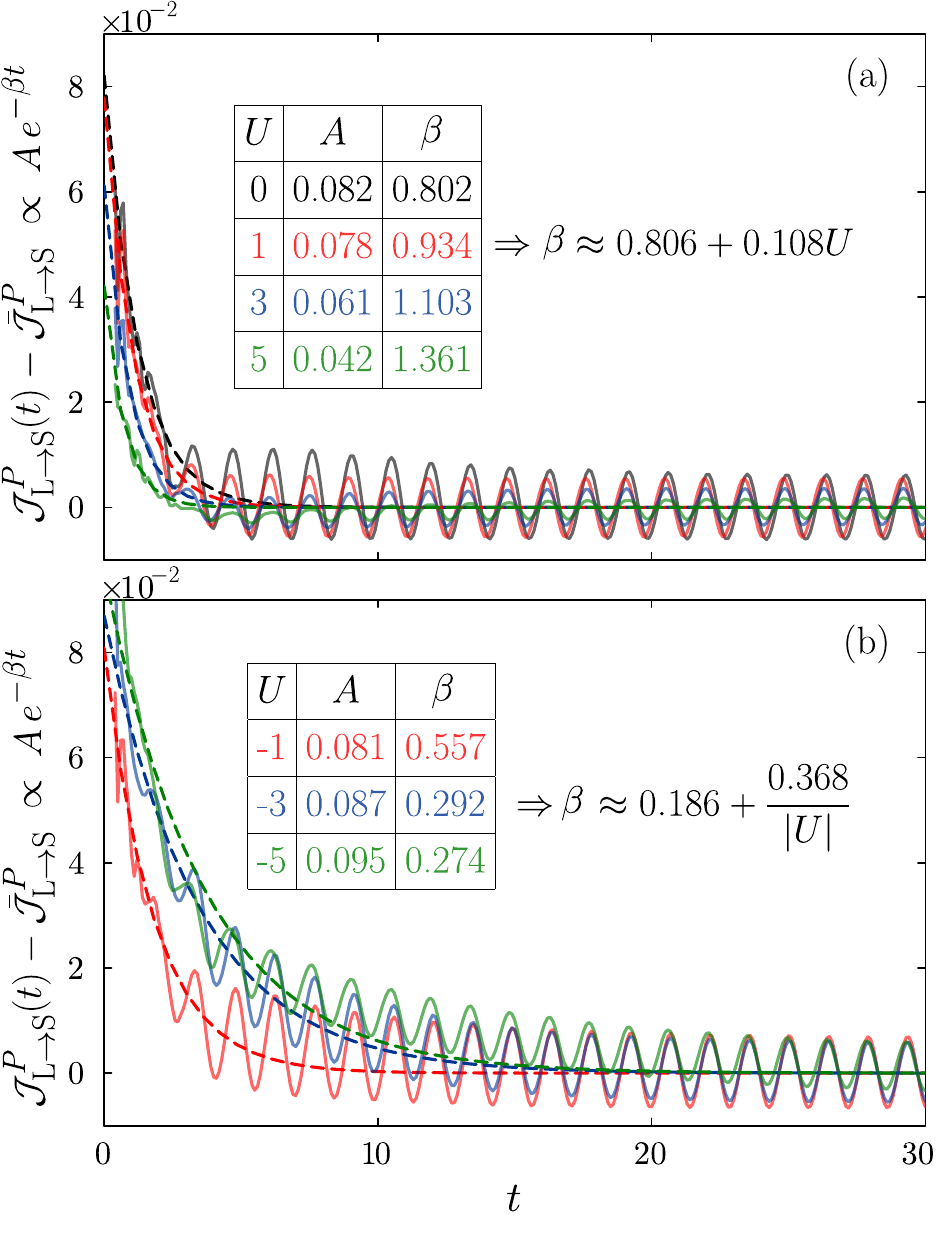} 
        \caption{Interaction impact on the relaxation rate $\beta$ extracted from Eq.~\eqref{eq:relaxation_fit}, for (a) $U \geq 0$ and (b) $U < 0$. For both figures, $E = 2$, $D=4$, $\omega = 4.35$, $\mu_{\text{L}, \text{R}} = 1$, $T_{\text{L}, \text{R}} = 10$, and $N = 80$. Solid lines show the deviation of the particle current from its cycle-averaged value. Dashed lines show the corresponding exponential fits, evaluated for the same values of $U$. The x-axis is limited to $t \in [0,30]$ for clarity; the fits, however, were performed over the full time range needed to reach convergence.}        
        \label{Fig: relaxation_fit}
    \end{figure}  

    To identify the microscopic mechanism behind this asymmetry, we use the results of Sec.~\ref{Interacting transport channels}. The left lead creates mainly unpaired configurations for $U>0$ and mainly paired configurations for $U<0$. For $U>0$, the injected fermion is unpaired and hops directly between nearest neighbors at the bare rate $t_s$. For $U<0$, it is paired with its neighbor, so moving it requires moving the pair through an energetically costly unpaired intermediate state ($|1100\rangle \rightarrow |1010\rangle \rightarrow |0110\rangle$). This superexchange process has the reduced effective rate $t_s^2/|U|$~\cite{Anderson1959-iq, Winkler2006-hq} and applies to bound pairs for either sign of $U$. However, the left lead creates mainly pairs only for $U<0$, since paired transitions then have the lower addition energies (Sec.~\ref{Interacting transport channels}).  For $|U|>t_s$, which holds for all nonzero values of $U$ in Fig.~\ref{Fig: Effect of U on JP and JE}, the rate $t_s^2/|U|$ is less than $t_s$. Fermions in paired configurations thus move more slowly than unpaired ones, which is consistent with the slower relaxation at $U<0$. This description can be confirmed by extracting the relaxation rate $\beta$ from a fit
    \begin{equation}
        \mathcal{J}^P_{{\rm L}\to{\rm S}}(t) - \bar{\mathcal{J}}^P_{{\rm L}\to{\rm S}}
        \;\propto\; A\, e^{- \beta t}
        \label{eq:relaxation_fit}
    \end{equation}
    performed independently for each value of $U$. This process is shown in Fig.~\ref{Fig: relaxation_fit}. For $U<0$, the extracted relaxation rate is well described by 
    \begin{equation}
        \beta (U<0) \approx 0.186(23) + 0.368(37) |U|^{-1},
        \label{eq:gamma_negative}
    \end{equation}
    where the second term is consistent with the $t_s^2/|U|$ scaling expected from the two-hop virtual process discussed above. The numbers in parentheses denote the standard errors of the fit parameters in units of the last digits shown. For example, $0.186(23)$ means $0.186 \pm 0.023$. In contrast, for $U\geq0 $ direct hopping between unpaired configurations remains efficient, and $\beta$ instead grows approximately linearly,
    \begin{equation}
        \beta (U \geq0) \approx 0.806(20) + 0.108(7) U,
        \label{eq:gamma_positive}
    \end{equation}
    where the offset $\beta(0) \approx0.806(20)$ is of the same order as the bare hopping rate $t_s=1$. We emphasize that this is a heuristic drawn from the many-body spectrum of the isolated, cycle-averaged system, and should be treated as a qualitative organizing principle rather than a quantitative theory of the full problem. 
    
\bibliography{ref}

@article{Landi_Introduction_Review,
  title = {Nonequilibrium boundary-driven quantum systems: Models, methods, and properties},
  author = {Landi, Gabriel T. and Poletti, Dario and Schaller, Gernot},
  journal = {Rev. Mod. Phys.},
  volume = {94},
  pages = {045006},
  numpages = {58},
  year = {2022},
  publisher = {American Physical Society},
  doi = {10.1103/RevModPhys.94.045006}
}

@article{PRX_JJ,
  title = {Giant Rectification in Strongly Interacting Driven Tilted Systems},
  author = {Mendoza-Arenas, Juan Jos\'e and Clark, Stephen R.},
  journal = {PRX Quantum},
  volume = {5},
  issue = {1},
  pages = {010341},
  numpages = {20},
  year = {2024},
  month = {Mar},
  publisher = {American Physical Society},
  doi = {10.1103/PRXQuantum.5.010341}
}

@article{ML_Tensor_Networks,
  title = {Tensor-Network Method to Simulate Strongly Interacting Quantum Thermal Machines},
  author = {Brenes, Marlon and Mendoza-Arenas, Juan Jos\'e and Purkayastha, Archak and Mitchison, Mark T. and Clark, Stephen R. and Goold, John},
  journal = {Phys. Rev. X},
  volume = {10},
  issue = {3},
  pages = {031040},
  numpages = {29},
  year = {2020},
  month = {Aug},
  publisher = {American Physical Society},
  doi = {10.1103/PhysRevX.10.031040}
}

@article{ML_Entropy_Production,
  title = {Entropy production in the mesoscopic-leads formulation of quantum thermodynamics},
  author = {Lacerda, Artur M. and Kewming, Michael J. and Brenes, Marlon and Jackson, Conor and Clark, Stephen R. and Mitchison, Mark T. and Goold, John},
  journal = {Phys. Rev. E},
  volume = {110},
  issue = {1},
  pages = {014125},
  numpages = {11},
  year = {2024},
  month = {Jul},
  publisher = {American Physical Society},
  doi = {10.1103/PhysRevE.110.014125}
}

@article{ML_Fast_Driving,
  title = {Quantum thermodynamics with fast driving and strong coupling via the mesoscopic leads approach},
  author = {Lacerda, Artur M. and Purkayastha, Archak and Kewming, Michael and Landi, Gabriel T. and Goold, John},
  journal = {Phys. Rev. B},
  volume = {107},
  issue = {19},
  pages = {195117},
  numpages = {14},
  year = {2023},
  month = {May},
  publisher = {American Physical Society},
  doi = {10.1103/PhysRevB.107.195117}
}

@article{ML_Subleties,
  title = {Subtleties in the pseudomodes formalism},
  author = {Alford, Wynter and Bettmann, Laetitia P. and Landi, Gabriel T.},
  journal = {Phys. Rev. B},
  volume = {113},
  issue = {20},
  pages = {205143},
  numpages = {21},
  year = {2026},
  month = {May},
  publisher = {American Physical Society},
  doi = {10.1103/79dp-vz29},
}

@article{ML_Swolak,
  title = {Transport in a periodically driven tilted lattice via the extended reservoir approach: Stability criterion for recovering the continuum limit},
  author = {De, Bitan and W\'ojtowicz, Gabriela and Zakrzewski, Jakub and Zwolak, Michael and Rams, Marek M.},
  journal = {Phys. Rev. B},
  volume = {107},
  issue = {23},
  pages = {235148},
  numpages = {13},
  year = {2023},
  month = {Jun},
  publisher = {American Physical Society},
  doi = {10.1103/PhysRevB.107.235148}
}

@incollection{Floquet_Tsuji,
  title = {{Floquet} states},
  editor = {Chakraborty, Tapash},
  booktitle = {Encyclopedia of Condensed Matter Physics (Second Edition)},
  publisher = {Academic Press},
  edition = {2nd},
  address = {Oxford},
  pages = {967-980},
  year = {2024},
  isbn = {978-0-323-91408-6},
  doi = {10.1016/B978-0-323-90800-9.00241-9},
  author = {Tsuji, Naoto}
  }

@article{Floquet_Tutorial,
  title = {Quantum Master Equations: Tips and Tricks for Quantum Optics, Quantum Computing, and Beyond},
  author = {Campaioli, Francesco and Cole, Jared H. and Hapuarachchi, Harini},
  journal = {PRX Quantum},
  volume = {5},
  issue = {2},
  pages = {020202},
  numpages = {33},
  year = {2024},
  month = {Jun},
  publisher = {American Physical Society},
  doi = {10.1103/PRXQuantum.5.020202}
}

@book{datta2005quantum,
  title = {Quantum transport: Atom to Transistor},
  author = {Datta, Supriyo},
  year = {2005},
  publisher = {Cambridge University Press}
}

@book{Stefanucci_van_Leeuwen_2013,
  place = {Cambridge},
  title = {Nonequilibrium Many-Body Theory of Quantum Systems: A Modern Introduction},
  publisher = {Cambridge University Press},
  author = {Stefanucci, Gianluca and van Leeuwen, Robert},
  year = {2013}
}

@book{nazarov2009quantum,
  title = {Quantum transport: Introduction to Nanoscience},
  author = {Nazarov, Yuli V. and Blanter, Yaroslav M.},
  year = {2009},
  publisher = {Cambridge University Press}
}

@book{ryndyk2016theory,
  title = {Theory of quantum transport at nanoscale},
  author = {Ryndyk, Dmitry A.},
  journal = {Springer Series in Solid-State Sciences},
  volume = {184},
  year = {2016},
  publisher = {Springer}
}

@book{osgood2019fourier,
  title     = {Lectures on the {Fourier} Transform and Its Applications},
  author    = {Osgood, Brad G.},
  series    = {Pure and Applied Undergraduate Texts},
  volume    = {33},
  publisher = {American Mathematical Society},
  address   = {Providence, Rhode Island},
  year      = {2019},
  isbn      = {978-1-4704-4191-3}
}

@article{PhysRevLett.93.040502,
  title = {Efficient Simulation of One-Dimensional Quantum Many-Body Systems},
  author = {Vidal, Guifr\'e},
  journal = {Phys. Rev. Lett.},
  volume = {93},
  issue = {4},
  pages = {040502},
  numpages = {4},
  year = {2004},
  month = {Jul},
  publisher = {American Physical Society},
  doi = {10.1103/PhysRevLett.93.040502}
}

@article{PhysRevLett.93.207205,
  title = {Mixed-State Dynamics in One-Dimensional Quantum Lattice Systems: A Time-Dependent Superoperator Renormalization Algorithm},
  author = {Zwolak, Michael and Vidal, Guifr\'e},
  journal = {Phys. Rev. Lett.},
  volume = {93},
  issue = {20},
  pages = {207205},
  numpages = {4},
  year = {2004},
  month = {Nov},
  publisher = {American Physical Society},
  doi = {10.1103/PhysRevLett.93.207205}
}

@article{PhysRevLett.98.070201,
  title = {Classical Simulation of Infinite-Size Quantum Lattice Systems in One Spatial Dimension},
  author = {Vidal, G.},
  journal = {Phys. Rev. Lett.},
  volume = {98},
  issue = {7},
  pages = {070201},
  numpages = {4},
  year = {2007},
  month = {Feb},
  publisher = {American Physical Society},
  doi = {10.1103/PhysRevLett.98.070201}
}

@article{RevModPhys.93.015008,
  title = {Simulation methods for open quantum many-body systems},
  author = {Weimer, Hendrik and Kshetrimayum, Augustine and Or\'us, Rom\'an},
  journal = {Rev. Mod. Phys.},
  volume = {93},
  issue = {1},
  pages = {015008},
  numpages = {24},
  year = {2021},
  month = {Mar},
  publisher = {American Physical Society},
  doi = {10.1103/RevModPhys.93.015008}
}

@article{PAECKEL2019167998,
  title = {Time-evolution methods for matrix-product states},
  journal = {Ann. Phys. (N.Y.)},
  volume = {411},
  pages = {167998},
  year = {2019},
  doi = {10.1016/j.aop.2019.167998},
  author = {Paeckel, Sebastian and Köhler, Thomas and Swoboda, Andreas and Manmana, Salvatore R. and Schollwöck, Ulrich and Hubig, Claudius}
}

@article{R_energy_experimental,
  doi = {10.1088/2632-959X/ad2a17},
  year = {2024},
  month = {feb},
  publisher = {IOP Publishing},
  volume = {5},
  number = {1},
  pages = {011001},
  author = {Rurali, Riccardo},
  title = {A one-way street for phonon transport: past, present and future of solid-state thermal rectification},
  journal = {Nano Express}
}

@article{Hirata2025Enhanced,
  title = {Enhanced rectification effect in silver chalcogenide-based thermal diode by using precipitation/dissolution of {Ag} impurity across the structure phase transition},
  author = {Hirata, K. and Goto, Yusuke and Takeuchi, Tsunehiro},
  journal = {Sci. Technol. Adv. Mater.},
  year = {2025},
  volume = {26},
    pages = {2549674},
  doi = {10.1080/14686996.2025.2549674}
}

@article{martinez2015rectification,
  title = {Rectification of electronic heat current by a hybrid thermal diode},
  author = {Mart{\'\i}nez-P{\'e}rez, Maria Jos{\'e} and Fornieri, Antonio and Giazotto, Francesco},
  journal = {Nat. Nanotechnol.},
  volume = {10},
  number = {4},
  pages = {303--307},
  year = {2015},
  publisher = {Nature Publishing Group UK London},
    doi = {10.1038/nnano.2015.11}
}

@article{10.1063/5.0283849,
  author = {Li, Feng and Sato, Taichi and Tanaka, Masamitsu and Fujimaki, Akira},
  title = {Practical field-free superconducting diode for {AC–DC} rectification},
  journal = {Appl. Phys. Lett.},
  volume = {127},
  number = {7},
  pages = {072601},
  year = {2025},
  month = {08},
  doi = {10.1063/5.0283849}
}

@article{Bauriedl2021Supercurrent,
  title = {Supercurrent diode effect and magnetochiral anisotropy in few-layer {NbSe2}},
  author = {Bauriedl, Lorenz and Bäuml, Christian and Fuchs, L. and Baumgartner, C. and Paulik, Nicolas and Bauer, J. M. and Lin, Kai-Qiang and Lupton, J. and Taniguchi, T. and Watanabe, Kenji and Strunk, C. and Paradiso, N.},
  journal = {Nat. Commun.},
  year = {2022},
  volume = {13},
  pages = {4266},
  doi = {10.1038/s41467-022-31954-5}
}

@article{RevModPhys.93.025003,
  title = {Finite-temperature transport in one-dimensional quantum lattice models},
  author = {Bertini, B. and Heidrich-Meisner, F. and Karrasch, C. and Prosen, T. and Steinigeweg, R. and {\v{Z}}nidari{\v{c}}, M.},
  journal = {Rev. Mod. Phys.},
  volume = {93},
  issue = {2},
  pages = {025003},
  numpages = {71},
  year = {2021},
  month = {May},
  publisher = {American Physical Society},
  doi = {10.1103/RevModPhys.93.025003}
}

@article{RevModPhys.92.035001,
  title = {Advances and challenges in single-molecule electron transport},
  author = {Evers, Ferdinand and Koryt\'ar, Richard and Tewari, Sumit and van Ruitenbeek, Jan M.},
  journal = {Rev. Mod. Phys.},
  volume = {92},
  issue = {3},
  pages = {035001},
  numpages = {64},
  year = {2020},
  month = {Jul},
  publisher = {American Physical Society},
  doi = {10.1103/RevModPhys.92.035001}
}

@article{chen2017molecular,
  title = {Molecular diodes with rectification ratios exceeding {$10^5$} driven by electrostatic interactions},
  author = {Chen, Xiaoping and Roemer, Max and Yuan, Li and Du, Wei and Thompson, Damien and Del Barco, Enrique and Nijhuis, Christian A.},
  journal = {Nat. Nanotechnol.},
  volume = {12},
  number = {8},
  pages = {797--803},
  year = {2017},
  publisher = {Nature Publishing Group UK London},
  doi = {10.1038/nnano.2017.110}
}

@article{PhysRevLett.103.140405,
  title = {Atomtronic Circuits of Diodes and Transistors},
  author = {Pepino, R. A. and Cooper, J. and Anderson, D. Z. and Holland, M. J.},
  journal = {Phys. Rev. Lett.},
  volume = {103},
  issue = {14},
  pages = {140405},
  numpages = {4},
  year = {2009},
  month = {Sep},
  publisher = {American Physical Society},
  doi = {10.1103/PhysRevLett.103.140405}
}

@article{PhysRevLett.123.020603,
  title = {Energy Current Rectification and Mobility Edges},
  author = {Balachandran, Vinitha and Clark, Stephen R. and Goold, John and Poletti, Dario},
  journal = {Phys. Rev. Lett.},
  volume = {123},
  issue = {2},
  pages = {020603},
  numpages = {6},
  year = {2019},
  month = {Jul},
  publisher = {American Physical Society},
  doi = {10.1103/PhysRevLett.123.020603}
}

@article{PhysRevE.99.032116,
  title = {Perfect thermal rectification in a many-body quantum {Ising} model},
  author = {Pereira, Emmanuel},
  journal = {Phys. Rev. E},
  volume = {99},
  issue = {3},
  pages = {032116},
  numpages = {8},
  year = {2019},
  month = {Mar},
  publisher = {American Physical Society},
  doi = {10.1103/PhysRevE.99.032116}
}

@article{PhysRevE.99.032136,
  title = {Heat current rectification in segmented {XXZ} chains},
  author = {Balachandran, Vinitha and Benenti, Giuliano and Pereira, Emmanuel and Casati, Giulio and Poletti, Dario},
  journal = {Phys. Rev. E},
  volume = {99},
  issue = {3},
  pages = {032136},
  numpages = {6},
  year = {2019},
  month = {Mar},
  publisher = {American Physical Society},
  doi = {10.1103/PhysRevE.99.032136}
}

@article{PhysRevResearch.2.043343,
  title = {Rectification in nonequilibrium steady states of open many-body systems},
  author = {Yamamoto, Kazuki and Ashida, Yuto and Kawakami, Norio},
  journal = {Phys. Rev. Res.},
  volume = {2},
  issue = {4},
  pages = {043343},
  numpages = {12},
  year = {2020},
  month = {Dec},
  publisher = {American Physical Society},
  doi = {10.1103/PhysRevResearch.2.043343}
}

@article{PhysRevE.103.032108,
  title = {Rectification induced by geometry in two-dimensional quantum spin lattices},
  author = {Chioquetta, Alessandra and Pereira, Emmanuel and Landi, Gabriel T. and Drumond, Raphael C.},
  journal = {Phys. Rev. E},
  volume = {103},
  issue = {3},
  pages = {032108},
  numpages = {6},
  year = {2021},
  month = {Mar},
  publisher = {American Physical Society},
  doi = {10.1103/PhysRevE.103.032108}
}

@article{PhysRevE.104.054137,
  title = {Heat rectification by two qubits coupled with {Dzyaloshinskii-Moriya} interaction},
  author = {Upadhyay, Vipul and Naseem, M. Tahir and Marathe, Rahul and M{\"u}stecapl{\i}o{\u g}lu, {\"O}zg{\"u}r E.},
  journal = {Phys. Rev. E},
  volume = {104},
  issue = {5},
  pages = {054137},
  numpages = {10},
  year = {2021},
  month = {Nov},
  publisher = {American Physical Society},
  doi = {10.1103/PhysRevE.104.054137}
}

@article{PhysRevResearch.5.013129,
  title = {Characterizing the performance of heat rectifiers},
  author = {Khandelwal, Shishir and Perarnau-Llobet, Mart\'{\i} and Seah, Stella and Brunner, Nicolas and Haack, G\'eraldine},
  journal = {Phys. Rev. Res.},
  volume = {5},
  issue = {1},
  pages = {013129},
  numpages = {10},
  year = {2023},
  month = {Feb},
  publisher = {American Physical Society},
  doi = {10.1103/PhysRevResearch.5.013129}
}

@article{PhysRevLett.120.200603,
  title = {Perfect Diode in Quantum Spin Chains},
  author = {Balachandran, Vinitha and Benenti, Giuliano and Pereira, Emmanuel and Casati, Giulio and Poletti, Dario},
  journal = {Phys. Rev. Lett.},
  volume = {120},
  issue = {20},
  pages = {200603},
  numpages = {5},
  year = {2018},
  month = {May},
  publisher = {American Physical Society},
  doi = {10.1103/PhysRevLett.120.200603}
}

@article{e22111311,
  author = {Lee, Kang Hao and Balachandran, Vinitha and Tan, Ryan and Guo, Chu and Poletti, Dario},
  title = {Giant Spin Current Rectification Due to the Interplay of Negative Differential Conductance and a Non-Uniform Magnetic Field},
  journal = {Entropy},
  volume = {22},
  year = {2020},
  number = {11},
  pages = {1311},
  doi = {10.3390/e22111311}
}

@article{PhysRevE.103.052143,
  title = {Giant rectification in segmented, strongly interacting spin chains despite the presence of perturbations},
  author = {Lee, Kang Hao and Balachandran, Vinitha and Poletti, Dario},
  journal = {Phys. Rev. E},
  volume = {103},
  issue = {5},
  pages = {052143},
  numpages = {8},
  year = {2021},
  month = {May},
  publisher = {American Physical Society},
  doi = {10.1103/PhysRevE.103.052143}
}

@article{PhysRevE.105.024120,
  title = {Transport and spectral properties of the {XX}+{XXZ} diode and stability to dephasing},
  author = {Lee, Kang Hao and Balachandran, Vinitha and Guo, Chu and Poletti, Dario},
  journal = {Phys. Rev. E},
  volume = {105},
  issue = {2},
  pages = {024120},
  numpages = {10},
  year = {2022},
  month = {Feb},
  publisher = {American Physical Society},
  doi = {10.1103/PhysRevE.105.024120}
}

@article{PhysRevA.105.052605,
  title = {Entanglement-enhanced quantum rectification},
  author = {Poulsen, Kasper and Santos, Alan C. and Kristensen, Lasse B. and Zinner, Nikolaj T.},
  journal = {Phys. Rev. A},
  volume = {105},
  issue = {5},
  pages = {052605},
  numpages = {12},
  year = {2022},
  month = {May},
  publisher = {American Physical Society},
  doi = {10.1103/PhysRevA.105.052605}
}

@article{tesser2022heat,
  title = {Heat rectification through single and coupled quantum dots},
  author = {Tesser, Ludovico and Bhandari, Bibek and Erdman, Paolo Andrea and Paladino, Elisabetta and Fazio, Rosario and Taddei, Fabio},
  journal = {New J. Phys.},
  volume = {24},
  number = {3},
  pages = {035001},
  year = {2022},
  publisher = {IOP Publishing},
  doi = {10.1088/1367-2630/ac53b8}
}

@article{su2024spin,
  title = {Spin-polarization and {Coulomb} interaction dependent thermal rectification in a quantum dot system},
  author = {Su, Yufeng and Qi, Fenghua and Jin, Guojun},
  journal = {New J. Phys.},
  volume = {26},
  number = {7},
  pages = {073027},
  year = {2024},
  publisher = {IOP Publishing},
  doi = {10.1088/1367-2630/ad60ef}
}

@article{PhysRevE.80.041103,
  title = {Nonlinear quantum heat transfer in hybrid structures: Sufficient conditions for thermal rectification},
  author = {Wu, Lian-Ao and Yu, Claire X. and Segal, Dvira},
  journal = {Phys. Rev. E},
  volume = {80},
  issue = {4},
  pages = {041103},
  numpages = {12},
  year = {2009},
  month = {Oct},
  publisher = {American Physical Society},
  doi = {10.1103/PhysRevE.80.041103}
}

@article{PhysRevLett.102.095503,
  title = {Sufficient Conditions for Thermal Rectification in Hybrid Quantum Structures},
  author = {Wu, Lian-Ao and Segal, Dvira},
  journal = {Phys. Rev. Lett.},
  volume = {102},
  issue = {9},
  pages = {095503},
  numpages = {4},
  year = {2009},
  month = {Mar},
  publisher = {American Physical Society},
  doi = {10.1103/PhysRevLett.102.095503}
}

@article{boolakee2022light,
  title = {Light-field control of real and virtual charge carriers},
  author = {Boolakee, Tobias and Heide, Christian and Garz{\'o}n-Ram{\'\i}rez, Antonio and Weber, Heiko B and Franco, Ignacio and Hommelhoff, Peter},
  journal = {Nature},
  volume = {605},
  number = {7909},
  pages = {251--255},
  year = {2022},
  publisher = {Nature Publishing Group UK London},
  doi = {10.1038/s41586-022-04565-9}
}

@article{PhysRevLett.134.186301,
  title = {Scattering Theory of Thermal and Bipolar Thermoelectric Diodes},
  author = {Balduque, Jos\'e and S\'anchez, Rafael},
  journal = {Phys. Rev. Lett.},
  volume = {134},
  issue = {18},
  pages = {186301},
  numpages = {8},
  year = {2025},
  month = {May},
  publisher = {American Physical Society},
  doi = {10.1103/PhysRevLett.134.186301}
}

@article{Rudner2020-ip,
  title = {Band structure engineering and non-equilibrium dynamics in {Floquet} topological insulators},
  author = {Rudner, Mark S. and Lindner, Netanel H.},
  journal = {Nat. Rev. Phys.},
  publisher = {Springer Science and Business Media LLC},
  volume = {2},
  number = {5},
  pages = {229--244},
  month = {may},
  year = {2020},
  doi = {10.1038/s42254-020-0170-z}
}

@article{PhysRevLett.99.220403,
  title = {Dynamical Control of Matter-Wave Tunneling in Periodic Potentials},
  author = {Lignier, H. and Sias, C. and Ciampini, D. and Singh, Y. and Zenesini, A. and Morsch, O. and Arimondo, E.},
  journal = {Phys. Rev. Lett.},
  volume = {99},
  issue = {22},
  pages = {220403},
  numpages = {4},
  year = {2007},
  month = {Nov},
  publisher = {American Physical Society},
  doi = {10.1103/PhysRevLett.99.220403}
}

@article{PhysRevA.75.063424,
  title = {Periodically driven quantum ratchets: Symmetries and resonances},
  author = {Denisov, S. and Morales-Molina, L. and Flach, S. and H\"anggi, P.},
  journal = {Phys. Rev. A},
  volume = {75},
  issue = {6},
  pages = {063424},
  numpages = {10},
  year = {2007},
  month = {Jun},
  publisher = {American Physical Society},
  doi = {10.1103/PhysRevA.75.063424}
}

@article{PhysRevE.99.032126,
  title = {Dynamically induced heat rectification in quantum systems},
  author = {Riera-Campeny, Andreu and Mehboudi, Mohammad and Pons, Marisa and Sanpera, Anna},
  journal = {Phys. Rev. E},
  volume = {99},
  issue = {3},
  pages = {032126},
  numpages = {10},
  year = {2019},
  month = {Mar},
  publisher = {American Physical Society},
  doi = {10.1103/PhysRevE.99.032126}
}

@article{Oliveira2024Voltage-Driven,
  title = {Voltage-driven breakdown of electronic order},
  author = {Oliveira, Miguel M. and Ribeiro, Pedro and Kirchner, Stefan},
  journal = {Phys. Rev. B},
  volume = {114},
  pages = {245101},
  year = {2026},
  doi = {10.1103/nqcs-jd2g}
}

@article{10.3389/fphy.2024.1474018,
  author = {Jacob, Samuel L. and Bettmann, Laetitia P. and Lacerda, Artur M. and Zawadzki, Krissia and Clark, Stephen R. and Goold, John and Mendoza-Arenas, Juan José},
  title = {Dephasing-assisted transport in a tight-binding chain with a linear potential},
  journal = {Front. Phys.},
  volume = {12},
  pages = {1474018},
  year = {2024},
  doi = {10.3389/fphy.2024.1474018}
}

@article{PhysRevLett.117.040601,
  title = {Diffusive and Subdiffusive Spin Transport in the Ergodic Phase of a Many-Body Localizable System},
  author = {{\v{Z}}nidari{\v{c}}, Marko and Scardicchio, Antonello and Varma, Vipin Kerala},
  journal = {Phys. Rev. Lett.},
  volume = {117},
  issue = {4},
  pages = {040601},
  numpages = {6},
  year = {2016},
  month = {Jul},
  publisher = {American Physical Society},
  doi = {10.1103/PhysRevLett.117.040601}
}

@article{https://doi.org/10.1002/andp.201600298,
  author = {Žnidarič, Marko and Mendoza-Arenas, Juan Jose and Clark, Stephen R. and Goold, John},
  title = {Dephasing enhanced spin transport in the ergodic phase of a many-body localizable system},
  journal = {Ann. Phys. (Berlin)},
  volume = {529},
  number = {7},
  pages = {1600298},
  doi = {10.1002/andp.201600298},
  year = {2017}
}

@article{PhysRevE.84.051115,
  title = {Transport properties of a boundary-driven one-dimensional gas of spinless fermions},
  author = {{\v{Z}}nidari{\v{c}}, Marko and {\v{Z}}unkovi{\v{c}}, Bojan and Prosen, Toma{\v{z}}},
  journal = {Phys. Rev. E},
  volume = {84},
  issue = {5},
  pages = {051115},
  numpages = {8},
  year = {2011},
  month = {Nov},
  publisher = {American Physical Society},
  doi = {10.1103/PhysRevE.84.051115}
}

@article{PhysRevLett.106.220601,
  title = {Spin Transport in a One-Dimensional Anisotropic {Heisenberg} Model},
  author = {{\v{Z}}nidari{\v{c}}, Marko},
  journal = {Phys. Rev. Lett.},
  volume = {106},
  issue = {22},
  pages = {220601},
  numpages = {4},
  year = {2011},
  month = {May},
  publisher = {American Physical Society},
  doi = {10.1103/PhysRevLett.106.220601}
}

@article{PhysRevB.90.115156,
  title = {Anomalous nonequilibrium current fluctuations in the {Heisenberg} model},
  author = {{\v{Z}}nidari{\v{c}}, Marko},
  journal = {Phys. Rev. B},
  volume = {90},
  issue = {11},
  pages = {115156},
  numpages = {6},
  year = {2014},
  month = {Sep},
  publisher = {American Physical Society},
  doi = {10.1103/PhysRevB.90.115156}
}

@article{Znidaric2010-th,
  title = {Exact solution for a diffusive nonequilibrium steady state of an open quantum chain},
  author = {{\v Z}nidari{\v c}, Marko},
  journal = {J. Stat. Mech.},
  publisher = {IOP Publishing},
  volume = {2010},
  number = {05},
  pages = {L05002},
  month = {may},
  year = {2010},
  doi = {10.1088/1742-5468/2010/05/L05002}
}

@article{PhysRevB.80.035110,
  title = {Charge and spin transport in strongly correlated one-dimensional quantum systems driven far from equilibrium},
  author = {Benenti, Giuliano and Casati, Giulio and Prosen, Toma{\v{z}} and Rossini, Davide and {\v{Z}}nidari{\v{c}}, Marko},
  journal = {Phys. Rev. B},
  volume = {80},
  issue = {3},
  pages = {035110},
  numpages = {18},
  year = {2009},
  month = {Jul},
  publisher = {American Physical Society},
  doi = {10.1103/PhysRevB.80.035110}
}

@article{PhysRevB.87.235130,
  title = {Dephasing enhanced transport in nonequilibrium strongly correlated quantum systems},
  author = {Mendoza-Arenas, J. J. and Grujic, T. and Jaksch, D. and Clark, S. R.},
  journal = {Phys. Rev. B},
  volume = {87},
  issue = {23},
  pages = {235130},
  numpages = {15},
  year = {2013},
  month = {Jun},
  publisher = {American Physical Society},
  doi = {10.1103/PhysRevB.87.235130}
}

@article{Hirohata2020Review,
  title = {Review on spintronics: Principles and device applications},
  author = {Hirohata, A. and Yamada, Keisuke and Nakatani, Y. and Prejbeanu, I. and Di{\'e}ny, B. and Pirro, P. and Hillebrands, B.},
  journal = {J. Magn. Magn. Mater.},
  year = {2020},
  volume = {509},
  pages = {166711},
  doi = {10.1016/j.jmmm.2020.166711}
}

@article{Dieny2020Opportunities,
  title = {Opportunities and challenges for spintronics in the microelectronics industry},
  author = {Di{\'e}ny, B. and Prejbeanu, I. and Garello, K. and Gambardella, P. and Freitas, P. and Lehndorff, R. and Raberg, W. and Ebels, U. and Demokritov, S. and Åkerman, J. and Deac, A. and Pirro, P. and Adelmann, C. and Anane, A. and Chumak, A. and Hirohata, A. and Mangin, S. and Valenzuela, S. and Onbaşlı, M. and d’Aquino, M. and Prenat, G. and Finocchio, G. and Lopez-Diaz, L. and Chantrell, R. and Chubykalo‐Fesenko, O. and Bortolotti, P.},
  journal = {Nat. Electron.},
  year = {2020},
  volume = {3},
  pages = {446-459},
  doi = {10.1038/s41928-020-0461-5}
}

@article{Uchida2021Transport,
  title = {Transport phenomena in spin caloritronics},
  author = {Uchida, K.},
  journal = {Proc. Jpn. Acad. Ser. B},
  year = {2021},
  volume = {97},
  pages = {69-88},
  doi = {10.2183/pjab.97.004}
}

@article{Uchida2026Spin,
  title = {Spin caloritronics: history and future prospects of experiments},
  author = {Uchida, K. and Hirai, T.},
  journal = {J. Magn. Magn. Mater.},
  year = {2026},
  volume = {652},
  pages = {174195},
  doi = {10.1016/j.jmmm.2026.174195}
}

@article{Morong2021-yg,
  title = {Observation of {Stark} many-body localization without disorder},
  author = {Morong, W. and Liu, F. and Becker, P. and Collins, K. S. and Feng, L. and Kyprianidis, A. and Pagano, G. and You, T. and Gorshkov, A. V. and Monroe, C.},
  journal = {Nature},
  publisher = {Springer Science and Business Media LLC},
  volume = {599},
  number = {7885},
  pages = {393--398},
  month = {nov},
  year = {2021},
  doi = {10.1038/s41586-021-04271-y}
}

@article{Guardado-Sanchez2020-zp,
  title = {Subdiffusion and Heat Transport in a Tilted Two-Dimensional {Fermi-Hubbard} System},
  author = {Guardado-Sanchez, Elmer and Morningstar, Alan and Spar, Benjamin M. and Brown, Peter T. and Huse, David A. and Bakr, Waseem S.},
  journal = {Phys. Rev. X},
  volume = {10},
  issue = {1},
  pages = {011042},
  numpages = {8},
  year = {2020},
  month = {Feb},
  publisher = {American Physical Society},
  doi = {10.1103/PhysRevX.10.011042},
}

@article{PRXQuantum.2.040325,
  title = {Benchmarking a Novel Efficient Numerical Method for Localized {1D} {Fermi-Hubbard} Systems on a Quantum Simulator},
  author = {Hebbe Madhusudhana, Bharath and Scherg, Sebastian and Kohlert, Thomas and Bloch, Immanuel and Aidelsburger, Monika},
  journal = {PRX Quantum},
  volume = {2},
  issue = {4},
  pages = {040325},
  numpages = {24},
  year = {2021},
  month = {Nov},
  publisher = {American Physical Society},
  doi = {10.1103/PRXQuantum.2.040325}
}

@article{PhysRevLett.127.240502,
  title = {{Stark} Many-Body Localization on a Superconducting Quantum Processor},
  author = {Guo, Qiujiang and Cheng, Chen and Li, Hekang and Xu, Shibo and Zhang, Pengfei and Wang, Zhen and Song, Chao and Liu, Wuxin and Ren, Wenhui and Dong, Hang and Mondaini, Rubem and Wang, H.},
  journal = {Phys. Rev. Lett.},
  volume = {127},
  issue = {24},
  pages = {240502},
  numpages = {7},
  year = {2021},
  month = {Dec},
  publisher = {American Physical Society},
  doi = {10.1103/PhysRevLett.127.240502}
}

@article{Bordia2017-km,
  title = {Periodically driving a many-body localized quantum system},
  author = {Bordia, Pranjal and L{\"u}schen, Henrik and Schneider, Ulrich and Knap, Michael and Bloch, Immanuel},
  journal = {Nat. Phys.},
  publisher = {Springer Science and Business Media LLC},
  volume = {13},
  number = {5},
  pages = {460--464},
  month = {may},
  year = {2017},
  doi = {10.1038/nphys4020}
}

@article{doi:10.1126/science.adh9932,
  author = {{Google Quantum AI}},
  journal = {Science},
  volume = {383},
  number = {6689},
  pages = {1332-1337},
  year = {2024},
  doi = {10.1126/science.adh9932}
}

@article{RevModPhys.94.041001,
  title = {Colloquium: Atomtronic circuits: From many-body physics to quantum technologies},
  author = {Amico, Luigi and Anderson, Dana and Boshier, Malcolm and Brantut, Jean-Philippe and Kwek, Leong-Chuan and Minguzzi, Anna and von Klitzing, Wolf},
  journal = {Rev. Mod. Phys.},
  volume = {94},
  issue = {4},
  pages = {041001},
  numpages = {30},
  year = {2022},
  month = {Nov},
  publisher = {American Physical Society},
  doi = {10.1103/RevModPhys.94.041001}
}

@article{10.1063/1.2191730,
  author = {Li, Baowen and Wang, Lei and Casati, Giulio},
  title = {Negative differential thermal resistance and thermal transistor},
  journal = {Appl. Phys. Lett.},
  volume = {88},
  number = {14},
  pages = {143501},
  year = {2006},
  month = {04},
  doi = {10.1063/1.2191730}
}

@article{PhysRevB.99.035129,
  title = {Quantum-dot circuit-{QED} thermoelectric diodes and transistors},
  author = {Lu, Jincheng and Wang, Rongqian and Ren, Jie and Kulkarni, Manas and Jiang, Jian-Hua},
  journal = {Phys. Rev. B},
  volume = {99},
  issue = {3},
  pages = {035129},
  numpages = {9},
  year = {2019},
  month = {Jan},
  publisher = {American Physical Society},
  doi = {10.1103/PhysRevB.99.035129}
}

@article{ikeda_nonequilibrium_2021,
  title = {Nonequilibrium steady states in the {Floquet}-{Lindblad} systems: van {Vleck}'s high-frequency expansion approach},
  volume = {4},
  doi = {10.21468/SciPostPhysCore.4.4.033},
  number = {4},
  journal = {SciPost Phys. Core},
  author = {Ikeda, Tatsuhiko and Chinzei, Koki and Sato, Masahiro},
  month = {dec},
  year = {2021},
  pages = {033}
}

@misc{viebahn_introduction_2020,
  title = {Introduction to {Floquet} theory},
  author = {Viebahn, Konrad},
  year = {2020},
  note = {Lecture notes},
  url = {https://boulderschool.yale.edu/sites/default/files/files/floquetlecture_Konrad%20Viebahn.pdf}
}

@article{SciPostPhysCodeb.4,
  title = {The {ITensor} Software Library for Tensor Network Calculations},
  volume = {4},
  author = {Fishman, Matthew and White, Steven and Stoudenmire, Edwin Miles},
  journal = {SciPost Phys. Codebases},
  year = {2022},
  publisher = {SciPost},
  doi = {10.21468/SciPostPhysCodeb.4}
}

@article{SciPostPhysCodeb.4-r0.3,
  title = {Codebase release 0.3 for {ITensor}},
  volume = {4-r0.3},
  author = {Fishman, Matthew and White, Steven and Stoudenmire, Edwin Miles},
  journal = {SciPost Phys. Codebases},
  year = {2022},
  publisher = {SciPost},
  doi = {10.21468/SciPostPhysCodeb.4-r0.3}
}

@misc{chen_non-equilibrium_2026,
  title = {Non-{Equilibrium} {Phase} {Transition} in a {Boundary}-{Driven} {Dissipative} {Fermionic} {Chain}},
  doi = {10.48550/arXiv.2601.20938},
  eprint = {2601.20938},
  archiveprefix = {arXiv},
  author = {Chen, Hao and Zhang, Wucheng and Kulkarni, Manas and Prem, Abhinav},
  month = {jan},
  year = {2026}
}

@article{Chen2005A,
  title = {A quantum corrected energy-transport model for nanoscale semiconductor devices},
  author = {Chen, Ren-Chuen and Liu, Jinn-Liang},
  journal = {J. Comput. Phys.},
  year = {2005},
  volume = {204},
  pages = {131-156},
  doi = {10.1016/j.jcp.2004.10.006}
}

@article{Chen2023Quantum,
  title = {Quantum interference enhances the performance of single-molecule transistors},
  author = {Chen, Zhixing and Grace, I. and Woltering, Steffen L. and Chen, Lina and Gee, A. and Baugh, J. and Briggs, G. and Bogani, L. and Mol, J. and Lambert, Colin J. and Anderson, H. and Thomas, James O.},
  journal = {Nat. Nanotechnol.},
  year = {2024},
  volume = {19},
  pages = {986-992},
  doi = {10.1038/s41565-024-01633-1}
}

@phdthesis{Chu2019Predictive,
  title = {Predictive Electro and Thermal Quantum Transport in Nanoscale Devices},
  author = {Chu, Yuanchen},
  school = {Purdue University},
  year = {2019},
  doi = {10.25394/pgs.11307146.v1}
}

@article{Lee2023Atomistic,
  title = {Atomistic Simulation of Nanoscale Devices},
  author = {Lee, Youseung and Cao, Jiang and Luisier, M.},
  journal = {IEEE Nanotechnol. Mag.},
  year = {2023},
  volume = {17},
  pages = {4-14},
  doi = {10.1109/mnano.2023.3278968}
}

@article{Pereira2019Thermal,
  title = {Thermal rectification in classical and quantum systems: Searching for efficient thermal diodes},
  author = {Pereira, E.},
  journal = {Europhys. Lett.},
  year = {2019},
  volume = {126},
  pages = {14001},
  doi = {10.1209/0295-5075/126/14001}
}

@misc{Day2024Nonperturbative,
  title = {Nonperturbative Nonlinear Transport in a {Floquet-Weyl} Semimetal},
  author = {Day, Matthew W. and Kusyak, Kateryna and Sturm, Felix and Aranzadi, J. and Bretscher, Hope M. and Fechner, M. and Matsuyama, T. and Michael, Marios H. and Schulte, B. and Li, Xinyu and Hagelstein, Jesse and Herrmann, Dorothee and Kipp, Gunda and Potts, Alexander M. and DeStefano, J. and Hu, Chaowei and Huang, Yunfei and Taniguchi, T. and Watanabe, Kenji and Meier, G. and Shin, Dongbin and Rubio, Angel and Chu, J. and Kennes, D. and Sentef, M. A. and McIver, J.},
  year = {2024},
  doi = {10.48550/arXiv.2409.04531},
  eprint = {2409.04531},
  archiveprefix = {arXiv}
}

@article{weitenberg_tailoring_2021,
  title = {Tailoring quantum gases by {Floquet} engineering},
  volume = {17},
  doi = {10.1038/s41567-021-01316-x},
  number = {12},
  journal = {Nat. Phys.},
  author = {Weitenberg, Christof and Simonet, Juliette},
  month = {dec},
  year = {2021},
  pages = {1342--1348}
}

@article{Ito2023Build-up,
  title = {Build-up and dephasing of {Floquet–Bloch} bands on subcycle timescales},
  author = {Ito, S. and Schüler, M. and Meierhofer, M. and Schlauderer, S. and Freudenstein, J. and Reimann, J. and Afanasiev, D. and Kokh, K. and Tereshchenko, O. and Güdde, J. and Sentef, Michael A. and Höfer, U. and Huber, R.},
  journal = {Nature},
  year = {2023},
  volume = {616},
  pages = {696-701},
  doi = {10.1038/s41586-023-05850-x}
}

@article{Choi2024Observation,
  title = {Observation of {Floquet–Bloch} states in monolayer graphene},
  author = {Choi, Dongsung and Mogi, Masataka and de Giovannini, U. and Azoury, Doron and Lv, Baiqing and Su, Yifan and Hübener, H. and Rubio, A. and Gedik, N.},
  journal = {Nat. Phys.},
  year = {2025},
  volume = {21},
  pages = {1100-1105},
  doi = {10.1038/s41567-025-02888-8}
}

@article{Shan2021Giant,
  title = {Giant modulation of optical nonlinearity by {Floquet} engineering},
  author = {Shan, Jun-Yi and Ye, M. and Chu, H. and Lee, Sungmin and Park, Je-Guen and Balents, L. and Hsieh, D.},
  journal = {Nature},
  year = {2021},
  volume = {600},
  pages = {235-239},
  doi = {10.1038/s41586-021-04051-8}
}

@article{Aditya2023Dynamical,
  title = {Dynamical localization and slow thermalization in a class of disorder-free periodically driven one-dimensional interacting systems},
  author = {Aditya, Sreemayee and Sen, D.},
  journal = {SciPost Phys. Core},
  year = {2023},
  volume = {6},
  pages = {083},
  doi = {10.21468/scipostphyscore.6.4.083}
}

@article{Castro2022Floquet,
  title = {{Floquet} engineering the band structure of materials with optimal control theory},
  author = {Castro, A. and de Giovannini, U. and Sato, Shunsuke A. and Hübener, H. and Rubio, {\'A}.},
  journal = {Phys. Rev. Res.},
  year = {2022},
  volume = {4},
  pages = {033213},
  doi = {10.1103/PhysRevResearch.4.033213}
}

@misc{Koenigsmann2025Nonequilibrium,
  title = {Nonequilibrium {DC} Current Generation in a Driven Dissipative {Haldane} Model},
  author = {Koenigsmann, Konrad and Bakshi, Sankha Subhra and Schauss, Peter and Chern, G.},
  year = {2025},
  doi = {10.48550/arXiv.2510.14157},
  eprint = {2510.14157},
  archiveprefix = {arXiv}
}

@article{PhysRevLett.115.075301,
  title = {Effective {Hamiltonians} for Rapidly Driven Many-Body Lattice Systems: Induced Exchange Interactions and Density-Dependent Hoppings},
  author = {Itin, A. P. and Katsnelson, M. I.},
  journal = {Phys. Rev. Lett.},
  volume = {115},
  issue = {7},
  pages = {075301},
  numpages = {6},
  year = {2015},
  month = {Aug},
  publisher = {American Physical Society},
  doi = {10.1103/PhysRevLett.115.075301}
}

@misc{Kusari2025Periodic,
  title = {Periodic Drive Induced Half-Metallic Phase in Insulators and Correlated Metals},
  author = {Kusari, Suryashekhar and Das, Arnab and Krishnamurthy, H. R. and Garg, Arti},
  year = {2025},
  doi = {10.48550/arXiv.2507.05935},
  eprint = {2507.05935},
  archiveprefix = {arXiv}
}

@article{Tamang2020Floquet,
  title = {{Floquet} engineering of low-energy dispersions and dynamical localization in a periodically kicked three-band system},
  author = {Tamang, L. and Nag, T. and Biswas, T.},
  journal = {Phys. Rev. B},
  year = {2021},
  volume = {104},
  pages = {174308},
  doi = {10.1103/PhysRevB.104.174308}
}

@misc{stefanini_is_2025,
  title = {Is {Lindblad} for me?},
  doi = {10.48550/arXiv.2506.22436},
  eprint = {2506.22436},
  archiveprefix = {arXiv},
  author = {Stefanini, Martino and Ziolkowska, Aleksandra A. and Budker, Dmitry and Poschinger, Ulrich and Schmidt-Kaler, Ferdinand and Browaeys, Antoine and Imamoglu, Atac and Chang, Darrick and Marino, Jamir},
  month = {jun},
  year = {2025}
}

@article{PhysRevB.94.155142,
  title = {{Lindblad}-driven discretized leads for nonequilibrium steady-state transport in quantum impurity models: Recovering the continuum limit},
  author = {Schwarz, F. and Goldstein, M. and Dorda, A. and Arrigoni, E. and Weichselbaum, A. and von Delft, J.},
  journal = {Phys. Rev. B},
  volume = {94},
  issue = {15},
  pages = {155142},
  numpages = {21},
  year = {2016},
  month = {Oct},
  publisher = {American Physical Society},
  doi = {10.1103/PhysRevB.94.155142}
}

@article{asboth_chiral_2014,
  title = {Chiral symmetry and bulk-boundary correspondence in periodically driven one-dimensional systems},
  volume = {90},
  doi = {10.1103/PhysRevB.90.125143},
  number = {12},
  journal = {Phys. Rev. B},
  author = {Asb{\'o}th, J. K. and Tarasinski, B. and Delplace, P.},
  month = {sep},
  year = {2014},
  pages = {125143}
}

@article{mendoza-arenas_coexistence_2015,
  title = {Coexistence of energy diffusion and local thermalization in nonequilibrium {XXZ} spin chains with integrability breaking},
  volume = {91},
  doi = {10.1103/PhysRevE.91.042129},
  number = {4},
  journal = {Phys. Rev. E},
  author = {Mendoza-Arenas, J. J. and Clark, S. R. and Jaksch, D.},
  month = {apr},
  year = {2015},
  pages = {042129}
}

@article{znidaric_thermalization_2010,
  title = {Thermalization and ergodicity in one-dimensional many-body open quantum systems},
  volume = {81},
  doi = {10.1103/PhysRevE.81.051135},
  number = {5},
  journal = {Phys. Rev. E},
  author = {{\v{Z}}nidari{\v{c}}, Marko and Prosen, Tomaž and Benenti, Giuliano and Casati, Giulio and Rossini, Davide},
  month = {may},
  year = {2010},
  pages = {051135}
}

@article{minganti_spectral_2018,
  title = {Spectral theory of {Liouvillians} for dissipative phase transitions},
  volume = {98},
  doi = {10.1103/PhysRevA.98.042118},
  number = {4},
  journal = {Phys. Rev. A},
  author = {Minganti, Fabrizio and Biella, Alberto and Bartolo, Nicola and Ciuti, Cristiano},
  month = {oct},
  year = {2018},
  pages = {042118}
}

@article{chen_periodically_2024,
  title = {Periodically driven open quantum systems: {Spectral} properties and nonequilibrium steady states},
  volume = {109},
  doi = {10.1103/PhysRevB.109.184309},
  number = {18},
  journal = {Phys. Rev. B},
  author = {Chen, Hao and Hu, Yu-Min and Zhang, Wucheng and Kurniawan, Michael Alexander and Shao, Yuelin and Chen, Xueqi and Prem, Abhinav and Dai, Xi},
  month = {may},
  year = {2024},
  pages = {184309}
}

@article{Anderson1959-iq,
  title   = {New approach to the theory of superexchange interactions},
  author  = {Anderson, P. W.},
  journal = {Phys. Rev.},
  volume  = {115},
  issue   = {1},
  pages   = {2--13},
  month   = {jul},
  year    = {1959},
  doi     = {10.1103/PhysRev.115.2}
}

@article{Winkler2006-hq,
  title   = {Repulsively bound atom pairs in an optical lattice},
  author  = {Winkler, K. and Thalhammer, G. and Lang, F. and Grimm, R. and Hecker Denschlag, J. and Daley, A. J. and Kantian, A. and B{\"u}chler, H. P. and Zoller, P.},
  journal = {Nature},
  volume  = {441},
  issue   = {7095},
  pages   = {853--856},
  month   = {jun},
  year    = {2006},
  doi     = {10.1038/nature04918}
}

@article{10.1021/jp505771f,
  title   = {Simple and Accurate Method for Time-Dependent Transport along Nanoscale Junctions},
  author  = {Chen, Liping and Hansen, Thorsten and Franco, Ignacio},
  journal = {J. Phys. Chem. C},
  volume  = {118},
  issue   = {34},
  pages   = {20009--20017},
  month   = {08},
  year    = {2014},
  doi     = {10.1021/jp505771f}
}

@article{Bhattacharjee2021Quantum,
  title = {Quantum thermal machines and batteries},
  author = {Bhattacharjee, Sourav and Dutta, Amit},
  journal = {Eur. Phys. J. B},
  volume = {94},
  pages = {239},
  year = {2021},
  doi = {10.1140/epjb/s10051-021-00235-3}
}

@article{Guzman2024Key,
  title = {Key issues review: useful autonomous quantum machines},
  author = {Guzm\'an, J. A. M. and Erker, P. and Gasparinetti, S. and Huber, M. and Halpern, N. Y.},
  journal = {Rep. Prog. Phys.},
  volume = {87},
  pages = {122001},
  year = {2024},
  doi = {10.1088/1361-6633/ad8803}
}

@article{Prosen_Znidaric_2009,
  title = {Matrix product simulations of non-equilibrium steady states of quantum spin chains},
  author = {Prosen, Toma{\v{z}} and \v{Z}nidari\v{c}, Marko},
  journal = {J. Stat. Mech.},
  volume = {2009},
  number = {02},
  pages = {P02035},
  year = {2009},
  month = {feb},
  doi = {10.1088/1742-5468/2009/02/P02035}
}

@article{Mendoza-Arenas_Znidaric_2019,
  title = {Asymmetry in energy versus spin transport in certain interacting disordered systems},
  author = {Mendoza-Arenas, J. J. and {\v{Z}}nidari{\v{c}}, M. and Varma, V. K. and Goold, J. and Clark, S. R. and Scardicchio, A.},
  journal = {Phys. Rev. B},
  volume = {99},
  issue = {9},
  pages = {094435},
  numpages = {6},
  year = {2019},
  month = {Mar},
  publisher = {American Physical Society},
  doi = {10.1103/PhysRevB.99.094435}
}

@article{Bao2025Universal,
  title = {Universal trade-off between irreversibility and intrinsic timescale in thermal relaxation with applications to thermodynamic inference},
  author = {Bao, Ruicheng and Du, Chaoqun and Cao, Zhiyu and Hou, Zhonghuai},
  journal = {Phys. Rev. E},
  volume = {112},
  issue = {4},
  pages = {044134},
  numpages = {14},
  year = {2025},
  month = {Oct},
  publisher = {American Physical Society},
  doi = {10.1103/fmsz-rdbj}
}

@misc{Tejero2025Asymmetric,
  title = {The role of entropy production and thermodynamic uncertainty relations in the asymmetric thermalization of open quantum systems},
  author = {Tejero, {\'A}lvaro},
  year = {2025},
  eprint = {2510.05072},
  archiveprefix = {arXiv},
  doi = {10.48550/arXiv.2510.05072}
}

@misc{Guo2026Information,
  title = {Information-geometric bounds on nonequilibrium relaxation in quantum {Markov} dynamics},
  author = {Guo, Xiao-Kan and Huang, Zhiqiang},
  year = {2026},
  eprint = {2609.09599},
  archiveprefix = {arXiv},
  doi = {10.48550/arXiv.2609.09599}
}

@article{Golod2022Demonstration,
  title = {Demonstration of a superconducting diode-with-memory, operational at zero magnetic field with switchable nonreciprocity},
  author = {Golod, Taras and Krasnov, Vladimir M.},
  journal = {Nat. Commun.},
  volume = {13},
  pages = {3658},
  year = {2022},
  doi = {10.1038/s41467-022-31256-w}
}

@article{De2024Confluence,
  title = {Confluence of fractured resonances at points of dynamical many-body flare},
  author = {De, Bitan and W{\'o}jtowicz, Gabriela and Rams, Marek M. and Zwolak, Michael and Zakrzewski, Jakub},
  journal = {Phys. Rev. B},
  volume = {110},
  pages = {155146},
  year = {2024},
  doi = {10.1103/PhysRevB.110.155146}
}

@article{Rams2020Breaking,
  title = {Breaking the entanglement barrier: Tensor network simulation of quantum transport},
  author = {Rams, Marek M. and Zwolak, Michael},
  journal = {Phys. Rev. Lett.},
  volume = {124},
  pages = {137701},
  year = {2020},
  doi = {10.1103/PhysRevLett.124.137701}
}

@inproceedings{Boerner2023ACCESS,
  title     = {{ACCESS}: Advancing Innovation: {NSF}'s Advanced Cyberinfrastructure Coordination Ecosystem: Services \& Support},
  author    = {Boerner, Timothy J. and Deems, Stephen and Furlani, Thomas R. and Knuth, Shelley L. and Towns, John},
  booktitle = {Practice and Experience in Advanced Research Computing (PEARC '23)},
  pages     = {173--176},
  year      = {2023},
  publisher = {ACM},
  address   = {New York, NY, USA},
  doi       = {10.1145/3569951.3597559}
}

@article{Rivera2016Shrinking,
  title   = {Shrinking light to allow forbidden transitions on the atomic scale},
  author  = {Rivera, Nicholas and Kaminer, Ido and Zhen, Bo and Joannopoulos, John D. and Solja\v{c}i\'{c}, Marin},
  journal = {Science},
  volume  = {353},
  number  = {6296},
  pages   = {263--269},
  year    = {2016},
  doi     = {10.1126/science.aaf6308}
}

@article{Benz2016Picocavities,
  title   = {Single-molecule optomechanics in ``picocavities''},
  author  = {Benz, Felix and Schmidt, Mikolaj K. and Dreismann, Alexander and Chikkaraddy, Rohit and Zhang, Yao and Demetriadou, Angela and Carnegie, Cloudy and Ohadi, Hamid and de Nijs, Bart and Esteban, Ruben and Aizpurua, Javier and Baumberg, Jeremy J.},
  journal = {Science},
  volume  = {354},
  number  = {6313},
  pages   = {726--729},
  year    = {2016},
  doi     = {10.1126/science.aah5243}
}

\end{document}